\documentclass[18pt]{article}
\usepackage{epsf}
\usepackage{color}
\usepackage{framed}
\usepackage{amsmath,amsfonts,amsbsy,amssymb}
\usepackage{indentfirst}
\usepackage{mathtools}
\usepackage{arydshln}

\newcommand{\nn}{\nonumber}
\def\rrangle{\rangle \hspace{-0.08cm} \rangle}
\def\llangle{\langle \hspace{-0.08cm} \langle}

\def\be{\begin{equation}}
\def\ee{\end{equation}}
\def\bef{\begin{framed}}
\def\eef{\end{framed}}
\def\bse{\begin{subequations}}
\def\ese{\end{subequations}}
\def\bal{\begin{align}}
\def\ealn{\end{align}}
\def\tr{\text{tr}}
\def\bs{\boldsymbol}
\def\ll{\langle}
\def\rl{\rangle}
\def\rrangle{\rangle\!\rangle}
\def\llangle{\langle\!\langle}
\def\mR{\mathbb{R}}
\def\mH{\mathbb{H}}

\begin{document}

\begin{titlepage}

\def\slash#1{{\rlap{$#1$} \thinspace/}}

\begin{flushright} 

\end{flushright} 

\vspace{0.1cm}

\begin{Large}
\begin{center}


{\bf $Sp(4;\mathbb{R})$  
Squeezing for Bloch Four-Hyperboloid  
\\
 via \\
The Non-Compact Hopf Map
}
\end{center}
\end{Large}

\vspace{1cm}

\begin{center}
{\bf Kazuki Hasebe}   \\ 
\vspace{0.5cm} 
\it{
National Institute of Technology, Sendai College,  
Ayashi, Sendai, 989-3128, Japan} \\ 

\vspace{0.5cm} 

{\sf
khasebe@sendai-nct.ac.jp} 

\vspace{0.8cm} 

{\today}

\end{center}

\vspace{1.0cm}

\begin{abstract}
\noindent

\baselineskip=18pt

We explore the hyperbolic geometry of squeezed states in the perspective of the non-compact Hopf map. Based on analogies between squeeze operation and $Sp(2,\mathbb{R})$ hyperbolic rotation,  two types of the squeeze operators, the (usual) Dirac- and the Schwinger-types, are introduced.  We clarify  the underlying hyperbolic geometry and $SO(2,1)$ representations of the squeezed states   along the line of the 1st non-compact Hopf map. 
Following to the geometric hierarchy of the non-compact Hopf maps, we extend the $Sp(2; \mathbb{R})$ analysis  to  $Sp(4; \mathbb{R})$ --- the isometry of an split-signature four-hyperboloid.     
We explicitly  construct the $Sp(4; \mathbb{R})$ squeeze operators in the Dirac- and Schwinger-types and investigate the  physical meaning of the four-hyperboloid coordinates in the context of  the Schwinger-type squeezed states.  
It is shown that the Schwinger-type $Sp(4;\mathbb{R})$ squeezed one-photon state  is equal to an entangled superposition state of two $Sp(2;\mathbb{R})$ squeezed states  and the corresponding concurrence has  a clear geometric meaning.   
Taking advantage of the group theoretical formulation,  basic properties of the $Sp(4;\mathbb{R})$ squeezed coherent states are also investigated. 
In particular, we show that the $Sp(4; \mathbb{R})$ squeezed vacuum naturally realizes a generalized  squeezing in a 4D manner.

\end{abstract}

\end{titlepage}

\newpage 

\tableofcontents

\newpage

\section{Introduction}

Qubit is a most fundamental object in the study of quantum information and quantum optics. Polarization of the qubit is specified by a point of the  Bloch sphere \cite{Bloch-1946}, and,   
in the Lie group  language of Perelomov \cite{Perelomov-1972}, the qubit is  the $SU(2)$  spin coherent state (of spin magnitude $1/2$) \cite{Arecchi-Courtens-Gilmore-Thomas-1972}.  
It is well known that the geometry of the Bloch sphere is closely related to the Hopf map 
\cite{Hopf-1931}: Qubit is a two-component normalized spinor geometrically representing $S^3$ and its overall $U(1)$ phase is not relevant to physics, so the physical space of the qubit is given by the projected space of the 1st Hopf map, $ S^3/U(1) \simeq S^3/S^1\simeq S^2$.   
 It is also reported that the 2nd and 3rd  Hopf maps that represent topological maps from spheres to spheres in different dimensions \cite{Hopf-1935}\footnote{As a review of the Hopf maps, see Ref.\cite{Hasebe-2010}  for instance.}  are sensitive to the entanglement of  qubits \cite{Mosseri-Dandoloff-2001,Bernevig-Chen-2003, Mosseri-2003}. 
Spherical geometries thus play important roles in describing the geometry  of quantum states. Beyond spheres,   one can find many applications of $\it{compact}$ manifolds in the  geometry of quantum states \cite{Bengtsson-Zyczkowski-2006}.    
Meanwhile,  hyperboloids or more generally $\it{non}$-$\it{compact}$ manifolds have been elusive in  applications  to  the study of geometry of quantum states,   
although a hyperbolic nature inherent to quantum mechanics is glimpsed in the Bogoliubov canonical  transformation that keeps the bosonic canonical commutation relations.\footnote{It is also recognized that the hyperbolic geometries naturally appear  in the  holographic interpretation of   
MERA \cite{Swingle-2009,Beny-2013}.}   
For $n$ species of bosonic operators, the  Bogoliubov transformation  is described by the symplectic group  $Sp(2n; \mathbb{R})$ \cite{Itzykson-1967,Berezin-1978,Perelomov-book}.\footnote{For $n$ species of fermionic operators, the canonical transformation is given by the special orthogonal group, $SO(2n)$ (Appendix \ref{subsec:bogtrans}).}  The simplest symplectic group is $Sp(2; \mathbb{R}) \simeq SU(1,1)$, which is  the double cover of  the $SO(2,1)$ isometry group of two-hyperboloid. 
Since $SU(1,1)$ is a non-compact counterpart of $SU(2)$, one can mathematically develop an argument similar to $SU(2)$: 
The  $SU(1,1)$ hyperbolic ``rotation'' gives rise to the  pseudo-spin coherent state \cite{Perelomov-1972, Perelomov-book, Gerry-1991, Sanders-2002, Combescure-Robert-book-2012}, and 
the $SU(1,1)$ pseudo-spin coherent state is specified by a position on the Bloch two-hyperboloid, $H^{2,0}$. 
What is interesting is that  the hyperbolic rotation is not a purely mathematical concept  but closely related to quantum optics as squeeze operation \cite{Gilmore-Yuan-1987, Dattoli-Dipace-Torre-1986, Gerry-1987}.   The squeeze operator or squeezed state  has more than forty year history, since its theoretical proposal in quantum optics \cite{Yuen-1976, Hollenhorst-1979, Caves-1981, Walls-1983, Schumaker-Caves-1985-1,Schumaker-Caves-1985-2}.  There are a number of  literatures about the squeezed state.   For instance,  $n$-mode generalization of the squeezed state  
was  investigated in Refs.
\cite{Milburn-1984, Gilmore-Yuan-1987, Bishop-Vourdas-1988, Ma-Rhodes-1990, Han-Kim-Noz-Yeh-1993,  Arvind-Dutta-Mukunda-Simon-1995-1,Arvind-Dutta-Mukunda-Simon-1995-2, Han-Kim-Noz-1995, Yukawa-Nemoto-2016}, and also  fermionic and  supersymmetric  squeezed states 
in \cite{Balantekin-Schmitt-Halse-1988, Buzano-Rasetti-Rastelli-1989, Balantekin-Schmitt-Halse-1989, Svozil-1990, Schmitt-Mufti-1991, Schmitt-1993}. Interested readers may consult Ref.\cite{Dodonov-2002} as a nice review of the history of squeezed states and references therein.    
  Here, we may encapsulate the above observation as    
\begin{align}
~~~~\text{Qubit state} ~~~~~~&\rightarrow~~~~~~~~~\text{Point on the Bloch sphere}~~~~~~~~~\rightarrow~~SU(2)~\text{spin coherent state} , \nn\\
&~~~~~~~~~~~~~~~~~~~~~~~~~~~~~~~~~~~~~~~~~~~~~~~~~~~~~~~~~~~~~~~~~~~~~~~~~~~~~~~~\downarrow\nn\\
\text{Squeezed state} ~~&\leftarrow~~~~~~\text{Point on the Bloch hyperboloid}~~~~~\leftarrow~~SU(1,1)~\text{pseudo-spin coherent state}. \nn
\end{align}
Interestingly, the hyperbolic Berry phase associated with the squeezed state was pointed out in \cite{Chiao-Jordan-1988, Kitano-Yabuzaki-1989}, and subsequently the hyperbolic Berry phase was  observed in  experiments \cite{Svensmark-Dimon-1994}. 
The geometry behind the hyperbolic Berry phase  
is 
 the 1st non-compact Hopf map, $H^{2,1}/U(1)\simeq H^{2,0}$. 

About a decade ago, the author proposed a non-compact version of the Hopf maps  based on the split algebras \cite{Hasebe-2009,Hasebe-2012}: 
\begin{center}
\begin{tabular}{ccccccc}
\\ 
~~~~~~ & & $H^{2,1}$ &   $\overset{H^{0,1}= S^{1}}\longrightarrow $ & $H^{2,0}$ & & ~~~~~~~~~~(1st)\\
~~~~~~ &  $H^{4,3}$ & $\longrightarrow$ & $H^{2,2}$ &  & & ~~~~~~~~~~(2nd) \\
~~~~~~ $H^{8,7}$ & $\longrightarrow$ &   $H^{4,4}$ &&  & & ~~~~~~~~~~(3rd) \\  
\end{tabular}
\end{center}
Just as in the original Hopf maps, the non-compact Hopf maps exhibit a dimensional hierarchy  in a hyperbolic manner. 
Taking advantage of  such hierarchical structure, we extend the  formulation of the squeezed states previously restricted to the $Sp(2; \mathbb{R})$ group to the $Sp(4; \mathbb{R})$ group based on the 2nd non-compact Hopf map.  
The base-manifold of the 2nd Hopf map is a split-signature four-hyperboloid, $H^{2,2}$, with isometry group  $SO(2,3)$ whose double cover  is  $Spin(2,3)~\simeq ~Sp(4; \mathbb{R})$ --- the next-simplest symplectic group of the Bogoliubov transformation for two bosonic operators \cite{Han-Kim-1998, Wunsche-2000}.   
The main goal of the present work is to construct the $Sp(4; \mathbb{R})$ squeezed state explicitly 
 and clarify its basic properties.  
To begin with,  we rewrite the single-mode and two-mode operators of $Sp(2; \mathbb{R})$  in a  perspective of the $SO(2, 1)$ group representation theory.   
We then observe   the following correspondences: 

\vspace{0.3cm}
~~~~~~~ {$Sp(2;\mathbb{R})$~one-/two-mode~squeezing} ~~$\longleftrightarrow$~~$SO(2,1)~\text{Majorana/Dirac~representation}$.  
\vspace{0.3cm}\\
For two-mode squeezing,   the $Sp(4; \mathbb{R})$ background symmetry  has been suggested in Refs.\cite{Milburn-1984, Gilmore-Yuan-1987, Bishop-Vourdas-1988, Ma-Rhodes-1990, Han-Kim-Noz-Yeh-1993,  Arvind-Dutta-Mukunda-Simon-1995-1,Arvind-Dutta-Mukunda-Simon-1995-2, Han-Kim-Noz-1995}. 
We will discuss that the $Sp(4; \mathbb{R})$ symmetry  is naturally realized in the context of  the Majorana representation of $SO(2,3)$. 
In a similar manner to the $Sp(2; \mathbb{R})$ case,  we introduce a four-mode squeeze operator  as Dirac representation of $SO(2; 3)$, 

\vspace{0.3cm}
~~~~~~ {$Sp(4;\mathbb{R})$~two-/four-mode~squeezing} ~~$\longleftrightarrow$~~$SO(2,3)~\text{Majorana/Dirac~representation}$,   
\vspace{0.3cm}\\
and   investigate their particular properties.   
 We introduce two types of squeeze operator, the (usual) Dirac- and Schwinger-type.\footnote{The ``Dirac-type'' of squeezing  has nothing to do with the ``Dirac representation'' of orthogonal group. The ``Schwinger-type'' of squeezing  has also  nothing to do with the ``Schwinger operator''.} In the case of $Sp(2; \mathbb{R})$ squeezing, the Dirac-  and the Schwinger-type squeeze operators generate  physically equivalent squeezed vacua, while in the case of  $Sp(4; \mathbb{R})$,  two types of  squeezing  generate physically distinct squeezed vacua.

It may be worthwhile to mention peculiar properties of  hyperboloids not observed in spheres. 
We can simply switch from  spherical geometry to  hyperbolic geometry by flipping several signatures of metric, but  hyperboloids have unique properties intrinsic to their non-compactness.      
First,  the non-compact isometry groups, such as $SO(2,1)$ and $SO(2,3)$,    
accommodate  Majorana representation, while  their  compact counterparts, $SO(3)$ and $SO(5)$, do not. 
Second, 
 unitary representations of non-compact groups are infinite dimensional and  very distinct from  finite unitary representations of  compact groups.  
Third,  non-compact groups exhibit more involved topological structures than those of their compact counterparts. 
For instance, the compact $USp(2)\simeq Spin(3)\simeq S^3$ is  simply connected, 
while $Sp(2; \mathbb{R})\simeq Spin(2,1) \simeq H^{2,1} \simeq R^2 \otimes S^1$ is not and leads to the  projective representation called  the metaplectic representation \cite{Simon-Mukunda-1993,Arvind-Dutta-Mehta-Mukunda-1994}. A similar relation  holds  for  $Sp(4; \mathbb{R})\simeq Spin(2,3)$ and  $USp(4)\simeq Spin(5)$.  

This paper is organized as follows. Sec.\ref{sec:unitarynoncomp} presents Hermitian realization of  non-compact algebra with pseudo-Hermiticity. The topology of symplectic groups is also reviewed.   We discuss the  $Sp(2; \mathbb{R})$ squeezing in the context of the 1st non-compact Hopf map and identify $Sp(2; \mathbb{R})$ one- and two-mode  operators with the $SO(2, 1)$ Majorana and Dirac representations in Sec.\ref{sec:sp2r}. Sec.\ref{sec:sp4r} gives the Majorana and Dirac representations of the $SO(2,3)$ group and the  factorization of  the $Sp(4; \mathbb{R})$ non-unitary coset matrix with emphasis on its relation to the non-compact 2nd Hopf map.  In Sec.\ref{sec:sp4rsqop}, we explicitly construct the $Sp(4;\mathbb{R})$ squeezed states and investigate their  properties.  We also extend the analysis to the $Sp(4; \mathbb{R})$ squeezed coherent states in Sec.\ref{sec:squcohstate}. Sec.\ref{sec:summary} is devoted to summary and discussions. 

\section{Pseudo-Hermitian matrices and symplectic group }\label{sec:unitarynoncomp}

 We develop a Schwinger boson  construction of  unitary operators for non-compact groups with pseudo-Hermiticity.\footnote{Non-compact group generally accommodates continuous representation as well as discrete representation. We focus on the discrete representation constructed by the Schwinger boson operator.} Topological structures of the symplectic groups and  ultra-hyperboloids are  also briefly reviewed.  

\subsection{Hermitian operators made of  the Schwinger bosons}\label{sec:hermschwin}

While unitary representations of  non-compact groups are not finite dimensional,  non-unitary representations are finite dimensional.    
Suppose that $t^a$ are  non-Hermitian matrices that satisfy the algebra 
\be
[t^a , t^b] =if^{abc}t_c, 
\label{algebratas}
\ee
where $f^{abc}$ denote the structure constants of the non-compact algebra. 
In the following, we assume that there exists a matrix $k$ that makes $kt^a$ be hermitian, 
\be
(k t^a)^{\dagger}=kt^a  
\label{kthermcond}
\ee
or 
\be
(t^a)^{\dagger}=k~ t^a ~(k^{\dagger})^{-1}. 
\label{pseudohermiticity}
\ee
Needless to say, it is not generally guaranteed  about the existence of such a matrix $k$. If there exists  $k$ satisfying (\ref{kthermcond}), the matrices $t^a$ are referred to as the pseudo-Hermitian matrices \cite{Mostafazadeh-2008, Sato-Hasebe-Esaki-Kohmoto-2011}.  
With the pseudo-Hermitian matrices, it is straightforward to construct Hermitian operators sandwiching the pseudo-Hermitian matrices by  the Schwinger boson operator $\hat{\phi}_{\alpha}$ and its conjugate:  
\be
X^a = \hat{\phi}_{\alpha}^{\dagger}~(k t^a)_{\alpha\beta}~\hat{\phi}_{\beta}= \hat{\phi}^{\dagger} ~kt^a \hat{\phi}=\bar{\hat{\phi}}~t^a \hat{\phi}, \label{phibarpixa}
\ee
where
\be
\bar{\hat{\phi}}\equiv \hat{\phi}^{\dagger}k. 
\ee
 We determine the commutation relations of the components $\hat{\phi}_{\alpha}$ so that 
 $X^a$  satisfy the same algebra as (\ref{algebratas}): 
\be
[X^a , X^b] =if^{abc}X_c.  
\ee
The commutation relations among $\hat{\phi}_{\alpha}$ are thus determined as   
\be
[\hat{\phi}_{\alpha}, \bar{\hat{\phi}}_{\beta}]=\delta_{\alpha\beta}, 
\ee
or 
\be
[\hat{\phi}_{\alpha}, \hat{\phi}^{\dagger}_{\beta}]=(k^{-1})_{\alpha\beta}.
\ee
Notice that while $t^a$ are  non-Hermitian matrices, $X^a$ are Hermitian operators. 
With generators $X^a$,  it is straightforward to construct elements of  non-compact group:   
\be
S=e^{-i\omega_a X^a}, 
\ee
with $\omega_a$  being group parameters.  
Obviously, $S$ is a  unitary operator 
\be
S^{\dagger}=S^{-1}.
\ee
From the  non-Hermitian matrix $t^a$, we can construct the non-unitary matrix element of the non-compact group as  
\be
M=e^{-i\omega_a t^a},
\ee
which satisfies the pseudo-unitary condition:   
\be
M^{\dagger}= k M^{-1} (k^{\dagger})^{-1}.
\ee
$X^a$ act to $\hat{\phi}$ as 
\be
[X^a, \hat{\phi}_{\alpha}]=-(t^a)_{\alpha\beta} \hat{\phi}_{\beta} 
\label{xacttophi}
\ee
or 
\be
[X^a, \bar{\phi}_{\alpha}]=\bar{\phi}_{\beta} (t^a)_{\beta\alpha}, 
\ee
which means that $\hat{\phi}$ behaves as the spinor representation of the non-compact group generated by $X^a$. We then have 
\be
S^{\dagger}~\hat{\phi}~ S= M\hat{\phi},  \label{relmands}
\ee
and 
\be
S~\bar{\hat{\phi}} ~S^{\dagger} =\bar{\hat{\phi}} ~M^{-1}, 
\ee
where 
\be
M^{-1} =e^{i\omega_a t^a}=k^{-1} M^{\dagger} k^{\dagger}.  
\ee
Notice that while  $S$ is a $\it{unitary}$ operator,   $M$ is a $\it{non}$-$\it{unitary}$ matrix. Both of them are specified by the same parameters $\omega_a$, and so there exists  one-to-one mapping between them. When $S$ acts to a normalized state $|n\rangle$ $(\langle n|n\rangle=1)$, the magnitude does not change under the transformation of the non-compact  group as shown by $\langle n|S^{\dagger} S |n\rangle=1$. In the matrix notation, however, the transformation does not preserve the magnitude of a normalized vector $\bs{n}$ ($\bs{n}^{\dagger}\bs{n}=1$) as implied by $\bs{n}^{\dagger}M^{\dagger} M\bs{n}\neq \bs{n}^{\dagger}\bs{n}$. This does not occur in  usual discussions of quantum mechanics for compact Lie groups, since we can realize the group elements by a finite dimensional unitary matrix. 
In non-compact Lie groups, finite dimensional unitary representation does not exist, 
however, when we adopt the unitary operator $S$ made by the Hermitian operators $X_a$, the probability conservation still holds, and so we do not need to worry about going beyond the usual probability interpretation of quantum mechanics.

In this paper, we mainly utilize the real symplectic groups $Sp(2n; \mathbb{R})=U(n; \mathbb{H}')$, and  we here summarize the basic properties of $Sp(2n; \mathbb{R})$ [see Appendix \ref{appen:sympmeta} also]. The generators of $Sp(2n; \mathbb{R})$ are represented by a $2n\times 2n$ matrix of the following form (\ref{matrgenesp2nr}): 
\be
X=\begin{pmatrix}
H & S^* \\
-S & -H^*
\end{pmatrix}, 
\ee 
where $H$ is a $n\times n$ Hermitian matrix and $S$ a $n\times n$ symmetric complex matrix. Though $X$ itself is  non-Hermitian in general, there obviously exists a matrix 
\be
K=\begin{pmatrix}
1_n & 0 \\
0 & -1_n
\end{pmatrix},  
\ee
which makes $X$ be  Hermitian: 
\be
KX=\begin{pmatrix}
H & S^* \\
S & H^*
\end{pmatrix}. 
\ee
In this sense, the $sp(2n; \mathbb{R})$ matrix generators are pseudo-Hermitian, and  we can construct the Hermitian  $sp(2n; \mathbb{R})$ operators by following the general method  discussed above.

\subsection{Topology of the symplectic groups  and ultra-hyperboloids}\label{subsec:toposymple}

Here, we review  geometric properties of the symplectic groups. 
The polar decomposition  of $Sp(2n;\mathbb{R})$ group  is given by\cite{Littlejohn-1986} 
\be
Sp(2n; \mathbb{R}) ~\simeq~ U(n) \otimes \mathbb{R}^{n(n+1)} ~\simeq~ U(1)\otimes SU(n) \otimes \mathbb{R}^{n(n+1)},    \label{polardecompsymp}
\ee
where $U(n)$ is the maximal Cartan subgroup of $Sp(2n; \mathbb{R})$. 
In particular, we have\footnote{The polar decomposition of $Sp(2; \mathbb{R})$ is well investigated in \cite{Simon-Mukunda-1993,Arvind-Dutta-Mehta-Mukunda-1994}.}   
\bse
\begin{align}
&Sp(2; \mathbb{R}) \simeq U(1) \otimes \mathbb{R}^2 \simeq S^1 \otimes \mathbb{R}^2 , \\
&Sp(4; \mathbb{R}) \simeq U(1)\otimes SU(2) \otimes \mathbb{R}^6\simeq S^1 \otimes S^3 \otimes \mathbb{R}^6 .  \label{sp2r4rtop}
\end{align}
\ese
The decomposition (\ref{polardecompsymp}) implies that the symplectic group is not simply connected: 
\be
\pi_1(Sp(2n ; \mathbb{R})) \simeq \pi_1(U(1))\simeq \mathbb{Z}. 
\ee
The double covering of the symplectic group is called the metaplectic group $Mp(2n ; \mathbb{R})$: 
\be
Mp(2n; \mathbb{R})/\mathbb{Z}_2 \simeq Sp(2n; \mathbb{R}), 
\ee
and its representation  is referred to as  the metaplectic representation which is  the  projective representation of the symplectic group. Note that  projective representation does not exist in the  compact group  counterparts of $Sp(2n; \mR)$, $i.e.$,  $USp(2n)$.\footnote{$USp(2n)=U(n; \mathbb{H})$ and $\pi_1(USp(2n))=1$. For instance, $USp(2)=SU(2)=Spin(3)$, $USp(4)=Spin(5)$.}

The coset spaces between the symplectic groups are given by 
\be
Sp(2n+2; \mathbb{H}')/Sp(2n; \mathbb{H}') ~\simeq ~H^{2n+2,2n+1}, 
\ee
where $H^{p,q}$ is referred to as the ultra-hyperboloid $H^{p,q}$\footnote{The anti-de Sitter,  de Sitter and Euclidean anti-de Sitter spaces are realized as the special cases of the ultra-hyperboloids:  
\bse
\begin{align}
&H^{2,0}=EAdS^2, ~~H^{1,1}=dS^2=AdS^2, \\
&H^{4,0}=EAdS^4, ~H^{3,1}=AdS^4 , ~H^{1,3}=dS^4.  
\end{align}
\ese
}  that is a $(p+d)$ dimensional manifold embedded in $\mathbb{R}^{p, q+1}$ as 
\be
\sum_{i=1}^p {x^i}{x^i} - \sum_{j=1
}^{q+1}x^{p+j}x^{p+j} =-1.  \label{condulthyperb}
\ee
(\ref{condulthyperb}) implies that as long as $x^{p+j}$ $(j=1,\cdots, q+1)$ is subject to the  condition of $q$-dimensional sphere with radius $\sqrt{1+\sum_{i=1}^p {x^i}{x^i}}$,  the remaining $p$ real coordinates  $x^{i}$ $(i=1,\cdots, p)$ can take any real numbers. Therefore, the topology of $H^{p,q}$ is identified with  a fibre-bundle made of  base-manifold $\mathbb{R}^p$ with fibre $S^q$ : 
\be
H^{p,q} \simeq~\mathbb{R}^p\otimes S^q. \label{hpqtopo}
\ee
In low dimensions, (\ref{hpqtopo}) yields
\bse
\begin{align}
&H^{2,0}\simeq \mathbb{R}^2 \simeq \mathbb{R}_+\otimes S^1, ~~~~H^{1,1}\simeq \mathbb{R}\otimes S^1,~~~~H^{0,2}\simeq S^2, \label{topolo2d}\\
&H^{4,0}\simeq \mathbb{R}^4, ~H^{3,1}\simeq \mathbb{R}^3\otimes S^1, ~~~H^{2,2}\simeq \mathbb{R}^2\otimes S^2, ~~~H^{1,3}\simeq \mathbb{R}^1\otimes S^3, ~~~H^{0,4}\simeq S^4. 
\end{align}
\ese
(\ref{condulthyperb}) also implies  that $H^{p,q}$ can be given by a coset between  indefinite orthogonal groups: 
\be
H^{p,q} ~\simeq~SO(p, q+1)/SO(p,q). 
\ee

\section{$Sp(2; \mathbb{R})$ group and squeezing}\label{sec:sp2r}

The isomorphism $Sp(2;\mathbb{R})~\simeq ~Spin(2,1)$ suggests that the $Sp(2; \mathbb{R})$ one- and two-mode operators are equivalent to the Majorana and the Dirac spinor operators of $SO(2,1)$.  Based on the identification of the squeeze operator with the $SU(1,1)\simeq Spin(2,1)$ ``rotation'' operator, we introduce two types of squeeze operators, the (usual) Dirac- and Schwinger-types. We discuss how  the non-compact 1st Hopf map is embedded in the geometry of the $Sp(2; \mR)$ squeezed state.

\subsection{$sp(2; \mathbb{R})$ algebra}\label{subsec:sp2r}

From the result of Sec.\ref{subsec:toposymple}, we have\footnote{Verification of $SU(1,1) \simeq H^{2,1}$ (\ref{relasp2r}) is not difficult. 
Since the $SU(1,1)$ group elements  satisfy 
\be
g^{\dagger}\sigma_z g =\sigma_z, ~~\det(g)=1, 
\ee
 the $SU(1,1)$ group elements  
 $g=\begin{pmatrix} 
\alpha & \beta \\
\beta^* & \alpha^*
\end{pmatrix}$ must obey the condition 
\be
1=|\alpha|^2 -|\beta|^2 ={\alpha_R}^2 +  {\alpha_I}^2 -{\beta_R}^2 -  {\beta_I}^2,   
\ee
which geometrically represents $H^{2,1}$. }   
\be
Mp(2; \mathbb{R})/\mathbb{Z}_2~\simeq~ Sp(2; \mR) ~\simeq~ SU(1,1)~~ \simeq ~Spin(2,1) ~\simeq~ H^{2,1} ~\simeq ~\mathbb{R}^2\times S^1,  \label{relasp2r}
\ee
and use the terminologies,  $SU(1,1)$ and $Sp(2;\mathbb{R})$,  interchangeably.   
 The $su(1,1)$ algebra is defined as 
\be
[T^i, T^j]=-i\epsilon^{ijk}T_k ~~~~(i,j,k=1,2,3)
\label{su11algebra}
\ee
with 
\be
g_{ij}=g^{ij}\equiv \text{diag}(-1,-1,+1),~~~\epsilon^{123}\equiv 1.
\ee
We adopt the finite dimensional matrix representation of the $su(1,1)$ generators :   
\be
\{\frac{1}{2}\tau^1, \frac{1}{2}\tau^2, \frac{1}{2}\tau^3\} =\{i\frac{1}{2}\sigma_x, i\frac{1}{2}\sigma_y, \frac{1}{2}\sigma_z\},  \label{su11nonhermat}
\ee
which satisfy 
\be
[\tau^i, \tau^j]=-2i\epsilon^{ijk}\tau_k, ~~~~\{\tau^i, \tau^j\} =-2g^{ij}.  
\ee
Note that $\tau^1$ and $\tau^2$ are chosen to be non-Hermitian. The completeness relation  is given by 
\be
(\tau^i)_{\alpha\beta}(\tau_i)_{\gamma\delta} =2\delta_{\alpha\delta}\delta_{\beta\gamma}-\delta_{\alpha\beta}\delta_{\gamma\delta}. \label{compkappa}
\ee
For later convenience, we introduce  the split-quaternions $q^m$ $(m=1,2,3,4)$\footnote{See Appendix.\ref{appen:splitq} for details.} that are related to the $su(1,1)$ matrices as  
\be
q^{m}=\{q^i, 1\}=\{-i\tau^i, 1\}=\{ \sigma_x, \sigma_y, -i\sigma_z, 1  \}, 
\label{quaternionicq}
\ee
and its quaternionic conjugate 
\be
\bar{q}^{m} =\{-q^i , 1 \} =\{i\tau^i, 1\}.  
\label{quaternionicbarq}
\ee
The $Sp(2;\mR)$ is isomorphic to the split-quaternionic unitary  group $U(1; \mH')$, and in general the real symplectic group is isomorphic to the split-quaternionic unitary  group,  $Sp(2n; \mR)\simeq U(n; \mH')$ (see Appendix \ref{appen:sympmetasplith}). 

As mentioned in Sec.\ref{sec:hermschwin}, the $sp(2;\mR)\simeq su(1,1)$ finite dimensional matrix generators  
(\ref{su11nonhermat}) are pseudo-Hermitian matrices: With 
\be
\kappa= \sigma_z, 
\ee
we can construct the corresponding Hermitian matrices as 
\be
\kappa^i \equiv \kappa \tau^i=\{-\sigma_y,\sigma_x, 1\}.  
\ee
$\kappa^i$ have the following properties 
\be
\kappa^{[i}\sigma_z \kappa^{j]} =-2i\epsilon^{ijk}\kappa_k, ~~~(\kappa^i)_{\alpha\beta}(\kappa_i)_{\gamma\delta} =2(\sigma_z)_{\alpha\delta}(\sigma_z)_{\beta\gamma}-(\sigma_z)_{\alpha\beta}(\sigma_z)_{\gamma\delta} \label{compkapparels}
\ee
where $\kappa^{[i}\sigma_z \kappa^{j]} \equiv \kappa^i\sigma_z \kappa^j-\kappa^j\sigma_z \kappa^i$. 
Since $\kappa^i$ are Hermitian, one may immediately see that $g=e^{i\omega_i\frac{1}{2}\tau^i}$ satisfies 
\be
g^{\dagger}~\sigma_z ~g=\sigma_z, 
\ee
which is one of the  relations that the $SU(1,1)$ group elements should satisfy.  
Following the general prescription in Sec.\ref{sec:unitarynoncomp}, we construct the  $su(1,1)$  Hermitian operators. We introduce the  two-component Schwinger boson operator subject to the condition 
\be
[\hat{\phi}_{\alpha}, \hat{\phi}_{\beta}]=(\sigma_z)_{\alpha\beta}.  
\label{cophisu11h}
\ee
(\ref{cophisu11h}) is readily satisfied when we choose 
\be
\hat{\phi}=
\begin{pmatrix}
\hat{\phi}_1 \\
\hat{\phi}_2
\end{pmatrix} = \begin{pmatrix}
a \\
b^{\dagger}
\end{pmatrix}, \label{diracschwin}
\ee
with $a$ and $b$ being two independent Schwinger operators: 
\be
[a, a^{\dagger}]=[b, b^{\dagger}]=1, ~~~~[a,b]=[a,b^{\dagger}]=0.
\label{commretwomodes}
\ee
The Hermitian $su(1,1)$ operators are then constructed as 
\be
T^i =\frac{1}{2}~\hat{\phi}^{\dagger}~ \kappa^i ~\hat{\phi} , 
\label{su11opdirac}
\ee
or 
\be
T^x= i\frac{1}{2}(-ab +a^{\dagger}b^{\dagger}) ,~~T^y= \frac{1}{2}(ab +a^{\dagger}b^{\dagger})     ,    ~~~T^z= \frac{1}{2}(a^{\dagger}a  + b^{\dagger}b )+\frac{1}{2}.
\ee
In quantum optics, These operators are usually referred to as the two-mode $su(1,1)$ operators \cite{Wodkiewicz-Eberly-1985, Yurke-McCall-Klauder-1986}. Using (\ref{compkappa}) and (\ref{cophisu11h}), we can easily derive the corresponding $SU(1,1)$ Casimir operator:  
\be
C = -(K^1)^2-(K^2)^2+(K^3)^2= \frac{1}{4}~(\bar{\hat{\phi}} \hat{\phi})\cdot  (\bar{\hat{\phi}}\hat{\phi} +2). 
\ee
$\hat{\phi}$ transforms as a spinor representation of $SO(2,1)$: 
\be
e^{-i\omega_i T^i}~\hat{\phi} ~e^{i\omega_i T^i} = e^{i\omega_i\frac{1}{2}\tau^i}~\hat{\phi}.  \label{covaphisu11}
\ee
Since $\hat{\phi}$ is a complex spinor, $\hat{\phi}$ realizes the Dirac (spinor) representation of $SO(2,1)$. 

The  $SO(2,1)$ group also accommodates the Majorana representation. 
For $SO(2,1)$,  there exists a charge conjugation matrix 
\be
C= \sigma_x  
\ee
that satisfies the relation   
\be
-(\tau^i)^* =C \tau^i C. 
\ee
Imposing the Majorana condition on $\hat{\phi}$  
\be
\hat{\phi}^* =C~\hat{\phi},  
\ee
we obtain  the identification 
\be
b =a.   
\label{majoranaident}
\ee
The Majorana spinor operator is thus constructed as 
\be
\hat{\varphi}=\begin{pmatrix}
a \\
a^{\dagger}
\end{pmatrix}, \label{varphimaj}
\ee
which satisfies
\be
[\hat{\varphi}_{\alpha}, \hat{\varphi}_{\beta}] =\epsilon_{\alpha\beta}.  \label{commrevarphi}
\ee
Note that the previous commutation relations (\ref{commretwomodes}) do not change under the identification (\ref{majoranaident}) except for  
\be
[a, b^{\dagger}]=0~~\rightarrow~~[a, a^{\dagger}]=1. 
\label{changecommide}
\ee
From the Majorana operator (\ref{varphimaj}), we can construct the corresponding $su(1,1)$ generators (\ref{su11opdirac}) as      
\be
T^i =\frac{1}{4}~\hat{\varphi}^t ~m^i ~\hat{\varphi}, 
\label{majosu11}
\ee
where 
\be
m^i =\sigma_x\kappa^i = -i\sigma_y \tau^i=\{ -i\sigma_z,  1_2, \sigma_x\}.  \label{defmi}
\ee
(\ref{majosu11}) are explicitly given by 
\be
T^x =i\frac{1}{4}(-a^2+{a^{\dagger}}^2),~~~T^y =\frac{1}{4}(a^2+{a^{\dagger}}^2), ~~~T^z=\frac{1}{2}a^{\dagger}a+\frac{1}{4}. 
\label{tsfromonemode}
\ee
 In quantum optics, 
such Majorana spinor operator is referred to as the   one-mode $su(1,1)$ operator \cite{Wodkiewicz-Eberly-1985, Yurke-McCall-Klauder-1986}.   
 $m^i$ (\ref{defmi}) are symmetric matrices $((m^i)^t=m^i)$ that satisfy 
\be
i ~m^{[i} \sigma_y m^{j]} =-2i\epsilon^{ijk}m_k, ~~(m^i)_{\alpha\beta}(m_i)_{\gamma\delta} =-2\epsilon_{\alpha\delta}\epsilon_{\beta\gamma}-\epsilon_{\alpha\beta}\epsilon_{\gamma\delta}, 
\label{commmi}
\ee
where $m^{[i} \sigma_y m^{j]} \equiv m^i \sigma_y m^{j}-m^{j} \sigma_y m^{i}$.   
 It is not difficult to verify that (\ref{majosu11}) satisfies the $su(1,1)$ algebra (\ref{su11algebra}) with (\ref{commrevarphi}) and $m^i$ (\ref{commmi}). 
$\hat{\varphi}$ also transforms as the spinor representation of $SO(2,1)$: 
\be
e^{-i\omega_i T^i}~\hat{\varphi} ~e^{i\omega_i T^i} = e^{i\omega_i\frac{1}{2}\tau^i}~\hat{\varphi},  \label{covavaphisu11}
\ee
and the $SU(1,1)$ Casimir for the Majorana representation becomes a constant: 
\be
C=T^i T_i=-(T^x)^2-(T^y)^2+(T^z)^2=-\frac{3}{16} .  \label{consusu11cas}
\ee
(\ref{tsfromonemode}) realizes the generators of $Mp(2; \mathbb{R})$.  
Indeed, the independent operators of (\ref{tsfromonemode}) can be taken as the all possible symmetric combinations between $a$ and $a^{\dagger}$, $i.e.$, $\{a, a\}, \{a^{\dagger}, a^{\dagger} \}$ and $\{a, a^{\dagger}\}$, which are the $Mp(2;\mathbb{R})$ operators (see Appendix \ref{app:metgr}).  
Notice that the factor $1/4$ in the Majorana representation (\ref{majosu11}) is  half of the coefficient $1/2$ of the Dirac representation (\ref{su11opdirac}), which is needed to compensate the change  of the commutation relation (\ref{changecommide}).  
Since the 1/2 change of the scale of the  coefficients,    the parameter range for the $Mp(2; \mR)$ operators should be taken twice of that for the Dirac operator  implying that  $Mp(2; \mR)$ is the double cover of the $Sp(2; \mR)$.

\subsection{The squeeze operator and the 1st non-compact Hopf map}\label{subsec:1stnoncomsqueeze}

Using the $su(1,1)$  ladder operators 
\be
T^{\pm} \equiv T^y \mp i T^x , 
\ee
the  squeeze operator  is given by 
\be
S(\xi) =e^{-\xi T^+ +\xi^* T^-}, 
\ee
with an arbitrary complex parameter  $\xi$: 
\be
\xi =\frac{\rho}{2}e^{i\phi}. 
\ee
Here, $\rho \in [0, \infty)$ and $\phi=[-\pi, \pi)$. 
We will see that the two parameter of $\rho$ and $\phi$ are naturally interpreted as the coordinates on the Bloch two-hyperboloid $H^{2,0}$. 
For single-mode and two-mode operators, the ladder operators are respectively given by 
\be
T^+ =\frac{1}{2}{a^{\dagger}}^2, ~~~T^- =\frac{1}{2}a^2, 
\ee
and 
\be
T^+=a^{\dagger}b^{\dagger}, ~~~T^-=ab.
\ee
Recall that the squeeze operation  acts to the two- and one-mode operators  as  
\be
S^{\dagger}~\hat{\phi}~ S =M~ \hat{\phi}, ~~~~~~~S^{\dagger}~\hat{\varphi}~ S =M ~\hat{\varphi}.  \label{squonem}
\ee

\begin{figure}[tbph]
\center
\includegraphics*[width=60mm]{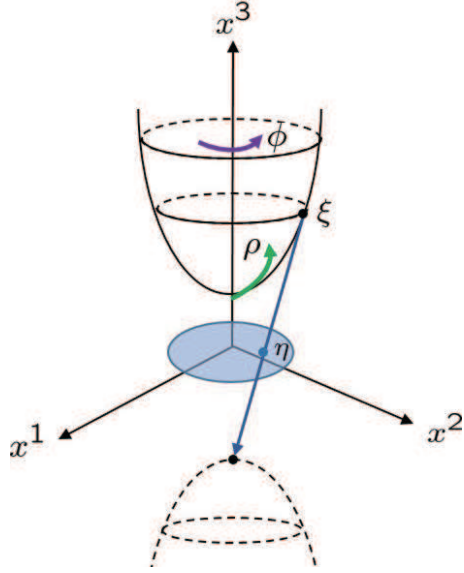}
\caption{The upper-leaf of Bloch two-hyperboloid $H^{2,0}$ : $-(x^1)^2-(x^2)^2+(x^3)^2=1$.  The regions of the parameters  are  $\rho \in [0,\infty)$ and $\phi \in [0, 2\pi)$ realizing $H^{2,0} \simeq \mathbb{R}_+ \otimes S^1$ (\ref{topolo2d}). The blue shaded region stands for the Poincar\'{e} disc.  }
\label{h20.fig}
\end{figure}

It is not convenient to handle the $su(1,1)$ ladder operators directly  to derive    factorization form of the squeeze operator $S$. A wise way to do so is  to utilize the non-unitary matrix $M$ that has  one-to-one correspondence to the squeeze operator. 
 Based on   simple $Sp(2; \mathbb{R})$ matrix manipulations,  it becomes feasible to obtain the factorization form of $M$, and once we were able to derive the factorization form, we could apply it to the squeeze operator according to the correspondence between the non-hermitian matrix generators and operators.  
For the squeeze operator $S(\xi)$, we introduce the non-unitary squeeze matrix: 
\be
M(\rho,\phi) = e^{-{\xi} t^+ + {\xi^*} t^-} , 
\label{nonunimmat}
\ee
where 
\be
t^{+} \equiv \frac{1}{2}(\tau^y - i \tau^x) =\begin{pmatrix}
0 & 1 \\
0 & 0 
\end{pmatrix}  , ~~~~ t^{-} \equiv \frac{1}{2}(\tau^y + i \tau^x )= -\begin{pmatrix}
0 & 0 \\
1 & 0 
\end{pmatrix}.   
\ee
$M$ is given by 
\be
M(\rho,\phi)
=
 e^{-i\frac{\rho}{2}\sum_{i=1,2}{{n}}_i{\tau}^i} =\begin{pmatrix}
\cosh\frac{\rho}{2} & -\sinh\frac{\rho}{2}~e^{i\phi} \\
-\sinh\frac{\rho}{2}~e^{-i\phi} & \cosh\frac{\rho}{2}
\end{pmatrix}, \label{cosetsu11hyp2}
\ee
where 
\be
{n}_1 =-\cos\phi, ~~{n}_2=\sin\phi ~~\in S^1.
\ee
The first expression on the right-hand side of (\ref{cosetsu11hyp2}) gives an intuitive interpretation of the squeezing: $M$ operators as a hyperbolic rotation by the ``angle'' $\rho$ around the axis ${\bs{n}}=-\cos\phi ~\bs{e}_x +\sin\phi ~\bs{e}_y$.  
For later convenience, we also mention  field theory technique  to realize a  matrix representation for the coset space  associated with the symmetry breaking $G~\rightarrow ~H$.  
Say  $t^i$ are 
the broken generators of the symmetry breaking, and   the coset manifold $G/H$ is represented by the matrix valued quantity\footnote{In  field theory, non-compact  manifolds with indefinite signature are usually not of interest,  because field theories on non-compact manifolds generally suffer from the existence of negative norm states, $i.e.$, the ghosts. In the present case, we are not dealing with field theory, and so either non-compactness or indefinite signature is not a problem.} 
\be
e^{-i\omega_i t^i}. \label{cosetmatgh}
\ee
In the perspective of $G/H$, the squeeze matrix (\ref{cosetsu11hyp2}) corresponds to (\ref{cosetmatgh}) when  the original symmetry is $G=SU(1,1)$ is spontaneously  broken to  $H =U(1)$, and the  broken generators are given by  $\frac{1}{2}\tau^1$ and $\frac{1}{2}\tau^2$. 
The squeeze matrix $M$  thus corresponds to  the coset 
\be
SU(1,1)/U(1) \simeq H^{2,0}.  
\label{groupnon1st}
\ee
Using hyperboloids,  (\ref{groupnon1st}) can be expressed as  
\be
H^{2,1}/S^1 \simeq H^{2,0}, 
\ee
which is exactly  the 1st non-compact Hopf map. We now discuss the geometric meaning of the parameters  $\rho$ and $\phi$ of (\ref{cosetsu11hyp2}). 
With $SU(1,1)$ group element $g$ satisfying 
$g^{\dagger}\sigma_z g =\sigma_z$ and $\det(g)=1$,  
the non-compact 1st Hopf map is realized as 
\be
g ~\in~ SU(1,1)~ \simeq ~H^{2,1}~~\rightarrow~~ x^i =\frac{1}{2}\tr(\sigma_z g^{-1} \tau^i g) =\frac{1}{2}\tr (g^{\dagger}\kappa^i g) ~\in~ H^{2,0}.   
\label{noncompact1st}
\ee
$x^i$ are invariant under the $U(1)$ transformation $g~\rightarrow~g ~e^{i\frac{\chi}{2} \tau^3}$,   
and automatically satisfy the condition of $H^{2,0}$: 
\be
x^i x_i=-(x^1)^2 -(x^2)^2+(x^3)^2=2(g^{\dagger}\sigma_z g)^2 -(g^{\dagger}\sigma_z g)^2 =1. 
\ee
In the analogy to the Euler angle decomposition of  $SU(2)$, the $SU(1,1)$ group element may be expressed as 
\begin{align}
g(\phi, \rho, \chi) =e^{i\frac{\phi}{2} \tau^z} e^{-i\frac{\rho}{2}\tau^x} e^{i\frac{\chi}{2} \tau^z}
=
\begin{pmatrix}
\cosh \frac{\rho}{2} ~ e^{i\frac{1}{2}(\phi+\chi)} &    \sinh \frac{\rho}{2} ~ e^{i\frac{1}{2}(\phi-\chi)} \\
 \sinh \frac{\rho}{2} ~ e^{-i\frac{1}{2}(\phi-\chi)}          & \cosh \frac{\rho}{2} ~ e^{-i\frac{1}{2}(\phi+\chi)}
\end{pmatrix}, 
\label{parasu11nonherm}
\end{align}
where 
\be
\rho =[0, \infty),~~\phi=[0, 2\pi),~~ \chi=[0, 4\pi). 
\ee
The coordinates on the two-hyperboloid (\ref{noncompact1st}) are explicitly derived as 
\be
x^1=\sinh\rho\sin\phi,~~~x^2=\sinh\rho\cos\phi,~~~x^3=\cosh\rho~(\ge 1). 
\ee
The parameters $\rho$ and $\phi$ thus represent the coordinates of the upper-leaf of the ``Bloch'' two-hyperboloid (Fig.\ref{h20.fig}). 
Notice that the squeeze matrix (\ref{cosetsu11hyp2}) is realized as a special case of $g$ (\ref{parasu11nonherm}): 
\be
M(\rho,\phi)= \begin{pmatrix}
\cosh\frac{\rho}{2} & -\sinh\frac{\rho}{2}~e^{i\phi} \\
-\sinh\frac{\rho}{2}~e^{-i\phi} & \cosh\frac{\rho}{2}
\end{pmatrix} =g(\phi, -\rho, -\phi) . \label{su11dirac}
\ee

In (\ref{parasu11nonherm}), the $U(1)$ fibre part $e^{i\frac{\chi}{2}\tau^3}$ represents the gauge degrees of freedom. Following the terminology of the $SU(2)$ case \cite{Felsager1998,Hasebe-2016}, we refer to the gauge $\chi =\phi$  as the Dirac-type and  $\chi=0$ as the Schwinger-type.   The Dirac-type $SU(1,1)$ element corresponds to the squeeze matrix  as demonstrated by (\ref{su11dirac}). 
Meanwhile for the Schwinger-type, we introduce a new squeeze matrix   
\be
\mathcal{M}(\rho,\phi) \equiv g(\phi, -\rho, 0)  =e^{i\frac{\phi}{2} \tau^z} \cdot e^{i\frac{\rho}{2}\tau^x}
=
\begin{pmatrix}
\cosh \frac{\rho}{2}~e^{i\frac{\phi}{2}}  &    -\sinh \frac{\rho}{2} ~ e^{i\frac{\phi}{2}} \\
 -\sinh \frac{\rho}{2} ~ e^{-i\frac{\phi}{2}}          & \cosh \frac{\rho}{2}~e^{-i\frac{\phi}{2}}
\end{pmatrix}.   \label{su11schwin}
\ee
Using the non-compact Hopf spinors\footnote{
(\ref{noncompa1sthopfspinos}) can be rewritten as 
\be
\psi_L = \frac{1}{\sqrt{2}} \begin{pmatrix}
\sqrt{x^3+1} \\
\sqrt{x^3-1} ~(y^2-iy^1)
\end{pmatrix},~~~~\psi_R = \frac{1}{\sqrt{2}} \begin{pmatrix}
\sqrt{x^3-1}~ (y^2+ iy^1) \\
\sqrt{x^3+1} 
\end{pmatrix},
\ee
with $y^1$ and $y^2$ being the normalized coordinates of the latitude of $H^{2,0}$~: 
\be
(y^{1}, y^2) \equiv  \frac{1}{\sqrt{(x^3)^2-1}} (x^1, x^2) =(\sin\phi, \cos\phi).  
\ee
} \cite{Hasebe-2009} 
\be
\psi_L =\frac{1}{\sqrt{2(x^3+1)}} \begin{pmatrix}
x^3+1 \\
x^2-ix^1
\end{pmatrix} , ~~~\psi_R =\sigma_x {\psi_L}^* = \frac{1}{\sqrt{2(x^3+1)}} \begin{pmatrix}
x^2+ix^1 \\
x^3+1 
\end{pmatrix}, \label{noncompa1sthopfspinos}
\ee
which satisfy ${\psi_L}^{\dagger}\kappa^i \psi_L ={\psi_R}^{\dagger}\kappa^i \psi_R  =x^i $, the Dirac-type squeeze matrix (\ref{su11dirac}) can be represented as 
\be
M =\begin{pmatrix}
\psi_L & \psi_R
\end{pmatrix}. 
\ee
Both $M$ and $\mathcal{M}$ are pseudo-unitary matrices:  
\bse
\begin{align}
&M(\rho, \phi)^{-1}  =\sigma_z~ M(\rho,\phi)^{\dagger}~\sigma_z = M(-\rho, \phi), \label{invnon1}\\
&\mathcal{M}(\rho, \phi)^{-1} =\sigma_z ~\mathcal{ M}(\rho,\phi)^{\dagger}~\sigma_z \neq  \mathcal{M}(-\rho, \phi) . \label{invnon2}
\end{align}
\ese

The replacement of the non-Hermitian matrices $t^i$ with the Hermitian operators $T^i$ transforms the squeeze matrix 
$M$ to the (usual) Dirac-type squeeze operator \cite{Hollenhorst-1979,Caves-1981,Walls-1983}:  
\be
S(\xi) =e^{-\xi T^+ +\xi^* T^-}=e^{i\phi T^3}e^{i\rho T^1}e^{-i\phi T^3}, 
\label{squeezediracope1}
\ee
which satisfies 
\be
S(\xi)^{\dagger} =S(-\xi)=S(\xi)^{-1}. 
\ee
 In deriving a number state expansion of the squeezed state, the Gauss decomposition  is quite useful \cite{Arecchi-Courtens-Gilmore-Thomas-1972}. The Gauss decomposition of the $Sp(2; \mathbb{R})$ squeeze operator  is given by\footnote{
The faithful ($i.e.$, one-to-one) matrix representation  of  the operator, $e^{\alpha T^+}e^{\beta T^3}e^{\gamma T^-}$, is given by  
\be
e^{\alpha t^+}e^{\beta t^3}e^{\gamma t^-}=
\begin{pmatrix}
1 & \alpha \\
0 & 1 
\end{pmatrix} \begin{pmatrix}
e^{\frac{\beta}{2}} & 0 \\
0 & e^{-\frac{\beta}{2}}
\end{pmatrix} 
\begin{pmatrix}
 1 & 0 \\
-\gamma & 1 
\end{pmatrix} =
\begin{pmatrix}
-\alpha~\gamma ~e^{-\frac{\beta}{2}} +e^{\frac{\beta}{2}} & \alpha ~e^{-\frac{\beta}{2}} \\
-\gamma~ e^{-\frac{\beta}{2}} & e^{-\frac{\beta}{2}}
\end{pmatrix} .  \label{faithsp2rmat}
\ee
The Gauss UDL decompositions, (\ref{diracnormasq}) and (\ref{schnormasq}),  are obtained by comparing (\ref{faithsp2rmat}) with (\ref{su11dirac}) and (\ref{su11schwin}), respectively. As emphasized in \cite{Arecchi-Courtens-Gilmore-Thomas-1972, Zhang-Feng-Gilmore-1990, Gilmore-2008}, the faithful representation preserves the group product, so the obtained matrix decompositions for the faithful  representation $\it{generally}$ hold in other  representations.   
} 
\be
S(\xi) = e^{-\eta T^+} ~e^{\ln (1-|\eta|^2) T^3}~ e^{\eta^* T^-}=e^{-\eta T^+} e^{-2\ln (\cosh\frac{\rho}{2})~ T^3 } e^{ \eta^* T^-} . \label{diracnormasq}
\ee
Here, $\eta$ is 
\be
\eta 
\equiv \tanh|\xi| ~\frac{\xi}{|\xi|}=\tanh\frac{\rho}{2} ~ e^{i\phi}=\frac{x^2+ix^1}{1+x^3},  
\ee
which also has a geometric meaning as  the stereographic coordinates on the Poincar\'e disc from $H^{2,0}$ (see Fig.\ref{h20.fig}). 

\subsection{Squeezed states}

We introduce the squeeze operator corresponding to the Schwinger-type squeeze matrix $\mathcal{M}$ (\ref{su11schwin}) : 
\be
\mathcal{S}(\xi) =e^{i{\phi} T^3} e^{i\rho T^1} ,  
\label{squeezeschope2}
\ee
which is a unitary operator 
\be
\mathcal{S}(\xi)^{\dagger} =\mathcal{S}(\xi)^{-1} \neq \mathcal{S}(-\xi). 
\ee
The Gauss decomposition is derived as  
\be
\mathcal{S}(\xi) =e^{-\eta T^+} e^{(\ln (1-|\eta|^2) +i\text{arg}(\eta) )T^3} e^{|\eta|T^-}  = e^{-\eta T^+} e^{-2\ln (\cosh\frac{\rho}{2})~ T^3 +i\phi ~ T^3} e^{|\eta|T^-} . \label{schnormasq}
\ee
The two types of the squeeze operator, (\ref{squeezediracope1}) and (\ref{squeezeschope2}), are related as 
\be
S(\xi) =\mathcal{S}(\xi)~e^{-i\phi T^z}. 
\ee
In literature, the Dirac-type squeeze operator $S$ is usually adopted, but there may be no special reason not to adopt $\mathcal{S}$, since at the level of non-unitary squeeze matrix,   both $M$ and $\mathcal{M}$ denote the coset $H^{2,0}$.  

Since $T^z$ is diagonalized for the number-basis states, the one-mode Dirac- and Schwinger-type squeezed number states\footnote{
The number state expansions of the  single-mode  squeezed vacuum and squeezed one-photon state are respectively given by 
\be 
|\xi\rangle_{(0)} = \frac{1}{\sqrt{\cosh \frac{\rho}{2}}} \sum_{n=0}^{\infty} (-\frac{\eta}{2})^n \frac{\sqrt{(2n)!}}{n!}~|2n\rangle,~~|\xi\rangle_{(1)} = \frac{1}{\sqrt{\cosh \frac{\rho}{2}}^3} \sum_{n=0}^{\infty} (-\frac{\eta}{2})^n \frac{\sqrt{(2n+1)!}}{n!}~|2n+1\rangle.  \label{zroonesques1}
\ee
} 
\be
|\xi\rangle_{(n)} \equiv S(\xi) |n\rangle, ~~~~~|\xi\rrangle_{(n)} \equiv \mathcal{S}(\xi) |n\rangle,
\ee
are merely different by a $U(1)$ phase: 
\be
|\xi\rangle_{(n)} =  e^{-i\frac{\phi}{4}}e^{-i\frac{\phi}{2} n} \cdot|\xi\rrangle_{(n)}, 
\label{u1diffxionemo}
\ee
where $|n\rangle =\frac{1}{\sqrt{n!}}{a^{\dagger}}^n|0\rangle.$ 
Similarly for two-mode, 
the squeezed number states are related as\footnote{
For two-modes, 
the squeezed number states are  given by\cite{Gerry-1991,Schumaker-Caves-1985-1,Schumaker-Caves-1985-2} 
\be
\hspace{-0.5cm}|\xi\rangle_{(n, 0 )}  = \biggl(\frac{1}{{\cosh \frac{\rho}{2}}}\biggr)^{n+1}\sum_{m=0}^{\infty}(-\eta)^m \sqrt{\frac{(n+m )!}{~n!~m! }}~ |n+m, m\rangle, ~~|\xi \rangle_{(0, n )}  = \biggl(\frac{1}{{\cosh \frac{\rho}{2}}}\biggr)^{n+1}\sum_{m=0}^{\infty}(-\eta)^m \sqrt{\frac{(n+m )!}{~n!~m! }}~ |m, n+m\rangle. \label{twomodesp2rsqunumb}
\ee
In particular for the squeezed vacuum state, we have  
\be
|\xi\rangle_{(0,0)} =\frac{1}{{\cosh \frac{\rho}{2}}} \sum_{m=0}^{\infty}(-\eta)^m  |m, m\rangle. 
\ee
} 
\be
|\xi\rangle_{(n_a, n_b)}  = e^{-i\frac{\phi}{2}}e^{-i\frac{\phi}{2} (n_a+n_b)} \cdot  |\xi\rrangle_{(n_a, n_b)}
\ee
where 
\be
|\xi\rangle_{(n_a, n_b)} \equiv S(\xi) |n_a, n_b \rangle , ~~~~~
|\xi\rrangle_{(n_a, n_b)} \equiv \mathcal{S}(\xi) |n_a, n_b\rangle, 
\label{sqnu2} 
\ee
with 
$|n_a,n_b\rangle =\frac{1}{\sqrt{n_a ! n_b !}} {a^{\dagger}}^{n_a}{b^{\dagger}}^{n_b} |0,0\rangle.$  
As the overall phase has nothing to do with the physics, 
the two type squeezed number states are physically identical.

Next, we consider the squeezed coherent state \cite{Yuen-1976,Hollenhorst-1979,Caves-1981}. 
Since the coherent state is a superposition of  number states 
\be
|\alpha\rangle = e^{\frac{1}{2}|\alpha|^2} \sum_{n=0}^{\infty} \frac{1}{\sqrt{n!}} \alpha^n |n\rangle, 
\ee
the squeezed coherent state can be expressed  by the superposition of the squeezed number states : 
\be
|\xi,\alpha\rangle \equiv S(\xi)|\alpha\rangle=e^{\frac{1}{2}|\alpha|^2} \sum_{n=0}^{\infty} \frac{1}{\sqrt{n!}} \alpha^n |\xi\rangle_{n}, ~~~~~~~|\xi,\alpha\rrangle \equiv \mathcal{S}(\xi)|\alpha\rangle=e^{\frac{1}{2}|\alpha|^2} \sum_{n=0}^{\infty} \frac{1}{\sqrt{n!}} \alpha^n |\xi\rrangle_{(n)}.  
\ee
Recall that the Dirac-type squeezed number states and the Schwinger-type only differ by the $U(1)$ factor depending on the number $n$ (\ref{u1diffxionemo}), so we obtain the relation between the squeezed coherent states of the Dirac-type and Schwinger-type as  
\be
|\xi, \alpha_D\rangle =e^{-i\frac{\phi}{4}} |\xi, \alpha_S\rrangle \label{twosqucohret}
\ee
with 
\be
\alpha_{{D}}=\alpha_{S} e^{-i\frac{\phi}{2}}. \label{phasealp}
\ee
 The Dirac- and Schwinger-type squeezed coherent states  represent superficially different physical states  except for the squeezed vacuum case $\alpha_S=\alpha_D=0$. 
However as implied by (\ref{phasealp}), the difference between the two type squeezed states can be absorbed in the phase part of the displacement parameter $\alpha$. Since the displacement parameter indicates the position of the squeezed coherent state on the $x^1$-$x^2$ plane \cite{Schumaker-Caves-1985-1,Schumaker-Caves-1985-2}, the elliptical uncertainty regions representing the two squeezed coherent states on the $x^1$-$x^2$ plane merely  differ  by the rotation $\frac{\phi}{2}$. This is also suggested by the $U(1)$  part $e^{i{\phi}T^3}$  of  (\ref{squeezediracope1}), which denotes the rotation around the $x^3$-axis.     
Similarly for the two-modes, the Dirac- \cite{Schumaker-Caves-1985-1,Schumaker-Caves-1985-2} and Schwinger-type  squeezed coherent  states   
\begin{align}
&|\xi, \alpha,\beta\rangle \equiv S(\xi) |\alpha,\beta\rangle = e^{\frac{1}{2}(|\alpha|^2 +|\beta|^2)} \sum_{n_a, n_b}\frac{1}{\sqrt{n_a ! n_b !}} ~\alpha^{n_a}\beta^{n_b} |\xi\rangle_{(n_a, n_b)}, \nn\\
&|\xi, \alpha,\beta\rrangle \equiv \mathcal{S}(\xi) |\alpha,\beta\rangle = e^{\frac{1}{2}(|\alpha|^2 +|\beta|^2)} \sum_{n_a, n_b}\frac{1}{\sqrt{n_a ! n_b !}} ~\alpha^{n_a}\beta^{n_b} |\xi\rrangle_{(n_a,n_b)},   
\end{align}
 are related as 
\be
|\xi, \alpha_D, \beta_D\rangle =e^{-i\frac{\phi}{2}}~|\xi, \alpha_S,\beta_S\rrangle 
\ee
with 
\be
\alpha_D=\alpha_S e^{-i\frac{\phi}{2}}, ~~\beta_D=\beta_S e^{-i\frac{\phi}{2}}. 
\ee

\section{$Sp(4; \mathbb{R})$ squeeze matrices and the non-compact 2nd Hopf map}\label{sec:sp4r}

The next-simple symplectic group is $Sp(4; \mathbb{R})$. Among the real symplectic groups, only $Sp(2;\mathbb{R})$ and $Sp(4;\mathbb{R})$ are isomorphic to indefinite spin groups;
\be
Sp(2; \mathbb{R}) \simeq Spin(2,1), ~~~~Sp(4; \mathbb{R}) \simeq Spin(2,3). 
\ee 
Futhermore, the $SO(2,3)$ group is the isometry group of the four-hyperboloid with split-signature, $H^{2,2}$, -- the basemanifold of the non-compact 2nd Hopf map. 
Encouraged by these mathematical analogies, we explore an $Sp(4;\mathbb{R})$ extension of the previous  $Sp(2; \mR)$ analysis.  For details of $Sp(4; \mathbb{R})$ group, one may consult with  Ref.\cite{Kastrup-2003} for instance.

\subsection{$sp(4; \mathbb{R})$ algebra}

From the result of Sec.\ref{subsec:toposymple}, we  see 
\be
Mp(4; \mathbb{R})/\mathbb{Z}_2~\simeq~ Sp(4; \mR) ~\simeq ~Spin(2,3) ~\simeq~ S^{1}\times S^3 \times \mathbb{R}^6. 
\ee
The metaplectic group $Mp(4, \mathbb{R})$ 
is the double cover of the symplectic group $Sp(4, \mathbb{R})$.  
 As the metaplectic representation of $Sp(2; \mR)$ is constructed by  the Majorana representation of $SO(2,1)$,   the  $SO(2,3)$ Majorana representation is expected to realize  the $Sp(4; \mR)$ metaplectic representation.   

The $sp(4; \mathbb{R})$ algebra is isomorphic to $so(2,3)$ algebra (Appendix \ref{sec:transfsp4r}), which consists of ten generators $T^{ab} =-T^{ba}$ $(a,b=1,2,\cdots,5)$: 
\be
[T^{ab}, T^{cd}] =ig^{ac}T^{bd}-ig^{ad}T^{bc} +ig^{bd}T^{ac} -ig^{bc}T^{ad}
\ee
where 
\be
g_{ab}=g^{ab}=\text{diag}(-1,-1,+1,+1,+1). 
\ee
The quadratic $SO(2,3)$ Casimir operator is given by 
\begin{align}
C&=\sum_{a<b=1}^5 T^{ab}~T_{ab} \nn\\
&=(T^{12})^2 - (T^{13})^2 - (T^{14})^2 - (T^{15})^2 - (T^{23})^2- (T^{24})^2- (T^{25})^2+  (T^{34})^2 +  (T^{35})^2+  (T^{45})^2. 
\end{align}
It is not difficult to construct non-Hermitian matrix realization of the $so(2,3)$ generators. 
For this purpose, we first introduce  the $SO(2,3)$ gamma matrices $\gamma^a$ that satisfy 
\be
\{\gamma^a , \gamma^b\} =2 g^{ab}.  
\ee
Placing the split-quaternion (\ref{quaternionicq})  and its conjugate (\ref{quaternionicbarq}) in the off-diagonal components of gamma matrices,  we can construct the $SO(2,3)$ gamma matrices  as\footnote{See Appendix \ref{sec:split-so5} also.}  
\be
\gamma^a =\{ \gamma^{m}, \gamma^5\} =\{\begin{pmatrix}
0 & \bar{q}^{m} \\
q^{m} & 0 
\end{pmatrix}, ~~\begin{pmatrix}
1 & 0 \\
0 & -1 
\end{pmatrix}
  \}  
\ee
or 
\be
\gamma^{i}=\begin{pmatrix}
0 & i\tau^i \\
-i\tau^i & 0 
\end{pmatrix},~~\gamma^4=\begin{pmatrix}
0 & 1 \\
1 & 0 
\end{pmatrix},~~\gamma^5=\begin{pmatrix}
1 & 0 \\
0 & -1 
\end{pmatrix}.  \label{taugammaso23}
\ee
Notice that $\gamma^a$ are pseudo-Hermitian: 
\be
{\gamma^a}^{\dagger} =\gamma_a =k \gamma^a k,  
\ee
where  
\be
k\equiv i\gamma^1 \gamma^2 =\begin{pmatrix}
\sigma_z & 0 \\
0 & \sigma_z
\end{pmatrix}. 
\ee
The corresponding $so(2,3)$ matrices, 
$\sigma^{ab} =-i\frac{1}{4}[\gamma^a, \gamma^b]$, are derived as 
\be
\sigma^{mn} =-\frac{1}{2}\begin{pmatrix}
\bar{\eta}^{mn i}\tau_i & 0 \\
0 & {\eta}^{mn i} \tau_i 
\end{pmatrix}, ~~~
\sigma^{i 5}=-\frac{1}{2}
\begin{pmatrix}
0 & \tau^i \\
\tau^i & 0 
\end{pmatrix},~~\sigma^{4 5}=i\frac{1}{2}
\begin{pmatrix}
0 & 1 \\
-1 & 0 
\end{pmatrix}. \label{so23n1exmat}
\ee
Here, $\eta^{mn i}$ and $\bar{\eta}^{mn i}$ denote the 't Hooft symbols with the split signature: 
\be
\eta_{mn i}=\epsilon_{m n i 4} +g_{m i}g_{n 4} -g_{n i}g_{m 4}, ~~~\bar{\eta}_{mn i}=\epsilon_{m n i 4} -g_{m i}g_{n 4} +g_{n i}g_{m 4}. 
\ee
The $so(2,3)$ matices are also pseudo-hermitian : 
\be
(\sigma^{ab})^{\dagger}=\sigma_{ab}=k\sigma^{ab} k. 
\ee
Obviously $k$ is unitarily equivalent to  $K=\begin{pmatrix}
1_2 & 0 \\
0 & -1_2 
\end{pmatrix}$ for $Sp(4;\mR)$.\footnote{$k$ also acts as the role of the $SU(2,2)$ invariant matrix where  $\frac{1}{2}\gamma^a$ and $\sigma^{ab}$ constitute the $su(2,2)$ generators. The completeness relation for the $u(2,2)$ algebra is given by 
\be
\sum_{a}(\gamma^a)_{\alpha\beta} (\gamma_a)_{\gamma\delta} +4\sum_{a<b=1}^5 (\sigma^{ab})_{\alpha\beta} (\sigma_{ab})_{\gamma\delta}   =4\delta_{\alpha\delta}\delta_{\beta\gamma} -\delta_{\alpha\beta}\delta_{\gamma\delta}
\ee
or 
\be
\sum_{a}(k^a)_{\alpha\beta} (k_a)_{\gamma\delta} +4\sum_{a<b=1}^5 (k^{ab})_{\alpha\beta} (k_{ab})_{\gamma\delta} =4 k_{\alpha\delta} k_{\beta\gamma} -k_{\alpha\beta} k_{\gamma\delta}.  \label{kcompresu22}
\ee
From (\ref{kcompresu22}) and (\ref{comspspcom23}), we obtain the $SU(2,2)$ Casimir as 
\be
\sum_{a=1}^5 X^aX_a + 4\sum_{a<b=1}^5 X^{ab}X_{ab}=3(\bar{\hat{\psi}}\hat{\psi}) (\bar{\hat{\psi}}\hat{\psi}+4)  . 
\label{su2,2casimir}
\ee
(\ref{su2,2casimir}) is consistent with the results (\ref{casimirso23gamgen}). 
}  
From the general discussion of Sec.\ref{sec:unitarynoncomp},  the corresponding Hermitian matrices  are given by 
\be
k^a\equiv k\gamma^a ={k^{a}}^{\dagger}, ~~~~~k^{ab} \equiv k\sigma^{ab} ={k^{ab}}^{\dagger},  \label{kskabs}
\ee
and the Hermitian operators  are 
\be
X^a=\hat{\psi}^{\dagger}~k^a ~\hat{\psi}, ~~~~  X^{ab} =\hat{\psi}^{\dagger}~k^{ab}~\hat{\psi},  \label{xaxabherons}
\ee
where $\hat{\psi}$ denotes a four-component operator whose components satisfy  
\be
[\hat{\psi}_{\alpha}, \hat{\psi}_{\beta}^{\dagger}]=k_{\alpha\beta}.  ~~~~~(\alpha,\beta=1,2,3,4)\label{comspspcom23}
\ee
We can explicitly realize $\hat{\psi}$  as 
\be
\hat{\psi}=
\begin{pmatrix}
a & 
b^{\dagger} & c & 
d^{\dagger}
\end{pmatrix}^t .  \label{expdiracsp4rspi}
\ee
Here, $a$, $b$, $c$ and $d$ are independent Schwinger boson  operators, $i.e.$
$[a, a^{\dagger}] = [b, b^{\dagger}] = [c, c^{\dagger}] = [d, d^{\dagger}] = 1$ and $[a, b^{\dagger}]=[a,c] =[c,d^{\dagger}]=\cdots=0.$  
$X^a$ and $X^{ab}$ (\ref{xaxabherons}) read as   
\begin{align}
&X^1=-a^{\dagger}d^{\dagger}+bc-ad+b^{\dagger}c^{\dagger},~~X^2=ia^{\dagger}d^{\dagger}+ibc-iad-ib^{\dagger}c^{\dagger} ,~~X^3=ia^{\dagger}c+id^{\dagger}b-ic^{\dagger}a-ib^{\dagger}d ,\nonumber\\
&X^4=a^{\dagger}c-d^{\dagger}b+c^{\dagger}a-b^{\dagger}d,~~~X^5= a^{\dagger}a-bb^{\dagger}-c^{\dagger}c+dd^{\dagger} =a^{\dagger}a-b^{\dagger}b-c^{\dagger}c+d^{\dagger}d,  
\label{expressx1tox5}
\end{align}
and 
\begin{align}
&X^{12}=-\frac{1}{2}(a^{\dagger}a +bb^{\dagger}+c^{\dagger}c+dd^{\dagger}), ~~X^{13}=-\frac{1}{2}(a^{\dagger}b^{\dagger} +ab+c^{\dagger}d^{\dagger} +cd), ~~X^{14}=i\frac{1}{2}(a^{\dagger}b^{\dagger} -ab-c^{\dagger}d^{\dagger} +cd), \nn\\
&X^{15}=i\frac{1}{2}(-a^{\dagger}d^{\dagger} +ad -b^{\dagger}c^{\dagger}+bc),~~X^{23}=i\frac{1}{2}(a^{\dagger}b^{\dagger} -ab+c^{\dagger}d^{\dagger} -cd),~~~X^{24}=\frac{1}{2}(a^{\dagger}b^{\dagger} +ab-c^{\dagger}d^{\dagger} -cd), \nn\\
&X^{25}=-\frac{1}{2}(a^{\dagger}d^{\dagger} +ad +b^{\dagger}c^{\dagger}+bc),   ~~~X^{34}=\frac{1}{2}(a^{\dagger}a +bb^{\dagger}-c^{\dagger}c-dd^{\dagger}), ~~~~X^{35}=-\frac{1}{2}(a^{\dagger}c +a c^{\dagger}+d^{\dagger}b  +d b^{\dagger}), \nn\\
&X^{45}=i\frac{1}{2}(a^{\dagger}c -c^{\dagger}a-d^{\dagger}b  +b^{\dagger}d). \label{so23genediracexp}
\end{align}
With (\ref{expressx1tox5}) and (\ref{so23genediracexp}), we  can show
\be
\sum_{a=1}^5 X^a X_a = (\bar{\hat{\psi}}\hat{\psi}+2)(\bar{\hat{\psi}}\hat{\psi}-2), ~~~~~\sum_{a>b=1}^5 X^{ab} X_{ab} = \frac{1}{2}(\bar{\hat{\psi}}\hat{\psi})(\bar{\hat{\psi}}\hat{\psi}+6)+1, 
 \label{casimirso23gamgen}
\ee
where
\be
\bar{\hat{\psi}}\hat{\psi}\equiv \hat{\psi}^{\dagger}k \hat{\psi} =a^{\dagger}a -b^{\dagger}b+c^{\dagger}c-d^{\dagger}d-2.   
\ee
$\bar{\hat{\psi}}\hat{\psi}$ is a singlet under the $SU(2,2)$ transformation: 
\be
[X^a, \bar{\hat{\psi}}\hat{\psi}]=[X^{ab}, \bar{\hat{\psi}}\hat{\psi}] =0,  
\ee
and the sixteen operators, $X^a$, $X^{ab}$ and $\bar{\psi}\psi$, constitute the $u(2,2)$ algebra. 

As we shall see below, the Majorana representation of $SO(2,3)$ realizes the metaplectic representation of $Sp(4; \mR)$. The $SO(2,3)$ group has the charge conjugation matrix satisfying 
\be
-(\sigma^{ab})^* =C \sigma^{ab}C, 
\ee
where 
\be 
C =\begin{pmatrix}
\sigma_x & 0 \\
0 & \sigma_x
\end{pmatrix}. \label{chargeconjmat}
\ee 
The $SO(2,3)$ Majorana spinor operator subject to 
 the Majorana condition 
\be
\hat{\psi}^* =C\hat{\psi}  \label{majoconsp4r}
\ee
is given by 
\be
\hat{\psi}_\text{M} =\begin{pmatrix}
a & 
a^{\dagger} & 
b & 
b^{\dagger}
\end{pmatrix}^t    \label{expmajosp4rspi}
\ee
whose  components satisfy the commutation relations 
\be
[\hat{\psi}_{\text{M} \alpha},  \hat{\psi}_{\text{M} \beta} ]=\mathcal{E}_{\alpha\beta} 
\label{commurelametap}
\ee
with 
\be
\mathcal{E} =kC =-Ck=\begin{pmatrix}
i\sigma_y & 0 \\
0 & i\sigma_y
\end{pmatrix}. 
\ee
Just as in the case of $SO(2,1)$ (\ref{defmi}), 
using $\mathcal{E}$, we can introduce  symmetric matrices 
\be
m^{ab}\equiv -\mathcal{E}\sigma^{ab}, ~~~~((m^{ab})^t =m^{ab}) 
\label{defemab}
\ee
to construct the $so(2,3)$ generators 
\be
X^{ab} \equiv \frac{1}{2}{\hat{\psi}_{\text{M}}}^t ~m^{ab}~\hat{\psi}_{\text{M}},  \label{xabmajo}
\ee
which are 
\begin{align}
&X^{12}=-\frac{1}{2}(a^{\dagger}a+bb^{\dagger})=-\frac{1}{2}(a^{\dagger}a +b^{\dagger}b+1), ~~~X^{13}=-\frac{1}{4}(a^2+{a^{\dagger}}^2 +b^2 +{b^{\dagger}}^2), \nn\\
&X^{14}=i\frac{1}{4}(-a^2+{a^{\dagger}}^2 +b^2 -{b^{\dagger}}^2), ~~~X^{15}=i\frac{1}{2}(ab -a^{\dagger}b^{\dagger}), ~~~X^{23}=i\frac{1}{4}(-a^2+{a^{\dagger}}^2 -b^2 +{b^{\dagger}}^2), \nn\\
&X^{24}=\frac{1}{4}(a^2+{a^{\dagger}}^2 -b^2 -{b^{\dagger}}^2),  ~~~~X^{25}=-\frac{1}{2}(ab +a^{\dagger}b^{\dagger}),~~~X^{34}=\frac{1}{2}(a^{\dagger}a -b^{\dagger}b), \nn\\
&X^{35}=-\frac{1}{2}(a^{\dagger}b +a b^{\dagger}), ~~~~X^{45}=i\frac{1}{2}(a^{\dagger}b -b^{\dagger}a). \label{so23majooprss}
\end{align}
Comparing the Majorana representation generators (\ref{xabmajo}) with the Dirac representation generators (\ref{xaxabherons}),  one can find the coefficient on the right-hand side of (\ref{xabmajo}) is half of that of (\ref{xaxabherons}) just as in  the case of the $Sp(2; \mathbb{R})$ and $Mp(2; \mathbb{R})$.   
This implies that (\ref{xabmajo}) are the generators of the double covering group of $Sp(4; \mathbb{R})$, which is $Mp(4; \mathbb{R})$.  

We also  construct antisymmetric matrices (\ref{sexpso23mab}) as 
\be
m^a\equiv \mathcal{E}\gamma^a.   \label{antisymmamat}
\ee
One can easily check that the corresponding operators identically vanish:
\be
X^a \equiv {\hat{\psi}_{\text{M}}}^t ~m^a~\hat{\psi}_{\text{M}}=0 . 
\label{xvanso23}
\ee
The $U(2,2)$ completeness relation is represented as 
\be
(m^a)_{\alpha\beta}(m_a)_{\gamma\delta} +4\sum_{a<b} (m^{ab})_{\alpha\beta}(m_{ab})_{\gamma\delta} =-4\mathcal{E}_{\alpha\delta}\mathcal{E}_{\beta\gamma} -\mathcal{E}_{\alpha\beta}\mathcal{E}_{\gamma\delta}. \label{u22comprems}
\ee
Using (\ref{u22comprems}) and (\ref{xvanso23}), we can show that the corresponding $SO(2,3)$ Casimir becomes a constant: 
\be
\sum_{a>b=1}^5 X^{ab}~X_{ab} =-\frac{5}{4},  \label{sp4rmetacon}
\ee
where we used $\mathcal{E}_{\alpha\beta}(\psi_{\text{M}})_{\alpha}(\psi_{\text{M}})_{\beta} =[a, a^{\dagger}]+[b, b^{\dagger}]=2$. (\ref{sp4rmetacon}) should be compared with the previous $SU(1,1)$ result (\ref{consusu11cas}).  Notice that the independent operators of (\ref{so23majooprss}) are simply given by the symmetric combination of the two-mode operators $a_i=a, b$ : 
\be
\{a_i, a_j\}, ~~~\{a_i^{\dagger}, a_j^{\dagger}\},~~\{a_i, a_j^{\dagger}\},  
\ee
which are known to realize the generators of  $Mp(4;\mathbb{R})$ (see Appendix \ref{app:metgr}).  Also from this observation, one may see that (\ref{xabmajo}) realizes  $mp(4; \mathbb{R})$ generators. There are two $SO(2,3)$ metaplectic irreducible representations  referred to as the singletons with the Casimir (\ref{sp4rmetacon}) \cite{Dirac-1963}.

\subsection{Gauss decomposition}

\begin{figure}[tbph]
\center
\includegraphics*[width=80mm]{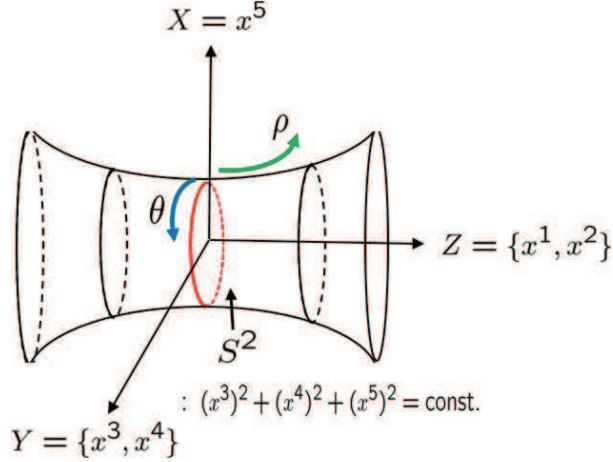}
\caption{Bloch four-hyperboloid $H^{2,2}$ : 
$-((x^1)^2 + (x^2)^2) +((x^3)^2 +(x^4)^2) +(x^5)^2=1$. The Bloch four-hyperboloid can be regarded as  a one-sheet hyperboloid $-Z^2 +Y^2+X^2=1$ with $Z=(x^1, x^2)$, $Y=(x^3, x^4)$ and $X=x^5$. Each of the  dimensions $Z$ and $Y$ has an internal $S^1$ structure. In the parametrization (\ref{x5defh22}) the range of  $x^5$ is $ [-1, 1]$, meaning that the parameterization does not  cover the whole surface of the Bloch four-hyperboloid.  }
\label{h22.fig}
\end{figure}

In the $Sp(2; \mathbb{R})$ case, we used the coset representation of $H^{2,0}$ 
\be
H^{2,0} ~\simeq~SO(2,1)/SO(2)~\simeq~SU(1,1)/U(1) ~\simeq~Sp(2,\mathbb{R})/U(1), 
\ee
which is equivalent to the 1st non-compact Hopf map
\be
H^{2,0}~\simeq~H^{2,1}/S^1. 
\ee
In the $Sp(4; \mR)$ case, the corresponding coset is  obviously given by 
\begin{align}
H^{2,2}&~\simeq~SO(2,3)/SO(2,2) \nn\\
&~\simeq~SO(2,3)/(SU(1,1)_L\otimes SU(1,1)_R)~\simeq~Sp(4,\mathbb{R})/(Sp(2;\mathbb{R})_L\otimes Sp(2; \mathbb{R})_R) , 
\label{groppdevidgroup}
\end{align}
which is  the basemanifold of  the 2nd non-compact Hopf map 
\be
H^{2,2}~\simeq~H^{4,3}/H^{2,1}. 
\label{htoh2ndhopf}
\ee
The coordinates $x^a$ $(a=1,2,3,4,5)$ on $H^{2,2}$ should satisfy 
\be
\sum_{a,b}g_{ab}x^ax^b=-x^1 x^1 -x^2x^2+x^3x^3+x^4x^4+x^5x^5=1. 
\ee
We parameterize $x^a$ as 
\begin{align}
&x^{m}=(x^1,x^2,x^3,x^4)=( \sin\theta \cos\chi\sinh\rho, ~\sin\theta \sin\chi\sinh\rho, ~\sin\theta \cos\phi\cosh\rho, ~\sin\theta \sin\phi\cosh\rho  ), \nn\\
&x^5=\cos\theta,  \label{x5defh22}
\end{align}
where the ranges of the parameters are given by (see Fig.\ref{h22.fig})   
\be
\rho, \theta ~\in ~\mathbb{R}_+\times S^1 \simeq H^{1,1}, ~~~\chi, \phi ~\in~S^1\times S^1.
\ee
As we have called $H^{2,0}$ associated with the $Sp(2; \mathbb{R})$ squeeze operator  the Bloch two-hyperboloid, we will refer to $H^{2,2}$  as the Bloch four-hyperboloid in the following.  

We also introduce ``normalized'' coordinates 
\be
y^{m} = (\cos\chi\sinh\rho, ~ \sin\chi\sinh\rho, ~\cos\phi\cosh\rho, ~\sin\phi\cosh\rho  ) \label{x5defh22nor}
\ee
which  satisfy 
$y^{m}y_{m} = -y^1 y^1 -y^2y^2+y^3y^3+y^4y^4=1$   
and denote the $H^{2,1}$-latitude of the Bloch four-hyperboloid with fixed $\theta$. 

Based on  the $G/H$ construction (\ref{groppdevidgroup}), we can easily derive a $Sp(4; \mR)$ squeeze matrix representing $H^{2,2}$.  
We take 
$\sigma^{mn}$, 
as the generators of $SO(2,2)$ group and $\sigma^{m 5}=\begin{pmatrix}
0 & -\bar{q}^{m} \\
q^{m} & 0 
\end{pmatrix}$ as 
the broken four generators. 
The squeeze matrix for $H^{2,2}$ is then given by 
\begin{align}
M&=e^{i\theta ~y_{m}\sigma^{m 5}}
=\begin{pmatrix}
 \cos\frac{\theta}{2} ~1_2 & -\sin\frac{\theta}{2}~y^{m}\bar{q}_{m} \\
 \sin\frac{\theta}{2}~y^{m}{q}_{m} & \cos\frac{\theta}{2}~1_2 
 \end{pmatrix}=\frac{1}{\sqrt{2(1+x^5)}}
\begin{pmatrix}
(1+x^5)1_2 & -\bar{q}_{m}x^{m} \\
q_{m}x^{m} & (1+x^5)1_2
\end{pmatrix}. \label{mascosetmatri2}
 \end{align}
In the polar coordinates, (\ref{mascosetmatri2}) is expressed as 
\be
M=\begin{pmatrix}
\cos\frac{\theta}{2} & 0 & -i\sin\frac{\theta}{2}~\cosh\rho~e^{-i\phi} & -\sin\frac{\theta}{2}~\sinh\rho~e^{-i\chi} \\
0 & \cos\frac{\theta}{2} & -\sin\frac{\theta}{2}~\sinh\rho~e^{i\chi} &  i\sin\frac{\theta}{2}~\cosh\rho~e^{i\phi} \\
-i\sin\frac{\theta}{2}~\cosh\rho~e^{i\phi}  & -\sin\frac{\theta}{2}~\sinh\rho~e^{-i\chi} & \cos\frac{\theta}{2} & 0 \\
-\sin\frac{\theta}{2}~\sinh\rho~e^{i\chi}  & i\sin\frac{\theta}{2}~\cosh\rho~e^{-i\phi} & 0 & \cos\frac{\theta}{2}
\end{pmatrix}.\label{coorh22xsspolar}
\ee
It is also possible to derive the $Sp(4; \mR)$ squeeze matrix (\ref{mascosetmatri2})  
 based on the 2nd non-compact Hopf map (\ref{htoh2ndhopf}). This construction will be important in the Euler angle decomposition  [Sec.\ref{subsec:eulerang}]. The 2nd non-compact Hopf map is explicitly given by   \cite{Hasebe-2009}  
\be
\psi~\in H^{4, 3}~\rightarrow~x^a =\psi^{\dagger}k^a \psi~\in~H^{2,2},  \label{2ndnoncomhopfmap}
\ee
where $\psi$ is subject to 
\be
\psi^{\dagger}k\psi = (\psi_1^*\psi_1+\psi_3^*\psi_3)- (\psi_2^*\psi_2+\psi_4^*\psi_4) =1,  
\label{orignorpsicond}
\ee
and  $x^a$ (\ref{2ndnoncomhopfmap}) automatically satisfy the condition of $H^{2,2}$:  
\be
g_{ab}x^a x^b = (\psi^{\dagger}k \psi)^2=1.  
\ee
We can express $\psi$  as 
\be
\psi= \Psi_L h\begin{pmatrix}
1 \\
0
\end{pmatrix} =\Psi_L \phi \label{2ndhopfexaplit}
\ee
where $\Psi_L$ denotes the following $4\times 2$ matrix
\footnote{While in (\ref{x5defh22}) the range of $x^5$ of $H^{2,2}$ is restricted to $[-1,1]$, we can adopt the range of $x^5$ as $[1, \infty)$ and 
\be
{x'}^{m} =\sqrt{(x^5)^2-1}~
{{y'}}^{m}=\sinh\tau ~{{y'}}^{m}, ~~~~x^5 =\cosh\tau.
\ee
with ${{y'}}^{m}{{y'}}_{m}=-1$, so  ${x'}^{m}{x'}_{m} +x^5x_5 =1$.  
The corresponding Hopf spinor for $\psi^{\dagger}k^a\psi=x^a$ is given by 
\be
\psi=\frac{1}{\sqrt{2(x^5+1)}}
\begin{pmatrix}
x^5 +1 \\
0  \\
-ix^3 +x^4 \\
-x^1-ix^2
\end{pmatrix},    
\ee
which realizes as the first column of the following matrix 
\be
e^{\tau {y'}_{m}\sigma^{m 5}} =\frac{1}{\sqrt{2(x^5+1)}} \begin{pmatrix}
(x^5+1) 1_2 & i x'_{m}\bar{q}^{m} \\
-ix'_{m} q^{m} & (x^5+1) 1_2
\end{pmatrix} \label{x5le1squeezem}
\ee
where $({x'}^1, {x'}^2, {x'}^3, -{x'}^4)\equiv (-x^2, x^1, -x^4, x^3)$. Replacement of the $so(2,3)$ matrices with the $so(2,3)$ Hermitian operators  transforms (\ref{x5le1squeezem}) to non-unitary operator $e^{\tau  {y'}_{m}X^{m 5}}$. 
Non-unitary operators generally violate the probability conservation, so we will not treat the parameterization for $x^5 \in [1,  \infty)$ in this paper. 
}  
\be
\Psi_L \equiv \frac{1}{\sqrt{2(1+x^5)}}\begin{pmatrix}
(1+x^5) 1_2\\
q_{m}x^{m}
\end{pmatrix}, \label{2ndhopfexaplitmat}
\ee
and 
$h$ is an arbitrary $SU(1,1)$ group element representing a $H^{2,1}$-fibre: 
\be
h = \begin{pmatrix} 
\phi & \sigma_x\phi^*
\end{pmatrix} = \begin{pmatrix}
\phi_1 & \phi_2^* \\
\phi_2 & \phi_1^*
\end{pmatrix}
\ee
subject to 
\be
\det h = |\phi_1|^2 -|\phi_2|^2 =\phi^{\dagger}\sigma_z\phi =1.
\ee
$\Psi_L$ is an eigenstate of the $x^a \gamma_a$ with positive chirality 
\be
x^{a}\gamma_a \Psi_L  = + \Psi_L.   
\ee
Similarly, a negative chirality matrix satisfying 
\be
x^{a}\gamma_a \Psi_R  = - \Psi_R   
\ee
is given by 
\be
\Psi_R=\frac{1}{\sqrt{2(1+x^5)}}\begin{pmatrix}
-\bar{q}_{m}x^{m}\\
(1+x^5)1_2
\end{pmatrix}.  \label{psirmat1}
\ee
With these two opposite chirality matrices, 
$M$ (\ref{mascosetmatri2}) can be simply expressed as 
\be
M=\begin{pmatrix}
\Psi_L & \Psi_R
\end{pmatrix}.
\ee
(See Appendix \ref{app:relhopfmat} for more details about relations between the squeeze matrix and the non-compact Hopf spinors.)     

Here, we mention the Gauss decomposition of $M$. Following to the general method of \cite{Gilmore-2008}, we may in principle derive the normal ordering of $M$.  However,   for the $Sp(4; \mathbb{R})$ group the ten generators are concerned, and the Gauss decomposition will be  a formidable task.  
Therefore instead of attempting the general method, we resort to an intuitive geometric structure of the Hopf maps to derive the Gauss decomposition.    
The hierarchical geometry of the Hopf maps 
 implies that the $U(1)$ part of the 1st non-compact Hopf map will be replaced with the $SU(1,1)$ group in the 2nd.  
 We then expect that the Gauss decomposition of $M$ will be given by\footnote{The Gauss (UDL) decomposition  is unique \cite{Horn-Johnson-1985}. }  
\begin{align}
M&=\text{Exp}\biggl({-\tan\frac{\theta}{2} ~\begin{pmatrix}
0 & y^{m}\bar{q}_{m} \\
0 & 0 
\end{pmatrix}} \biggr)  \cdot \text{Exp}\biggl({-\ln \biggl(\cos\frac{\theta}{2}\biggr)~ \begin{pmatrix}
1_2 & 0 \\
0 & -1_2
\end{pmatrix}} \biggr) \cdot \text{Exp}\biggl({\tan\frac{\theta}{2} 
~\begin{pmatrix}
0 &  0\\
y^{m}{q}_{m} & 0 
\end{pmatrix}}\biggr) \nn\\
&=\text{Exp} \biggl(-\tan\frac{\theta}{2} ~y^{m} \cdot (\frac{1}{2}\gamma_{m}-i\sigma_{m 5}) \biggr)
  \cdot \text{Exp}\biggl(-\ln \biggl(\cos\frac{\theta}{2}\biggr)~\gamma^5 \biggr) 
\cdot \text{Exp}\biggl( \tan\frac{\theta}{2} 
~y^{m}\cdot (\frac{1}{2}\gamma_{m}+i\sigma_{m 5})\biggr). \label{gaussdecompmat}
\end{align}
Substituting the matrices, we can demonstrate the validity of (\ref{gaussdecompmat}).  
Notice that, unlike the $Sp(2; \mathbb{R})$ case (\ref{diracnormasq}), the Gauss decomposition (\ref{gaussdecompmat}) cannot be expressed only within the ten generators of $Sp(4; \mathbb{R})$, but we need to utilize 
 the five $SO(2,3)$ gamma matrices as well. 
The fifteen matrices made of the $SO(2,3)$ gamma matrices and generators amount to the $so(2,4)\simeq su(2,2)$ algebra.

\subsection{Euler angle decomposition}\label{subsec:eulerang}

Here we derive  Euler angle decomposition of the $Sp(4; \mathbb{R})$ squeeze matrix based on the hierarchical geometry of the non-compact Hopf maps.  The Euler decomposition is crucial to perform the number state expansion of  $Sp(4; \mathbb{R})$ squeezed states. 

We first introduce a dimensionality reduction of the 2nd non-compact Hopf map, which we refer to as  the  non-compact chiral Hopf map:    
\be
H^{2,1}_L\otimes H^{2,1}_R ~~\overset{H^{2,1}_{\text{diag.}}}{\longrightarrow}~~H^{2,1}. 
\label{chiralmath}
\ee
(\ref{chiralmath}) is readily obtained by imposing one more constraint to the non-compact 2nd Hopf spinor:  
\be
\psi^{\dagger}k^5 \psi=1, \label{k5psinorcond}
\ee
 in addition to  the original constraint (\ref{orignorpsicond}). When we denote the non-compact Hopf spinor  as 
$
\psi=(\psi_L  ~~
\psi_R )^t 
$
 the two constraints, (\ref{orignorpsicond}) and (\ref{k5psinorcond}), are rephrased as ``normalizations'' for each of the two-component chiral Hopf  spinors, 
\be
{\psi_L}^{\dagger}\sigma_z\psi_L =1, ~~~{\psi_R}^{\dagger}\sigma_z\psi_R =1. 
\ee
$\psi_L$ and $\psi_R$ are thus the coordinates on $H^{2,1}\otimes H^{2,1}$, and  
(\ref{chiralmath}) is explicitly realized as   
\be
\psi_L,~\psi_R ~~\rightarrow~~y^{m} = \frac{1}{2}( {\psi_L}^{\dagger}\sigma_z \bar{q}^{m} \psi_R + {\psi_R}^{\dagger} \sigma_z q^{m}\psi_L),   \label{chiralhopfmapsexp}
\ee
and so $y^{m}$ automatically satisfy 
\be
y^{m}y_{m}=-(y^1)^2 -(y^2)^2+(y^3)^2+(y^4)^2=  ({\psi_L}^{\dagger} \sigma_z\psi_L )({\psi_R}^{\dagger} \sigma_z\psi_R ) =1, 
\ee
so $y^{m}$ stand for the coordinates on $H^{2,1}$. The simultaneous $SU(1,1)$ transformation of $\psi_L$ and $\psi_R$ has nothing to do with $y^{m}$ and  geometrically represents $H^{2,1}_{\text{diag}}$-fibre part which is projected out in (\ref{chiralmath}).  

We can express the chiral Hopf spinors as\footnote{Here, the imaginary unit $i$ is added on the right-hand side of $\psi_R$ for later convenience.  } 
\be
\psi_L =e^{-i\frac{\phi}{2}\sigma_z}~\begin{pmatrix}
\cosh\frac{\rho}{2} ~e^{-i\frac{1}{2}\chi} \\
-i\sinh\frac{\rho}{2} ~e^{i\frac{1}{2}\chi} 
\end{pmatrix}, ~~~\psi_R =-i~e^{i\frac{\phi}{2}\sigma_z}~ \begin{pmatrix}
\cosh\frac{\rho}{2} ~e^{-i\frac{1}{2}\chi} \\
-i\sinh\frac{\rho}{2} ~e^{i\frac{1}{2}\chi} 
\end{pmatrix}, 
\ee
and the resultant $y^{m}$ from (\ref{chiralhopfmapsexp}) are given by (\ref{x5defh22nor}). 
 Notice that when $\phi=0$, $\psi_L$ and $\psi_L$ are reduced to the 1st non-compact Hopf spinor and $y^{m}$ (\ref{x5defh22nor}) are also reduced to the coordinates on $H^{2,0}$. In this sense, the non-compact chiral Hopf map incorporates the structure of the 1st non-compact  Hopf map in a hierarchical manner of dimensions. 
The   $SU(1,1)$ group elements corresponding to $\psi_L$ and $\psi_R$ are given by  
\begin{align}
H_L &\equiv \begin{pmatrix}
\psi_L & \sigma_x {\psi_L}^*\end{pmatrix}
=\begin{pmatrix}
\cosh \frac{\rho}{2} ~e^{-i\frac{1}{2}(\chi+\phi)} & i\sinh \frac{\rho}{2} ~ e^{-i\frac{1}{2}(\chi+\phi)} \\
-i\sinh \frac{\rho}{2}  ~e^{i\frac{1}{2}(\chi+\phi)}  & \cosh \frac{\rho}{2} ~e^{i\frac{1}{2}(\chi+\phi)}
\end{pmatrix}=e^{-\frac{\rho}{2}\sigma_y} ~e^{-i\frac{1}{2}(\chi+\phi)\sigma_z}, \nn\\
H_R&\equiv i\begin{pmatrix}
\psi_R & -\sigma_x {\psi_R}^*\end{pmatrix}=\begin{pmatrix}
\cosh \frac{\rho}{2} ~e^{-i\frac{1}{2}(\chi-\phi)} & i\sinh \frac{\rho}{2} ~ e^{-i\frac{1}{2}(\chi-\phi)} \\
-i\sinh \frac{\rho}{2} ~ e^{i\frac{1}{2}(\chi-\phi)}  & \cosh \frac{\rho}{2} ~e^{i\frac{1}{2}(\chi-\phi)}
\end{pmatrix}=e^{-\frac{\rho}{2}\sigma_y}~ e^{-i\frac{1}{2}(\chi-\phi)\sigma_z}.  
\end{align}
From the chiral  Hopf spinors, we can reconstruct a non-compact 2nd Hopf spinor that satisfies the 2nd non-compact Hopf map (\ref{2ndnoncomhopfmap}) as 
\be
{\psi'} =\begin{pmatrix}
\sqrt{\frac{1+x^5}{2}}~ \psi_L \\
\sqrt{\frac{1-x^5}{2}}~ \psi_R  
\end{pmatrix}.  \label{psidashexp}
\ee
($\psi$ (\ref{2ndhopfexaplit}) and $\psi'$ (\ref{psidashexp}) are related by the $SU(1,1)$ gauge transformation as we shall see below.)  
One may find that the $x^5$ coordinate on $H^{2,2}$ determines  the weights of the chiral Hopf spinors in $\psi'$.  In particular at the ``north pole'' ($x^5=1$), $\psi'$ (\ref{psidashexp}) is reduced to $\psi_L$, while at the ``south pole'' ($x^5=-1$) $\psi_R$. The hierarchical geometry of the Hopf maps is summarized as 

\vspace{0.3cm}
The 1st Hopf map for   $H^{2,0}$  
~ $\rightarrow$ ~ The chiral  Hopf map for  $H^{2,1}$ ~ $\rightarrow$ ~  The  2nd  Hopf map for  $H^{2,2}$. 
\vspace{0.3cm}

From the chiral Hopf spinors,  we construct the following $4\times 2$  matrix    
\be
\Psi'_L = \begin{pmatrix}
 \sqrt{\frac{1+x^5}{2}} \begin{pmatrix}\psi_L & \sigma_x {\psi_L
 }^* \end{pmatrix} \\
  \sqrt{\frac{1-x^5}{2}} \begin{pmatrix}
\psi_R & \sigma_x {\psi_R
 }^* \end{pmatrix}
\end{pmatrix}. \label{psidashso23hopfs1}
\ee
A short calculation shows that (\ref{psidashso23hopfs1}) is given by 
\be
\Psi'_L=
\begin{pmatrix}
H_L & 0 \\
0 & H_R 
\end{pmatrix}  \begin{pmatrix}
\sqrt{\frac{1+x^5}{2}} 1_2 \\
-i\sqrt{\frac{1-x^5}{2}}\sigma_z 
\end{pmatrix}.
\ee
We also introduce 
\be
\Psi'_R \equiv  -i\begin{pmatrix}
 \sqrt{\frac{1-x^5}{2}} \begin{pmatrix}\psi_L & -\sigma_x {\psi_L }^* \end{pmatrix} \\
  \sqrt{\frac{1+x^5}{2}} \begin{pmatrix}
\psi_R & -\sigma_x {\psi_R
 }^* \end{pmatrix}
\end{pmatrix}
=\begin{pmatrix}
H_L & 0 \\
0 & H_R
\end{pmatrix}  \begin{pmatrix}
-i\sqrt{\frac{1-x^5}{2}}\sigma_z \\
\sqrt{\frac{1+x^5}{2}}1_2 
\end{pmatrix}.\label{psidashso23hopfs}
\ee
$\sigma^{ab}$ are the $so(2,3)$ matrices (\ref{so23n1exmat}). 
With $\Psi'_L$ and $\Psi'_R$, we construct the $4\times 4$ matrix $\mathcal{M}$, which we will refer to as the $\it{Schwinger\text{-}type}$ $Sp(4; \mR)$ squeeze matrix:\footnote{In the polar coordinates, (\ref{macadeco}) is expressed as 
\be 
\mathcal{M} =
\begin{pmatrix}
\cos\frac{\theta}{2}~\cosh\frac{\rho}{2}~e^{-i\frac{1}{2}(\chi+\phi)} & i\cos\frac{\theta}{2}~\sinh\frac{\rho}{2}~e^{-i\frac{1}{2}(\chi+\phi)} & -i\sin\frac{\theta}{2}~\cosh\frac{\rho}{2}~e^{-i\frac{1}{2}(\chi+\phi)} & -\sin\frac{\theta}{2}~\sinh\frac{\rho}{2}~e^{-i\frac{1}{2}(\chi+\phi)} \\
 -i\cos\frac{\theta}{2}~\sinh\frac{\rho}{2}~e^{i\frac{1}{2}(\chi+\phi)} & \cos\frac{\theta}{2}~\cosh\frac{\rho}{2}~e^{i\frac{1}{2}(\chi+\phi)} & -\sin\frac{\theta}{2}~\sinh\frac{\rho}{2}~e^{i\frac{1}{2}(\chi+\phi)} & i\sin\frac{\theta}{2}~\cosh\frac{\rho}{2} ~e^{i\frac{1}{2}(\chi+\phi)} \\
 -i\sin\frac{\theta}{2}~\cosh\frac{\rho}{2}~e^{-i\frac{1}{2}(\chi-\phi)} & -\sin\frac{\theta}{2}~\sinh\frac{\rho}{2}~e^{-i\frac{1}{2}(\chi-\phi)}  & \cos\frac{\theta}{2}~\cosh\frac{\rho}{2}~e^{-i\frac{1}{2}(\chi-\phi)} & i\cos\frac{\theta}{2}~\sinh\frac{\rho}{2}~e^{-i\frac{1}{2}(\chi-\phi)} \\
 -\sin\frac{\theta}{2}~\sinh\frac{\rho}{2}~e^{i\frac{1}{2}(\chi-\phi)} & i\sin\frac{\theta}{2}~\cosh\frac{\rho}{2} ~e^{i\frac{1}{2}(\chi-\phi)} & -i\cos\frac{\theta}{2}~\sinh\frac{\rho}{2}~e^{i\frac{1}{2}(\chi-\phi)} & \cos\frac{\theta}{2}~\cosh\frac{\rho}{2}~e^{i\frac{1}{2}(\chi-\phi)}
\end{pmatrix}. \label{polarmathcalmmat}
\ee
} 
\be
\mathcal{M} \equiv \begin{pmatrix}\Psi_L' & \Psi'_R \end{pmatrix} 
= \begin{pmatrix}
H_L & 0 \\
0 & H_R
\end{pmatrix}
\begin{pmatrix}
\sqrt{\frac{1+x^5}{2}} 1_2 & 
 -i\sqrt{\frac{1-x^5}{2}}\sigma_z \\
-i\sqrt{\frac{1-x^5}{2}}\sigma_z  & \sqrt{\frac{1+x^5}{2}}1_2 
\end{pmatrix}. 
\label{macadeco}
\ee
Here,\footnote{In the polar coordinates, $H$ is given by 
\be
H =\begin{pmatrix}
\cosh \frac{\rho}{2} ~e^{-i\frac{1}{2}(\chi+\phi)} & i\sinh \frac{\rho}{2} ~ e^{-i\frac{1}{2}(\chi+\phi)}  & 0 & 0 \\
-i\sinh \frac{\rho}{2} ~ e^{i\frac{1}{2}(\chi+\phi)}  & \cosh \frac{\rho}{2} ~e^{i\frac{1}{2}(\chi+\phi)} & 0 & 0 \\
0 & 0 & \cosh \frac{\rho}{2} ~e^{-i\frac{1}{2}(\chi-\phi)} & i\sinh \frac{\rho}{2} ~ e^{-i\frac{1}{2}(\chi-\phi)}  \\
0 & 0 & -i\sinh \frac{\rho}{2} ~ e^{i\frac{1}{2}(\chi-\phi)}  & \cosh \frac{\rho}{2} ~e^{i\frac{1}{2}(\chi-\phi)}
\end{pmatrix}.\label{hmatexpre}
\ee
} 
\be
H\equiv \begin{pmatrix}
H_L & 0 \\
0 & H_R
\end{pmatrix} 
= 
\begin{pmatrix}
e^{-i\frac{\phi}{2}\sigma_z} & 0 \\
0 & e^{i\frac{\phi}{2}\sigma_z}
\end{pmatrix} \begin{pmatrix}
e^{-i\frac{\chi}{2}\sigma_z} & 0 \\
0 & e^{-i\frac{\chi}{2}\sigma_z}
\end{pmatrix} \begin{pmatrix}
e^{-\frac{\rho}{2}\sigma_y} & 0 \\
0 & e^{-\frac{\rho}{2}\sigma_y}
\end{pmatrix}    = 
e^{-i\phi\sigma^{34}} e^{i\chi \sigma^{12}} e^{-i\rho\sigma^{13}}
\label{deflargehmat}
\ee 
and
\be 
\begin{pmatrix}
\sqrt{\frac{1+x^5}{2}} 1_2 & 
 -i\sqrt{\frac{1-x^5}{2}}\sigma_z \\
-i\sqrt{\frac{1-x^5}{2}}\sigma_z  & \sqrt{\frac{1+x^5}{2}}1_2 
\end{pmatrix} = \begin{pmatrix}
\cos\frac{\theta}{2}~1_2 & -i\sin\frac{\theta}{2}~\sigma_z \\
-i\sin\frac{\theta}{2}~\sigma_z & \cos\frac{\theta}{2}~1_2
\end{pmatrix} = \text{Exp}\biggl( -i\frac{\theta}{2}\begin{pmatrix}
0 & \sigma_z \\
\sigma_z & 0 
\end{pmatrix} \biggr) =e^{i\theta \sigma^{35}}. \label{middletermeul}
\ee
Hence, we have a concise expression for $\mathcal{M}$ as 
\be
\mathcal{M} = H \cdot  e^{i\theta \sigma^{35}} . 
\label{simplecalm}
\ee

The expression of $\Psi'_L$ (\ref{psidashso23hopfs1}) is distinct from that  of $\Psi_L$ (\ref{2ndhopfexaplitmat}), but this is not a problem because they are related by a $SU(1,1)$ gauge transformation.   
Indeed, the comparison between (\ref{2ndhopfexaplitmat}) and (\ref{psidashso23hopfs1}) implies   
\be
\Psi'_L =\Psi_L H_L.    
\ee
Similarly for (\ref{psirmat1}) and (\ref{psidashso23hopfs}), we have 
\be
{\Psi}'_R =\Psi_R H_R. 
\ee
As a result, we obtain the relation between   $M$ (\ref{mascosetmatri2}) and $\mathcal{M}$ (\ref{macadeco}) as 
\be
\mathcal{M} =\begin{pmatrix}
\Psi'_L & \Psi'_R 
\end{pmatrix} 
= \begin{pmatrix}
\Psi_L & \Psi_R 
\end{pmatrix} 
\begin{pmatrix}
H_L & 0 \\
0 & H_R 
\end{pmatrix} = {M} \cdot H . \label{mathcmandh}
\ee
(\ref{simplecalm}) and (\ref{mathcmandh}) yield a factorized form of $M$:    
\be
M =\mathcal{M}\cdot H^{-1} = H \cdot e^{i\theta\sigma^{35}} \cdot H^{-1}.  
\label{eulerdexsp4r} 
\ee
This is the Euler angle decomposition of the $Sp(4; \mathbb{R})$ squeeze matrix we have sought. 
In (\ref{eulerdexsp4r}),  the off-diagonal block matrix $e^{i\theta\sigma^{35}}$  is sandwiched by the diagonal block matrix $H$ and its inverse.  Recall that the Euler angle decomposition of the $Sp(2; \mR)$ squeeze operator (\ref{squeezediracope1}) exhibits the same structure,  
$S=e^{i{\phi} T^z} \cdot e^{i {\rho}T^x} \cdot (e^{i{\phi}T^z})^{-1}$. The squeeze parameter $\rho$ in the $Sp(2; \mR)$ case corresponds to $\theta$ in the $Sp(4; \mR)$ case.  Notice that at $\theta=0$ (``no squeeze'') the $Sp(4; \mR)$ squeeze matrix  (\ref{eulerdexsp4r}) becomes trivial.

Using the squeeze matrix, the non-compact 2nd Hopf map (\ref{2ndnoncomhopfmap}) can be realized as\footnote{In more detail, we have 
\be
 (M^{\dagger} k^a M)_{11}=  -(M^{\dagger} k^a M)_{22}=  -(M^{\dagger} k^a M)_{33}=  (M^{\dagger} k^a M)_{44}= x^a, 
\ee
and 
\be
 (\mathcal{M}^{\dagger} k^a \mathcal{M})_{11}=  -(\mathcal{M}^{\dagger} k^a \mathcal{M})_{22}=  -(\mathcal{M}^{\dagger} k^a \mathcal{M})_{33}=  (\mathcal{M}^{\dagger} k^a \mathcal{M})_{44}= x^a. 
\ee
} 
\be
x^a =\frac{1}{4}\tr (k^5 M^{\dagger} k^a M). 
\label{xafrommdagm}
\ee
Since $H_L$ and $H_R$ are  $SU(1,1)$ group elements and  $H$ (\ref{deflargehmat}) satisfies  
\be
H ~k^5~ H^{\dagger}= k^5, 
\ee
it is obvious that $x^a$ (\ref{xafrommdagm}) is invariant under the $SU(1,1)$ transformation 
\be
M~~\rightarrow~~M H'
\ee
with $H'$ subject to 
\be
\det(H')=1, ~~~~{H'}^{\dagger} k H'=k. 
\ee
At the level of matrix representation for  the basemanifold $H^{2,2}$,  $\mathcal{M}$ is no less legitimate than $M$, since their difference is only about the $SU(1,1)$-fibre part which is projected out in the 2nd non-compact Hopf map. However, as  we shall see below, 
the Dirac- and  Schwinger-type $Sp(4; \mR)$ squeeze operators 
yield physically distinct squeezed vacua unlike the previous $Sp(2; \mR)$ case.

\section{$Sp(4; \mathbb{R})$ squeezed states and their basic properties}\label{sec:sp4rsqop}

Replacement of the $Sp(4;\mathbb{R})$  non-Hermitian matrices with the corresponding operators 
yields the $Sp(4; \mathbb{R})$ squeeze operator: 
\be
M = e^{i\theta\sum_{m=1}^4 y_{m}\sigma^{m 5}} ~~~
{\rightarrow}~~~S= e^{i\theta\sum_{m=1}^4 y_{m}X^{m 5}} . \label{replacemtosqusp4r}
\ee
With four-mode representation (\ref{so23genediracexp}) and two-mode representation (\ref{so23majooprss}), (\ref{replacemtosqusp4r}) is  respectively given by  
\begin{subequations}
\begin{align}
&
S =\exp\biggl({-i\frac{\theta}{2} (\xi (ad+bc)+\xi^* (a^{\dagger}d^{\dagger} +b^{\dagger} c^{\dagger}) + \eta (a c^{\dagger} + b^{\dagger}d) + \eta^* (a^{\dagger}c+ bd^{\dagger}))}\biggr)   , \\
&
S=\exp\biggl({-i\frac{\theta}{2} (\xi ab+\xi^* a^{\dagger}b^{\dagger} + \eta a b^{\dagger} + \eta^* a^{\dagger} b)}\biggr), 
\end{align}
\end{subequations}
where 
\be
\xi \equiv \sinh\rho ~e^{i(\chi+\frac{\pi}{2})}, ~~~\eta \equiv \cosh\rho~e^{i\phi}. 
\ee
We now discuss properties of the $Sp(4; \mR)$ squeeze operators and $Sp(4; \mR)$ squeezed states.

\subsection{$Sp(4; \mathbb{R})$ squeeze operator}

From the Gauss decomposition (\ref{gaussdecompmat}), we have 
\be
S=\text{Exp} \biggl(-\tan\frac{\theta}{2} ~y^{m} \cdot (\frac{1}{2}X_{m}-iX_{m 5}) \biggr)
  \cdot \text{Exp}\biggl(\ln \biggl(\cos\frac{\theta}{2}\biggr)~\cdot X^5 \biggr) 
\cdot \text{Exp}\biggl(- \tan\frac{\theta}{2} 
~y^{m}\cdot (\frac{1}{2}X_{m}+iX_{m 5})\biggr). \label{gaussdecompsqueezeop}
\ee
The operators on the exponential of the most right component  are $\frac{1}{2}X^{m} +i{X}^{m 5}$  that are given by a linear combinations of the operators $ad$, $c^{\dagger}a$, $d^{\dagger}b$ and $b^{\dagger}c^{\dagger}$ as found in 
(\ref{so23genediracexp}). Because of the existence of  $b^{\dagger}c^{\dagger}$, it is not easy to derive the number state basis expansion even for the squeezed vacuum state. The situation is even worse when we utilize the Euler angle decomposition: 
\be
S= e^{-i\phi X^{34}} e^{i\chi X^{12}} e^{-i\rho X^{13}} e^{i\theta X^{35}} e^{i\rho X^{13}}e^{-i\chi X^{12}}e^{i\phi X^{34}},  \label{diractypesqueezeop}
\ee
since $X^{13}$ contains both $a^{\dagger}b^{\dagger}$ and $c^{\dagger}d^{\dagger}$. 
 Meanwhile the Schwinger-type squeeze operator  
\be
\mathcal{S}
= e^{-i\phi X^{34}} e^{i\chi X^{12}} e^{-i\rho X^{13}} e^{i\theta X^{35}} \label{squeezing1}
\ee
is much easier to handle. To obtain a better understanding of $Sp(4;\mR)$ squeezed states,  we will derive  number state expansion for several  Schwinger-type squeezed states.  

\subsection{Two-mode squeeze operator and $Sp(4 ; \mathbb{R})$ two-mode squeeze vacuum}

Representing $X^{34}$ and $X^{12}$ (\ref{so23majooprss})  by the number operators, $\hat{n}_a =a^{\dagger}a$ and $\hat{n}_b =b^{\dagger}b$, we express the Schwinger-type squeeze operator  (\ref{squeezing1})  as 
\be
\mathcal{S}  = e^{-i\frac{1}{2}\chi} e^{-i\frac{1}{2}(\chi+\phi)\hat{n}_a} e^{-i\frac{1}{2}(\chi-\phi)\hat{n}_b} e^{-i\rho X^{13}} e^{i\theta X^{35}}. 
\label{twomodesqueesp4r}
\ee
The operators of the last two terms, $X^{35} =-\frac{1}{2}(a^{\dagger}b +b^{\dagger}a)$ and $X^{13} =-\frac{1}{2}(a^2 +{a^{\dagger}}^2 +b^2 +{b^{\dagger}}^2)$, are respectively made of the ladder operators of the $su(2)$ and $su(1,1)$ algebra. 
We apply the  Gauss decomposition formula \cite{Zhang-Feng-Gilmore-1990, Gilmore-2008} to these terms to have 
\bse
\begin{align}
&e^{i\theta X^{35}}=e^{-i\tan \frac{\theta}{2} \cdot a^{\dagger}b } ~\biggl( \frac{1}{\cos\frac{\theta}{2}}  \biggr)^{n_a-n_b}~ e^{-i\tan \frac{\theta}{2} \cdot b^{\dagger}a } =e^{-i\tan \frac{\theta}{2} \cdot b^{\dagger}a } ~\biggl( \frac{1}{\cos\frac{\theta}{2}}  \biggr)^{-n_a+n_b}~ e^{-i\tan \frac{\theta}{2} \cdot a^{\dagger}b} , \\
&e^{-i\rho X^{13}} = \frac{1}{\cosh \frac{\rho}{2}} e^{i\frac{1}{2}\tanh \frac{\rho}{2} \cdot ({a^{\dagger}}^2 + {b^{\dagger}}^2)}~\biggl(\frac{1}{\cosh\frac{\rho}{2}}\biggr)^{n_a+n_b} ~ e^{i\frac{1}{2}\tanh \frac{\rho}{2} \cdot (a^2 + b^2)} . 
\end{align} 
\ese
Based on these decompositions, we investigate the $Sp(4;\mathbb{R})$  squeezing of two-mode number states 
\be
|\text{tm}\rrangle_{(n_a, n_b)} = \mathcal{S}|n_a, n_b\rangle. 
\ee
We can derive the $Sp(4;\mR)$ squeezed vacuum as  
\be
|\text{tm}\rrangle_{(0,0)} =e^{-i\frac{\chi}{2}}~|\xi_+\rangle_{(0)}  \otimes|\xi_- \rangle_{(0)} , \label{tmsvstr} 
\ee
where $|\xi_{\pm}\rangle_{(0)}$ denotes the $Sp(2;\mathbb{R})$ single-mode squeezed vacuum  (\ref{zroonesques1}) with   
\be
\xi_{\pm} \equiv  \frac{\rho}{2}e^{-i(\chi\pm\phi +\frac{\pi}{2})}. 
\ee
The Schwinger-type squeezed vacuum  does not depend on the parameter $\theta$ and  is given by a direct product of the two $Sp(2; \mathbb{R})$ single-mode squeezed vacua with phase difference, $\text{arg}(\xi_+)-\text{arg}(\xi_-)=-2\phi$.  We then find the physical meanings of the three parameters of the four-hyperboloid as follows.   The parameter $\rho$ signifies the squeezing parameter common to the two $Sp(2; \mathbb{R})$ squeezed vacua and $\chi$ stands for their overall rotation, and $\phi$ denotes the relative rotation  between them (see Sec.\ref{sec:sp4runcert} also). To see the physical meaning of the remaining parameter $\theta$, let us consider the $Sp(4; \mathbb{R})$ squeezed one-photon states. 
The squeezed one-photon states are similarly obtained as 
\bse
\begin{align}
|\text{tm}\rrangle_{(1,0)}&=e^{-i\chi} \biggl(e^{-i\frac{1}{2}\phi} \cos\frac{\theta}{2} |\xi_+ \rangle_{(1)} \otimes  |\xi_-\rangle_{(0)}  - i   e^{i\frac{1}{2}\phi} \sin\frac{\theta}{2} |\xi_+\rangle_{(0)} \otimes  |\xi_- \rangle_{(1)}    \biggr), \\
|\text{tm}\rrangle_{(0,1)}&=e^{-i\chi} \biggl(e^{i\frac{1}{2}\phi} \cos\frac{\theta}{2} |\xi_+\rangle_{(1)} \otimes  |\xi_-\rangle_{(0)}  - i   e^{-i\frac{1}{2}\phi} \sin\frac{\theta}{2} |\xi_+\rangle_{(0)} \otimes  |\xi_-\rangle_{(1)}    \biggr), 
\end{align} \label{twomodeonsqueezesta}
\ese
where $|\xi_{\pm}\rangle_{(1)}$ denotes the $Sp(2;\mathbb{R})$ single-mode squeezed one-photon state (\ref{zroonesques1}).  
Thus the $Sp(4; \mathbb{R})$ squeeze of the one-photon state represents a superposition of the tensor products of the  $Sp(2; \mathbb{R})$ squeezed vacuum and squeezed one-photon state. 
Let us focus on the $H^{2,1}$-latitude  at $\phi=0$ ($x^4=0$) on $H^{2,2}$. 
Both of (\ref{twomodeonsqueezesta}) are reduced to the same state 
\be
|\text{tm}\rrangle|_{\phi=0} 
=e^{-i\chi} \biggl( \cos\frac{\theta}{2} |\xi \rangle_{(1)} \otimes  |\xi\rangle_{(0)}  -i   \sin\frac{\theta}{2} |\xi\rangle_{(0)} \otimes  |\xi \rangle_{(1)}    \biggr)  
\label{phi0tm}
\ee
with $\xi \equiv -i\frac{\rho}{2}e^{-i\chi}$.  Interestingly, (\ref{phi0tm}) represents an entangled state of two squeezed states. 
Indeed, when we assign qubit states  $|1\rangle$ and $|0\rangle$  to the two squeezed states $|\xi\rangle_{(1)}$ and $|\xi\rangle_{(0)}$,  (\ref{phi0tm}) can be expressed as  
\be
|\text{tm}\rrangle|_{\phi=0}   =\sum_{i,j=1,0}Q_{ij}~|i\rangle \otimes |j\rangle  \label{qtwomodestate}
\ee
where  
\be
Q=e^{-i\chi} 
\begin{pmatrix}
0 & \cos\frac{\theta}{2} \\
-i  \sin\frac{\theta}{2} & 0 
\end{pmatrix}. 
\ee
The concurrence for the entanglement of two qubits \cite{Hill-Wootters-1997} is readily calculated as  
\be
c = \sqrt{2(1-\tr ((Q^{\dagger}Q)^2))} = |\sin\theta|=\sqrt{1-(x^5)^2}, 
\ee
which is exactly equal to the ``radius'' of the  $H^{2,0}$-latitude at $\theta$ on  $H^{2,1}$. 
Thus the  concurrence associated with the $Sp(4; \mR)$ squeezed state   has a clear geometrical meaning as the radius of  hyperbolic latitude on $H^{2,1}$, and the azimuthal angle $\theta$ specifies the degree of the entanglement.     
In particular, the two-mode squeezed state (\ref{qtwomodestate}) is   maximally entangled $c=1$ at the ``equator'' of $H^{2,1}$ ($\theta=\pi/2$), while  it becomes a product state $c=0$  at the ``north pole'' ($\theta=0$) or  the ``south pole'' ($\theta=\pi$).

\subsection{Four-mode squeeze operator and $Sp(4; \mathbb{R})$ squeezed vacuum}

In a similar fashion to the two-mode case, we can discuss the four-mode squeezed states. 
From the four-mode $Sp(4;\mR)$ operators (\ref{so23genediracexp}), the Schwinger-type squeeze operator is represented as 
\be
\mathcal{S} 
=e^{-i\chi} e^{-i\frac{1}{2}(\chi+\phi)(\hat{n}_a +\hat{n}_b)}e^{ -i\frac{1}{2}(\chi-\phi)(\hat{n}_c +\hat{n}_d)}
 e^{-i\rho X^{13}} e^{i\theta X^{35}} . \label{sp4rsqueezeopprods}
\ee
The Gaussian decompositions of the last two terms on the right-hand side of (\ref{sp4rsqueezeopprods}) are given by  
\bse
\begin{align}
&e^{i\theta X^{35}}=e^{-i\tan \frac{\theta}{2} \cdot (a^{\dagger}c+b^{\dagger}d) } ~\biggl( \frac{1}{\cos\frac{\theta}{2}}  \biggr)^{n_a-n_b+n_c-n_d}~ e^{-i\tan \frac{\theta}{2} \cdot (c^{\dagger}a +d^{\dagger}b)}, \label{expx35fact}\\
&e^{-i\rho X^{13}} = \frac{1}{\cosh^2 \frac{\rho}{2}} e^{i\tanh \frac{\rho}{2} \cdot ({a^{\dagger}}b^{\dagger} + {c^{\dagger}}d^{\dagger})}~\biggl(\frac{1}{\cosh\frac{\rho}{2}}\biggr)^{n_a+n_b+n_c+n_d} ~ e^{i\tanh \frac{\rho}{2} \cdot (ab + cd)}. 
\end{align}
\ese
Using these formulas, we can derive the number state expansion  of  $Sp(4; \mathbb{R})$ four-mode squeezed states: 
\be
|\text{fm}\rrangle_{(n_a, n_b, n_c, n_d)} \equiv \mathcal{S} |n_a, n_b, n_c, n_d\rangle. 
\ee
The Schwinger-type squeezed vacuum is derived as 
\be
|\text{fm}\rrangle_{(0,0,0,0)} 
=e^{-i\chi}|\xi_+\rangle_{(0,0)} \otimes |\xi_-\rangle_{(0,0)} ,  \label{fmsvstr}
\ee
where  $|\xi_{\pm}\rangle_{(0,0)}$ denotes the $Sp(2; \mathbb{R})$ squeezed vacuum with 
\be
\xi_{\pm} \equiv  \frac{\rho}{2}e^{-i(\chi\pm\phi +\frac{\pi}{2})}. 
\ee
Notice that the two-mode  (\ref{tmsvstr})  and the  four-mode (\ref{fmsvstr}) have the same structure. 
The one-photon squeezed states are similarly obtained as 
\bse
\begin{align}
|\text{fm}\rrangle_{(1,0,0,0)} 
&=e^{-i\frac{3}{2}\chi }\biggl(e^{-i\frac{1}{2}\phi} \cos\frac{\theta}{2}|\xi_+ \rangle_{(1,0)} \otimes |\xi_-\rangle_{(0,0)} -i e^{i\frac{1}{2}\phi} \sin\frac{\theta}{2}|\xi_+\rangle_{(0,0)} \otimes |\xi_-\rangle_{(1,0)}\biggr), \\
|\text{fm}\rrangle_{(0,1,0,0)} 
&=e^{-i\frac{3}{2}\chi }\biggl(e^{-i\frac{1}{2}\phi} \cos\frac{\theta}{2}|\xi_+ \rangle_{(0,1)} \otimes |\xi_-\rangle_{(0,0)} -i e^{i\frac{1}{2}\phi} \sin\frac{\theta}{2}|\xi_+\rangle_{(0,0)} \otimes |\xi_-\rangle_{(0,1)}\biggr), \\
|\text{fm}\rrangle_{(0,0,1,0)} 
&=e^{-i\frac{3}{2}\chi }\biggl(e^{i\frac{1}{2}\phi} \cos\frac{\theta}{2}|\xi_+ \rangle_{(0,0)} \otimes |\xi_-\rangle_{(1,0)} -i e^{-i\frac{1}{2}\phi} \sin\frac{\theta}{2}|\xi_+\rangle_{(1,0)} \otimes |\xi_-\rangle_{(0,0)}\biggr), \\
|\text{fm}\rrangle_{(0,0,0,1)} 
&=e^{-i\frac{3}{2}\chi }\biggl(e^{i\frac{1}{2}\phi} \cos\frac{\theta}{2}|\xi_+ \rangle_{(0,0)} \otimes |\xi_-\rangle_{(0,1)} -i e^{-i\frac{1}{2}\phi} \sin\frac{\theta}{2}|\xi_+\rangle_{(0,1)} \otimes |\xi_-\rangle_{(0,0)}\biggr), 
\end{align}
\ese
where $|\xi_{\pm}\rangle_{(1,0),~(0,1)}$ are the $Sp(2;\mathbb{R})$ two-mode one-photon squeezed states (\ref{twomodesp2rsqunumb}).

\subsection{$Sp(4; \mathbb{R})$ uncertainty relation}\label{sec:sp4runcert}

Next, we investigate uncertainty relation for the $Sp(4;\mR)$ squeezed vacua. 
Unlike the derivations of the number state expansion, what is needed to evaluate  uncertainty relations  is  only the $Sp(4;\mR)$ covariance of the spinor operators. The following  derivation of $Sp(4; \mR)$ uncertainty relations is a straightforward generalization of the $Sp(2;\mR)$ case \cite{Yuen-1976}.

For the $Sp(4; \mR)$ two-mode with two kinds of  annihilation operators,   we introduce  four operator coordinates:  
\bse
\begin{align}
&X^1= \frac{1}{2} (a+a^{\dagger}), ~~~X^2=-i\frac{1}{2} (a-a^{\dagger}), \\ 
&X^3= \frac{1}{2} (b+b^{\dagger}), ~~~X^4=-i\frac{1}{2} (b-b^{\dagger}), 
\end{align} \label{xfromatob}
\ese
which satisfy the 4D Heisenberg-Weyl algebra, 
\be
[X^1, X^2] =[X^3, X^4] =i\frac{1}{2}, ~~[X^1, X^3]=[X^1, X^4]=[X^2,X^3]=\cdots=0. 
\label{twononcomalplanes}
\ee
We thus have two independent sets of 2D non-commutative coordinate spaces constituting 4D non-commutative space,  in the terminology of non-commutative geometry,  $\mathbb{R}^2_{NC} \oplus \mathbb{R}^2_{NC}=\mathbb{R}^4_{NC}$.  
In a similar manner, in the case of the $Sp(4; \mR)$ four-mode,  four operator  coordinates  are introduced as  
\bse
\begin{align}
&X^1= \frac{1}{2\sqrt{2}} (a+a^{\dagger}+b+b^{\dagger}), ~~X^2=-i\frac{1}{2\sqrt{2}} (a-a^{\dagger}+b-b^{\dagger}), \\ 
&X^3= \frac{1}{2\sqrt{2}} (c+c^{\dagger}+d+d^{\dagger}), ~~X^4=-i\frac{1}{2\sqrt{2}} (c-c^{\dagger}+d-d^{\dagger}),
\end{align} \label{xfromatod}
\ese
which satisfy (\ref{twononcomalplanes}) again. In the following we evaluate the deviations of these  coordinates for the $Sp(4;\mathbb{R})$ squeezed vacua.

Let us denote the $Sp(4; \mathbb{R})$ squeezed vacuum as  
\be
|\text{sq}\rangle \equiv S |0\rangle,   
\ee
where $|0\rangle$ denotes  the vacuum of the Schwinger boson operators: 
\be
a|0\rangle =b|0\rangle =c|0\rangle =d|0\rangle =0. 
\ee
Obviously, the squeezed vacuum is the vacuum of the squeezed annihilation operator
\be
\tilde{a} \equiv S a S^{\dagger}. \label{tildesqueza}
\ee
Since the operator $\hat{\psi}$ (Dirac-type (\ref{expdiracsp4rspi}) and Majorana-type (\ref{expmajosp4rspi})) behaves as a spinor under  the $Sp(4; \mathbb{R})$ transformation (see Sec.\ref{sec:hermschwin} for general discussions), the Schwinger operator transforms as  
\be
S^{\dagger} ~\hat{\psi} ~S  = M~\hat{\psi}. 
\label{sdapsis}
\ee
For the Dirac-type, $M$ is given by (\ref{coorh22xsspolar}), while for the Schwinger-type by (\ref{polarmathcalmmat}). Notice that (\ref{sdapsis}) implies that the product of  the three operators on the left-hand side is simply equal to the linear combination of the components of $\hat{\psi}$ on right-hand side.  
By this relation (\ref{sdapsis}),  it becomes  feasible  to evaluate the expectation values of operator $O(\hat{\psi})$ for the squeezed vacuum as  
\be
\langle O(\hat{\psi}) \rangle_{\text{sq}}\equiv \langle \text{sq}| O(\hat{\psi})|\text{sq}\rangle =  \langle 0|S^{\dagger} O(\hat{\psi}) S |0\rangle = \langle 0|  O(S^{\dagger} \hat{\psi} S) |0 \rangle =  \langle 0|O(M\hat{\psi})|0\rangle,  
\label{generalexvalo}
\ee
where we assumed that $O(\hat{\psi})$ is a sum of  polynomials of the components of  $\hat{\psi}$.  Thus, the evaluation of the expectation values for the squeezed vacuum is boiled down to that for the usual vacuum.   

\begin{figure}[tbph]
\center
\includegraphics*[width=120mm]{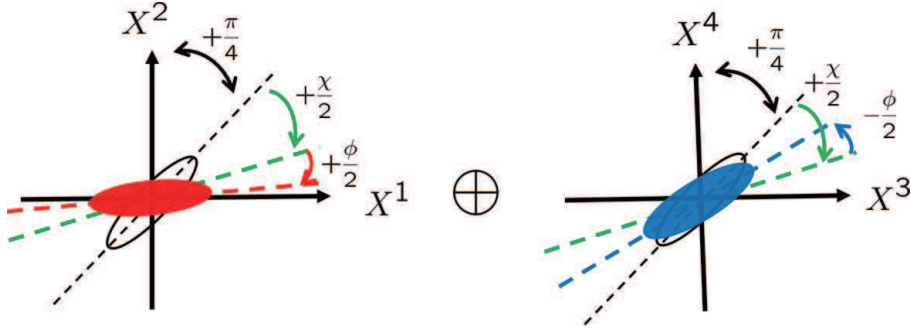}
\caption{ At $\theta=\pi$, the 4D uncertainty region for the $Sp(4; \mathbb{R})$ squeezed vacuum  is exactly equal to the ``direct sum'' of the two 2D uncertainty regions described by  two $Sp(2;\mathbb{R})$ squeezed vacua. The $Sp(4; \mathbb{R})$ squeezed  vacuum thus realizes the squeezing in  a
 4D manner.  The parameter $\rho$ denotes the degree of squeezing of both  $Sp(2; \mathbb{R})$ squeezed vacua, $\chi$ stands for their overall rotation, and $\phi$ signifies the relative rotation between them.   In particular at $(\chi, \phi)=(\frac{\pi}{2}, 0)$, both of the two squeezings are aligned to the ``same'' direction  (the squeezing on $X^1-X^2$ plane is $X^1$ direction, and that on $X^3-X^4$ is $X^3$ direction), while  at $(\chi, \phi)=(0, \frac{\pi}{2})$, two  squeezings are ``perpendicular'' to each other (the squeezing on $X^1-X^2$ plane is $X^1$ direction, while that on $X^3-X^4$ plane is $X^4$ direction).   
}
\label{4squeeze.fig}
\end{figure}

Since only the covariance of the operator is concerned here, the following discussions can be applied to both two-mode and four-mode.   
According to (\ref{generalexvalo}), we can readily derive  the squeezed vacuum expectation value of $\hat{\psi}$ : 
\be
\langle \hat{\psi} \rangle_{\text{sq}} 
=M \langle 0|\hat{\psi} |0\rangle=0,  
\ee
and, from (\ref{xfromatob}) or (\ref{xfromatod}), we  have 
\be
\langle X^1 \rangle_{\text{sq}}=\langle X^2 \rangle_{\text{sq}} =\langle X^3 \rangle_{\text{sq}}=\langle X^4 \rangle_{\text{sq}} =0. 
\ee
A bit of calculations shows\footnote{
Here, we performed  calculations such as 
\be
\langle ({\Delta X^1})^2\rangle_{\text{sq}} =\langle (X^1)^2 \rangle_{\text{sq}}-\langle { X^1}\rangle^2_{\text{sq}} =\langle (X^1)^2 \rangle_{\text{sq}}=\frac{1}{4}(\langle a^2 \rangle_{\text{sq}} + \langle aa^{\dagger} \rangle_{\text{sq}} +\langle  a^{\dagger} a\rangle_{\text{sq}} +\langle {a^{\dagger}}^2 \rangle_{\text{sq}}   ).  
\ee
To evaluate  $\langle a^2 \rangle_{\text{sq}}$ for instance,    we proceeded as $\langle a^2 \rangle_{\text{sq}} = \langle 0|S^{\dagger} a^2 S |0\rangle = \langle 0|(S^{\dagger}aS)^2|0\rangle$ and substituted $S^{\dagger}a S = M_{11}a +M_{12}a^{\dagger} +M_{13}b + M_{14} b^{\dagger} $  
 to derive $\langle a^2\rangle_{\text{sq}} = M_{11}M_{12}\langle 0| a a^{\dagger}|0\rangle + M_{13}M_{14}\langle 0| bb^{\dagger}|0\rangle =M_{11}M_{12}+ M_{13}M_{14} = i\sin^2\frac{\theta}{2} \cosh\rho \sinh\rho ~e^{-i(\chi+\phi)}$. In the last equation, we utilized (\ref{coorh22xsspolar}).   The other expectation values were also obtained in a similar way. }  
\bse
\begin{align}
&\langle (\Delta {X^{1/2}})^2 \rangle_{\text{sq}} =\frac{1}{4}\biggl(\cos^2(\frac{\theta}{2}) +\sin^2(\frac{\theta}{2}) (\cosh (2\rho) ~+\!/\!-~\sinh(2\rho) \sin(\chi+\phi)) \biggr),\\
&\langle (\Delta {X^{3/4}})^2 \rangle_{\text{sq}} =\frac{1}{4}\biggl(\cos^2(\frac{\theta}{2}) +\sin^2(\frac{\theta}{2}) (\cosh (2\rho) ~+\!/\!-~\sinh(2\rho) \sin(\chi-\phi)) \biggr) .   
\end{align} 
\label{deviationscoordiracex}
\ese
Consequently,  we have the uncertainty relations for the $Sp(4; \mathbb{R})$ squeezed vacuum: 
\bse
\begin{align}
&\ll (\Delta {X^1})^2 \rl_{\text{sq}}~\ll (\Delta {X^2})^2 \rl_{\text{sq}} = \frac{1}{16} (1+\sin^2\theta ~\sinh^2 \rho +\sin^4 (\frac{\theta}{2})\sinh^2 (2\rho) \cos^2(\chi+\phi))~~\ge ~\frac{1}{16},\\
&\ll (\Delta {X^3})^2 \rl_{\text{sq}}~\ll (\Delta {X^4})^2 \rl_{\text{sq}} =\frac{1}{16} (1+\sin^2\theta ~\sinh^2 \rho +\sin^4 (\frac{\theta}{2})\sinh^2 (2\rho) \cos^2(\chi-\phi))~~\ge ~\frac{1}{16}. 
\end{align}
\ese
The uncertainty bound is saturated at  
$(i)$ $\theta=0$ (the ``north pole'' of the Bloch four-hyperboloid) and $(ii)$ $\theta=\pi$  (the ``south pole''), at which, 
(\ref{deviationscoordiracex}) becomes   
\bse
\begin{align}
&\langle (\Delta {X^{1/2}})^2 \rangle_{\text{sq}} |_{\theta=\pi}=\frac{1}{4}(\cosh (2\rho) ~+\!/\!-~\sinh(2\rho) \sin(\chi+\phi)), \\
&\langle (\Delta {X^{3/4}})^2 \rangle_{\text{sq}} |_{\theta=\pi}=\frac{1}{4}(\cosh (2\rho) ~+\!/\!-~\sinh(2\rho) \sin(\chi-\phi)).  
\end{align} \label{southuncer4d}
\ese
Notice that (\ref{southuncer4d}) represents the uncertainty regions of two $Sp(2; \mathbb{R})$ squeezed vacua \cite{Gerry-Knight-2005}.  (See Fig.\ref{4squeeze.fig} also.)  
Since $\theta$ represents the squeezing parameter of the $Sp(4; \mathbb{R})$ squeeze operator, 
the case $(i)$ corresponds to the trivial vacuum and (\ref{deviationscoordiracex}) is reduced to 
 $\langle (\Delta {X^{m}})^2 \rangle_{\text{sq}}=\frac{1}{4}$ (no sum for $m=1,2,3,4$), and so  
 the case $(i)$ is  rather trivial.  Meanwhile for the case $(ii)$, at 
   $(\chi, \phi) =(\frac{\pi}{2}, 0)$ or  $(\chi, \phi) =(0, \frac{\pi}{2})$,  the  deviations (\ref{deviationscoordiracex}) become   
\be
\ll (\Delta {X^1})^2\rl_{\text{sq}} =\frac{1}{4}e^{2\rho},~~\ll (\Delta {X^2})^2\rl_{\text{sq}} =\frac{1}{4}e^{-2\rho},~~\ll (\Delta {X^3})^2\rl_{\text{sq}} =\frac{1}{4}e^{\pm 2\rho},~~\ll{(\Delta X^4})^2\rl_{\text{sq}} =\frac{1}{4}e^{\mp 2\rho},  \label{diracsquspe}
\ee
and non-trivially saturate the  uncertainty bound: 
\be
\ll (\Delta {X^1})^2 \rl_{\text{sq}}~\ll (\Delta {X^2})^2 \rl_{\text{sq}} = \ll (\Delta {X^3})^2 \rl_{\text{sq}}~\ll (\Delta {X^4})^2 \rl_{\text{sq}} =\frac{1}{16}.  
\ee
Performing similar calculations  for the Schwinger-type squeezed vacuum, 
 we obtain 
\bse
\begin{align}
&\llangle (\Delta {X^{1/2}})^2 \rrangle_{\text{sq}}=  \frac{1}{4} (\cosh\rho ~+/-~\sinh \rho \sin(\chi+\phi))   ~~\ge ~\frac{1}{16} , \\
&\llangle (\Delta {X^{3/4}})^2 \rrangle_{\text{sq}} =  \frac{1}{4} (\cosh\rho ~+/-~ \sinh \rho ~\sin(\chi-\phi)) ~~\ge ~\frac{1}{16},    
\end{align}\label{uncerdelxs}
\ese
where (\ref{polarmathcalmmat}) was used. 
Notice that the deviations do not depend on the parameter $\theta$ unlike the Dirac-type and are exactly equal to the Dirac-type at $\theta=\pi$ (\ref{southuncer4d}) with half squeezing.   
Therefore, (\ref{uncerdelxs}) is identical to  the uncertainty regions of two $Sp(2; \mathbb{R})$ squeezed vacua.   
This result is actually  expected,  since  the  Schwinger-type $Sp(4; \mathbb{R})$ squeezed vacuum  (\ref{tmsvstr}) does not depend on $\theta$ and is simply the direct product of the two $Sp(2; \mathbb{R})$ squeezed vacua.  

\section{$Sp(4; \mathbb{R})$ squeezed coherent states}\label{sec:squcohstate}

 The $Sp(4; \mathbb{R})$ squeezed coherent state is introduced as the $Sp(4; \mathbb{R})$ squeezed vacuum displaced on 4D plane and  exhibits a natural 4D generalization of  the properties of the original $Sp(2;\mR)$ squeezed coherent state. 

\subsection{Squeezed coherent state}

With the displacement operator $D_a(\alpha)=e^{\alpha a^{\dagger}-\alpha^*a}$, 
the two-mode and four-mode displacement operators are respectively given by 
\be
D(\alpha, \beta) =D_a(\alpha)D_b(\beta) 
,~~~~~
D(\alpha,\beta) =D_a(\alpha) D_b(\beta) D_c(\alpha)D_d(\beta) .   
\ee
It is straightforward to introduce a $Sp(4; \mR)$  version of the  squeezed coherent state as 
\be
|\alpha, \beta, \text{sq} \rangle =D(\alpha,\beta)~S|0\rangle. 
\ee
Each displacement operator acts to the two-mode $\hat{\psi}=(\hat{\psi}_1~\hat{\psi}_2~\hat{\psi}_3~\hat{\psi}_4)^t=(a~a^{\dagger}~b~b^{\dagger})^t$ and 
the four-mode   $\hat{\psi}=(\hat{\psi}_1~\hat{\psi}_2~\hat{\psi}_3~\hat{\psi}_4)^t=(a~b^{\dagger}~c~d^{\dagger})^t$  as 
\be
D(\alpha,\beta)~\hat{\psi}~ D(\alpha, \beta)^{\dagger} =\hat{\psi}-\varphi 
\ee
where 
\be
\varphi= \begin{pmatrix}
\tilde{\alpha} & 
\tilde{\alpha}^* & 
\tilde{\beta} &
\tilde{\beta}^*
\end{pmatrix}^t.
\ee
Relations  
\begin{align}
&D~S ~ ~\begin{pmatrix}
\hat{\psi}_1 \\
\hat{\psi}_3
\end{pmatrix} ~S^{\dagger}~D^{\dagger}~  |\alpha,\beta, \text{sq} \rangle=\begin{pmatrix}
0 \\
0 
\end{pmatrix}, \nn\\
&D~S ~ ~\begin{pmatrix}
\hat{\psi}_2^{\dagger} \\
\hat{\psi}_4^{\dagger}
\end{pmatrix} ~S^{\dagger}~D^{\dagger}~  |\alpha,\beta, \text{sq} \rangle=\begin{pmatrix}
0 \\
0 
\end{pmatrix},
\end{align}
immediately tell that the squeezed coherent state satisfies the following operator eigenvalue equations 
\begin{align}
&\hat{\psi}'_1  |\alpha,\beta, \text{sq}\rangle = \varphi'_1 |\alpha,\beta, \text{sq}\rangle, ~~~~~~~~\hat{\psi}'_3  |\alpha,\beta, \text{sq}\rangle = \varphi'_3 |\alpha,\beta, \text{sq}\rangle, \nn\\
&(\hat{\psi}'_2)^{\dagger}  |\alpha,\beta, \text{sq}\rangle = {\varphi'}^*_2 |\alpha,\beta, \text{sq}\rangle, ~~~(\hat{\psi}'_4)^{\dagger}  |\alpha,\beta, \text{sq}\rangle = {\varphi'}^*_4 |\alpha,\beta, \text{sq}\rangle,  \label{eigenvaeqeach}
\end{align}
where 
\be
\hat{\psi}'\equiv S\hat{\psi} S^{\dagger} =M^{-1}\hat{\psi}, ~~~~\varphi'\equiv  M^{-1}\varphi.
\ee
For instance,  the first equation of (\ref{eigenvaeqeach}) for the  two-mode Dirac-type  squeezed coherent state reads as  
\begin{align}
&\biggl(\cos\frac{\theta}{2}~ a +i\sin\frac{\theta}{2}~\cosh\rho~ e^{-i\phi}~ b +\sin\frac{\theta}{2}~\sinh\rho ~e^{-i\chi} ~b^{\dagger}\biggr)|\alpha,\beta; \text{sq}\rangle \nn\\
&~~~~~~~~~~~~~~~~~~~~~~~~~~~= \biggl(\cos\frac{\theta}{2}~ \alpha +i\sin\frac{\theta}{2}~\cosh\rho~ e^{-i\phi}~ \beta +\sin\frac{\theta}{2}~\sinh\rho ~e^{-i\chi} ~\beta^{*}\biggr)|\alpha,\beta; \text{sq}\rangle.
\end{align}

\subsection{Several properties}

\begin{itemize}
\item{Two-mode $Sp(4; \mathbb{R})$ squeezed coherent state}
\end{itemize}

For two-mode squeezed coherent state,  
\be
|\alpha, \beta, \text{sq}\rangle= D_a(\alpha)D_b(\beta)S~|0,0\rangle, 
\ee
the expectation values of $X$s are derived as 
\begin{align}
&\langle X^1 \rangle_{(\alpha, \beta, \text{sq})}=
\text{Re} (\alpha) =\langle X^1 \rangle_{\alpha}, ~~~\langle X^2 \rangle_{(\alpha, \beta, \text{sq})} 
=\text{Im} (\alpha)=\langle X^2 \rangle_{\alpha}, \nn\\
&\langle X^3 \rangle_{(\alpha, \beta, \text{sq})}=\text{Re} (\beta)=\langle X^3 \rangle_{\beta}, 
~~~\langle X^4 \rangle_{(\alpha, \beta, \text{sq})}=\text{Im} (\beta) =\langle X^4 \rangle_{\beta}. 
\label{twomodemean} 
\end{align}
The expectation values (\ref{twomodemean}) exactly coincide with those of the  coherent states.  
Similarly, the deviations of $X$s are obtained as 
\be
\langle (\Delta X^{m})^2 \rangle_{(\alpha, \beta, \text{sq})} =\langle (\Delta X^{m})^2 \rangle_{(\alpha, \text{sq})}   =  \langle (X^{m})^2 \rangle_{(\alpha, \text{sq})} -\langle X^{m} \rangle_{(\alpha, \text{sq})}^2  = \langle       (\Delta X^{m})^2 \rangle_{\text{sq}}. ~~~(\text{no sum for $m$}=1,2,3,4). \label{twomodedeviations}
\ee
The deviations (\ref{twomodedeviations})  are equal to   those of the squeezed vacuum, (\ref{deviationscoordiracex}) and (\ref{uncerdelxs}). Thus, the position of the squeezed coherent state is accounted for by its coherent state part, while the deviation is by its squeezed state part, implying that the $Sp(4;\mR)$ squeezed coherent vacuum is the squeezed vacuum displaced by $(\alpha, \beta)$ on $\mathbb{C}^2\simeq \mR^4$ plane. Obviously, this signifies a natural 4D generalization of the known properties  of the original $Sp(2; \mR)$ squeezed coherent state \cite{Yuen-1976}.

\begin{itemize}
\item{Four-mode $Sp(4; \mathbb{R})$ squeezed coherent state}
\end{itemize}

From the four-mode generators of $Sp(4; \mathbb{R})$, we can define two kinds of annihilation operators :  
\be 
A=X^1+iX^2=\frac{1}{\sqrt{2} }(a+b), 
~~B=X^3+iX^4=\frac{1}{\sqrt{2} }(c+d), 
\ee
which satisfy 
$[A, A^{\dagger}]=[B,B^{\dagger}]=1.$  
We construct the displacement operator as 
\be
D(\alpha, \beta) =D_A(\alpha)~ D_B(\beta)=
D_a(\frac{1}{\sqrt{2}}\alpha) ~ D_b(\frac{1}{\sqrt{2}}\alpha)  ~D_c(\frac{1}{\sqrt{2}}\beta)  ~D_d(\frac{1}{\sqrt{2}}\beta), 
\ee
and  introduce  four-mode squeezed coherent state as 
\be
|\alpha,\beta, \text{sq}\rangle= D(\alpha, \beta)S|0,0, 0,0\rangle. 
\ee
It is easy to see that the expectation values of the coordinates are given by 
\begin{align}
&\langle X^1 \rangle_{(\alpha, \beta, \text{sq})} =\text{Re} (\alpha) =\langle X^1 \rangle_{(\alpha,\beta)}  , ~~~\langle X^2 \rangle_{(\alpha, \beta, \text{sq})} =\text{Im} (\alpha)=\langle X^2 \rangle_{(\alpha,\beta)} , \nn\\
&\langle X^3 \rangle_{(\alpha, \beta, \text{sq})} =\text{Re} (\beta) =\langle X^3 \rangle_{(\alpha,\beta)}  , ~~~\langle X^4 \rangle_{(\alpha, \beta, \text{sq})} =\text{Im} (\beta)=\langle X^4 \rangle_{(\alpha,\beta)}, 
\end{align}
and the deviations are 
\be
\langle (\Delta X^{m})^2 \rangle_{(\alpha, \beta, \text{sq})}   =  \langle (X^{m})^2 \rangle_{(\alpha, \beta, \text{sq})} -\langle X^{m} \rangle_{(\alpha, \beta, \text{sq})}^2  = \langle       (\Delta X^{m})^2 \rangle_{\text{sq}}. ~~~(\text{no sum for $m$}=1,2,3,4) 
\ee
These results are equal to those of the two-mode case, (\ref{twomodemean}) and (\ref{twomodedeviations}).  Hence, also for the four-mode,  the $Sp(4;\mR)$ squeezed coherent vacuum is intuitively interpreted as the squeezed vacuum  displaced by $(\alpha, \beta)$ on  $\mathbb{C}^2\simeq \mR^4$ plane.

\section{Summary and discussions}\label{sec:summary}

We constructed the $Sp(4;\mR)$ squeezed coherent states and investigated their characteristic properties. We clarified the underlying hyperbolic geometry of the  $Sp(2; \mR)$ squeezed states in the context of the 1st non-compact Hopf map.  Taking advantage of the hierarchical geometry of the Hopf maps, we derived the $Sp(4; \mR)$ squeeze operator with Bloch four-hyperboloid geometry. Unlike the $Sp(2; \mR)$ case, the $Sp(4;\mR)$ squeezed vacua of the Dirac- and Schwinger-type are physically distinct.  
Based on the Euler angle decomposition of the $Sp(4; \mathbb{R})$ squeeze operator, we investigated   the Schwinger-type $Sp(4; \mR)$ squeezed states, and clarified the physical meaning of the four coordinates of the Bloch four-hyperboloid. In particular, the entanglement concurrence of the $Sp(4; \mR)$ squeezed one-photon state was shown to be a geometric quantity determined by the 5th axis of the Bloch four-hyperboloid. 
We evaluated  the mean values and  deviations of the 4D non-commutative coordinates for the $Sp(4;\mR)$ squeezed (coherent) states  and confirmed that they realize a natural 4D generalization of the original properties of the $Sp(2;\mR)$ squeezed states.   
 The next direction will be a construction of  an anharmonic oscillator Hamiltonian for the $Sp(4; \mathbb{R})$ squeezed state as in the   $Sp(2; \mathbb{R})$ case \cite{Gerry-1987} and its experimental realizations. Interestingly  in \cite{Gerry-Benmoussa-2000},  though not exactly same as the present case, Gerry and Benmoussa proposed  
 analogous $SU(1,1)\otimes SU(1,1)$ entangled state of two squeezed states and suggested the possibility of generation  in trapped ion experiments \cite{Leibfried-Blatt-Monroe-Wineland-2003}. 
Their indication about experimental realization may also hold for the present state.   
Besides, the $SO(2,3)$ pseudo-spin coherent state accompanies the $SU(1,1)$ Berry phase as the $SU(1,1)$ pseudo-spin coherent  state the $U(1)$ Berry phase. It is also interesting how such non-Abelian phase appears in optical experiments and brings pseudo-spin dynamics \cite{Jezek-Hernandez-1990}  particular  to its non-Abelian nature,  which may be compared to the exotic geometric phase of $SU(2)$ higher spins \cite{Lin-2004,Enriquez-Cruz-2018}.

The split quaternion was crucial  in constructing the non-compact 2nd Hopf map.  
The split quaternion is closely related to the time-reversal operation for bosons by the following identification: 
\be
(q_1,q_2,q_3) =(iT, T, i). \label{triptimerev}
\ee
Here $T$ stands for the time-reversal operator for boson, $T^2=+1$, and $i$ is the imaginary unit. Since the time reversal operator is an anti-linear operator, $T$ is anticommutative with the imaginary unit,  $Ti=-iT$,   
 and so the identification  $q_1=iT$ gives ${q_1}^2 =+1$. 
Therefore, the triplet (\ref{triptimerev}) can be regarded as a realization of the imaginary split quaternions, ${q_1}^2 ={q_2}^2=+1$, ${q_3}^2=-1$ and $q_iq_j=-q_j q_i$ $(i\neq j)$. 
In this way, the split quaternions naturally appear in the context of the time-reversal operation for bosons, just as the quaternions  for fermions. More in detail  as indicated in Table \ref{table:correspDNYM}, there are intriguing correspondences between fermion and boson sectors starting from the quaternions and split quaternions. 
\begin{table}
\center 
   \begin{tabular}{|c|c|c|}\hline
        &  Quantum information  &   Quantum optics    \\ \hline
    Time-reversal symmetry             &  $T^2=-1$  (Fermion)   &  $T^2=+1$ (Boson)       \\ \hline  
Algebra         & Quaternion  $\mathbb{H}$      &  Split-quaternion $\mathbb{H}'$                \\ \hline 
Bogoliubov trans.        &  $SO(2n)$       &  $Sp(2n; \mathbb{R}) =U(n; \mathbb{H}')$      \\ \hline  
Double covering group       &  $Spin(2n)$       &  $Mp(2n; \mathbb{R})$         \\ \hline
   Topological map    &   Hopf map   &  Non-compact Hopf map        \\ \hline
    Quantum manifold      &  Bloch sphere     &  Bloch hyperboloid     \\ \hline 
    Fundamental quantum state    &   Qubit state    &  Squeezed state      \\ \hline 
      Group coherent state     &   $SU(2)$ spin coherent state  &  $SU(1,1)$ pseudo-spin coherent state       \\ \hline 
    \end{tabular}       
\caption{ Comparison between quantum information sector of Bloch sphere and  quantum optics sector of Bloch hyperboloid. 
}
\label{table:correspDNYM}
\end{table}
The list of the boson sector of Table \ref{table:correspDNYM} may suggest that  the  non-compact (hyperbolic) geometry is no less important than  the compact (spherical) geometry for fermion sector already extensivley used  in quantum information. 
As a concrete demonstration,  we clarified the hyperbolic geometry of the squeezed states  and  applied it  to construct a generalized $Sp(4; \mR)$ formulation of the squeezed states  in the present work.     It is very tempting to excavate further hyperbolic structure in quantum mechanics and quantum information theory. 
As a straightforward study along this line,  
one may think of  applications of the non-compact 3rd Hopf map or more generally  indefinite complex projective spaces.   
It should also be  mentioned that the geometric structures of  non-compact manifolds are richer than those of the compact counterparts: Non-compact manifolds generally accommodate compact manifolds as their submanifolds, which make the geometry of non-compact manifolds to be more interesting than that of   compact manifolds.   It is expected that the study of non-compact geometry will spur the developments of   quantum information theory.

Though we focused on the squeezed states in this work, the non-compact Hopf map has begun to be applied  in various fields, such as  non-commutative geometry \cite{Hasebe-2012}, twistorial quantum Hall effect \cite{Hasebe-2010-2},   non-hermitian topological insulator  \cite{Sato-Hasebe-Esaki-Kohmoto-2011, Esaki-Sato-Hasebe-Kohmoto-2011}, and  indefinite signature matrix model of string theory \cite{Steinacker-2017,Sperling-Steinacker-2018,Stern-Xu-2018, Sperling-Steinacker-2019}.  Applications of the non-compact Hopf map may be ubiquitous. 
It may also be worthwhile to speculate its further possible applications.

\section*{Acknowledgement}

 I would like to thank Masahito Hotta and Taishi Shimoda for useful discussions. 
    This work was supported by JSPS KAKENHI Grant Number~16K05334 and 16K05138.

\appendix

\section{Symplectic algebra and metaplectic algebra}\label{appen:sympmeta}

\subsection{$U(n;\mathbb{H}')$}\label{appen:sympmetasplith}

We denote the split-quaternions  as 
\be
q_{m}=\{q_i, q_4\} =\{q_1,q_2,q_3,1\},  
\ee
which satisfy 
\begin{align}
&{q_1}^2={q_2}^2=-{q_3}^2=1, \nn\\
&q_1q_2=-q_2q_1=-q_3,~~q_3q_1=-q_1q_3=-q_2,~~q_2q_3=-q_3q_2=-q_1. \label{splitquaternonsbasalg}
\end{align}
 The quaternionic conjugate of $h=c^{m} ~q_{m}$ $(c^{m}: \text{real parameters})$ is defined as 
\be
\bar{h} =\overline{(c^{m} ~q_{m})}~\equiv~ c^{m}~\bar{q}_{m},  
\ee
with 
\be
\bar{q}_{m} =\{-q_i, 1\}. 
\ee
$GL(n; \mathbb{H}')$ is a group of split-quaternion valued $n\times n$ matrix  
\be
g=\begin{pmatrix}
g_{11} & g_{12} & \cdots & g_{1n} \\
g_{21} & g_{22} & \cdots & g_{2n} \\
\vdots & \vdots & \ddots & \vdots \\
g_{n1} & g_{n2} & \cdots & g_{nn} 
\end{pmatrix}  
\ee
where $g_{ij}$ are given by 
\be
g_{ij} =c_{ij} ^{m}~q_{m}
\ee
with $c_{ij}^{m}$ real numbers. 
The (split-)quaternionic Hermitian conjugate of $g$ is defined as 
\be
g^{\ddagger} \equiv (\bar{g})^t =\begin{pmatrix}
\overline{g_{11}} & \overline{g_{21}}& \cdots & \overline{g_{n1}} \\
\overline{g_{12}} & \overline{g_{22}} & \cdots & \overline{g_{n2}} \\
\vdots & \vdots & \ddots & \vdots \\
\overline{g_{1n}} & \overline{g_{2n}} & \cdots & \overline{g_{nn}} 
\end{pmatrix} ,  
\ee
where $\overline{q_{ij}} =c_{ij}^{m}\bar{q}_{m}$. 
The quaternionic conjugate and the quaternionic Hermitian conjugate have the following properties: 
\bse
\begin{align}
&\overline{(h_1~  h_2)} =\overline{h_2}~ \overline{h_1}, \\
&(g_1 \cdot g_2)^{\ddagger} ={g_2}^{\ddagger}\cdot {g_1}^{\ddagger}.  \label{g1g2ddagg}
\end{align}
\ese
Here, we consider the $GL(n; \mathbb{H}')$ transformation 
that keeps the inner product of split-quaternion vectors invariant,  
\begin{equation}
g^{\ddagger}~g=1, \label{condunhda}
\end{equation}
and such a transformation is called the split-quaternionic unitary transformation denoted by 
$U(n; \mathbb{H}')$.\footnote{ 
Since the inner product of split quaternion $h=\sum_{m=1}^4 h^{m}q_{m}$ 
is split signature, the overall signature of the inner product is not essential :  $-\bar{h} h = (h^1)^2 +(h^2)^2-(h^3)^2-(h^4)^2=+\bar{h'}h'$ with  
%
$h'=\sum_{m=1}^4 h^{5-m}q_{m}$. 
Hence, we find 
\be
U(n-m, m; \mathbb{H}') =U(n; \mathbb{H}). 
\ee
} 
When we introduce  $u(n; \mH')$ generator $X$ as 
\be
g=e^X, 
\ee
(\ref{condunhda}) imposes the following condition on   $X$:  
\begin{equation}
X^{\ddagger}=-X.  
\label{dashdefgeneuquaternions}
\end{equation}
The generators of $U(n; \mathbb{H}')$ are simply  split-quaternionic anti-Hermitian matrices. 
Then, the  $u(n; \mH')$ basis matrices are given by 
\be
\begin{pmatrix}
0 & 0 & 0 & 0 & 0 \\
0 & 0  &  \ddots  & 0 & 0 \\
0 & 0 &  q_i & 0 & 0 \\
0 & \ddots  &  0 & 0 & 0 \\
0 & 0 &  0 & 0 & 0 
\end{pmatrix},~~~\begin{pmatrix}
0 & 0 & 0 & 0 & 0 \\
0 & \ddots  &  0 & q_{m} & 0 \\
0 & 0 &  0 & 0 & 0 \\
0 & -\bar{q}_{m} &  0 & \ddots  & 0 \\
0 & 0 &  0 & 0 & 0 
\end{pmatrix}. 
\ee
The dimension of $u(n; \mathbb{H}')$ algebra is then counted as 
\be
\dim U(n; \mathbb{H}')
=n\times 3 +  \frac{n(n-1)}{2} \times 4  
=  n(2n+1). 
\ee

We can realize the split-quaternions by the $su(1,1)$ matrices\footnote{(\ref{quaternionicq}) gives another matrix realization of the spilt quaternions.}  
\be
\{q_1, q_2, q_3, 1\} =\{\sigma_x, \sigma_z, i\sigma_y, 1_2\}  
\label{matrixsu11splitqua}
\ee
and demonstrate the isomorphism $U(n; \mathbb{H}')\simeq Sp(2n; \mathbb{R})$ as follows. 
Notice that the matrices on the right-hand side of (\ref{matrixsu11splitqua}) are all real matrices, and so the $U(n; \mathbb{H}')$ group elements can be expressed by  real matrices, $g^*=g$. In the matrix realization, the split-quaternionic conjugate is not equal to the usual Hermitian conjugate but given by  
\be
\bar{q}_{m} =\{-\sigma_x, -\sigma_z, -i\sigma_y, 1_2\} =\epsilon^t ~{q_{m}}^{\dagger}~\epsilon=\epsilon^t ~{q_{m}}^{t}~\epsilon
\ee
where 
\be
\epsilon\equiv i\sigma_y. 
\ee
Consequently for the matrix realization of $U(n; \mH')$,  
we have 
\be
g^{\ddagger} =\mathcal{E}^t~ g^{t}~\mathcal{E}, 
\ee
 with 
\be
\mathcal{E} =\begin{pmatrix}
\epsilon & 0 & 0 & 0 \\
0 & \epsilon & 0 & 0 \\
0 & 0 & \ddots & 0 \\
0 & 0 & 0 & \epsilon
\end{pmatrix}.  
\ee
The $U(n; \mathbb{H}')$ condition (\ref{condunhda}) can be expressed as 
\be
g^t ~\mathcal{E}~g =\mathcal{E}. \label{presp2nrcond}
\ee
Under the following unitary transformation 
\be
g ~~\rightarrow~~U ~g~ U^t
\ee
where 
\be
U=\begin{pmatrix}
\bs{e}_1 & \bs{e}_3 & \cdots & \bs{e}_{2n-1} & \bs{e}_2 & \bs{e}_4 & \cdots & \bs{e}_{2n}
\end{pmatrix}
\ee
with $(\bs{e}_a)_b \equiv \delta_{ab}$ $(a,b=1,2,\cdots,2n)$,  (\ref{presp2nrcond}) is transformed as   
 \begin{equation}
 g^t ~J~ g=J. 
 \end{equation}
This is the very condition that defines the  $Sp(2n; \mathbb{R})$ group (\ref{sp2ndef}). We thus find  
\be
U(n; \mathbb{H}')\simeq Sp(2n; \mathbb{R}). 
\ee

\subsection{Symplectic algebra $sp(2n; \mathbb{R})$}

 Elements of the $Sp(2n; \mathbb{R})$ group are given by a real matrix  $g$ that satisfies  the condition\footnote{It is obvious that the symplectic form 
\begin{equation}
\sum_{i=1}^n dx_i \wedge dp_i =\frac{1}{2}\sum_{A,B=1}^{2n}J_{AB} dq_A \wedge dq_B ~~~~~~~(q_A\equiv (x_i, p_j))
\end{equation}
is invariant under the $Sp(2n; \mR)$ transformation 
\begin{equation}
dq_A \rightarrow g_{AB} dq_B 
\end{equation}
with $g$  subject to (\ref{sp2ndef}). 
}    
 \begin{equation}
 g^t ~J~ g=J, 
 \label{sp2ndef}
 \end{equation}
where $J$ is called the $Sp(2n; \mathbb{R})$ invariant matrix:
\begin{equation}
J=\begin{pmatrix}
0 & 1_n \\
-1_n & 0 
\end{pmatrix}. \label{sp2nrinvmat}
\end{equation}
With the generator $X$ 
\begin{equation}
g=e^X, 
\end{equation}
the relation (\ref{sp2ndef}) can be rewritten as 
\begin{equation}
X^t J+JX=0, 
\label{sprelatgene}
\end{equation}
or equivalently 
\begin{equation}
(JX)^t=JX. \label{xjjx}
\end{equation}
(\ref{xjjx}) determines the form of $X$ as   
\begin{equation}
X=
\begin{pmatrix}
M & S_1 \\
S_2 & -M^t 
\end{pmatrix}, \label{sp2ngeger}
\end{equation}
where $M$ denotes arbitrary $n\times n$ real matrix, and $S_1$ and $S_2$ are two arbitrary $n\times n$ symmetric real matrices.  
 The dimension of the symplectic algebra is readily obtained as 
\be
\dim (sp(2n; \mathbb{R})) =(\text{real degrees of }M) +(\text{real degrees of }S)= n^2+\frac{n(n+1)}{2}\times 2 =  n(2n+1).  \label{dimsp2nrreal}
\ee
From  (\ref{sp2ngeger}), we can choose $n(2n+1)$ $sp(2n; \mathbb{R})$ basis  matrices  as 
\begin{align}
&(X^i_j)_{ab} =\delta_{a,i }\delta_{b,j}-\delta_{n+i,b}\delta_{n+j,a}=(X^j_i)_{ba}, \nn\\
&(X^{ij})_{ab} =\delta_{a,i}\delta_{b,n+j} +\delta_{a,j}\delta_{b,n+j} =(X^{ji})_{ab},\nn \\
&(X_{ij})_{ab} =-\delta_{a,n+i}\delta_{b,j} -\delta_{a,n+j}\delta_{b,j} =(X_{ji})_{ab}, 
 \label{matrixreasp2nr}
\end{align}
where $i,j=1,2,\cdots, n$ and $a,b=1,2,\cdots, 2n$. 
They satisfy 
\begin{align}
&[X_{ij}, X_{kl}]=[X^{ij}, X^{kl}]=0, \nn\\
&[X_{ij}, X^{kl}] =X_i^l \delta_j^k +X_j^l \delta_i^k +X_i^k \delta_j^l +X_j^k \delta_i^l ,\nn\\
&[X_{ij}, X_k^l] =X_{ik}\delta_j^l +X_{jk}\delta_i^l, \nn\\
&[X^{ij}, X^{l}_k] =-X^{il}\delta_k^j -X^{jl}\delta_k^i, \nn\\
&[X_j^i, X_k^l] =X^i_k \delta_{j}^l -X_j^l\delta_k^i. \label{sp2nalgebraex}
\end{align}

The $Sp(2n, \mR)$ invariant matrix (\ref{sp2nrinvmat}) 
is diagonalized by the following unitary transformation 
\be
\Omega ~J ~\Omega^{\dagger} =iK, 
\ee
where $K$ is a  diagonal matrix with neutral  components: 
\be
K \equiv \begin{pmatrix}
1_n & 0_n \\
0_n & -1_n
\end{pmatrix}, ~~~~~~~(K^{-1}=K^{\dagger}=K) \label{kmatsympano}
\ee
and $\Omega$ can be taken as 
\be
\Omega =\frac{1}{\sqrt{2}}
\begin{pmatrix}
 R & -iR \\
R & iR
\end{pmatrix} ~~~~~~~~~~(\Omega^{\dagger} =\Omega^{-1})\label{expomega}
\ee
with $n\times n$ matrix $R$
\be
R =\begin{pmatrix}
0 & 0 & 0 & \cdots & 0 & 1 \\
0 & 0 & 0 & \cdots & 1 & 0 \\
0 & 0 & 0 & 1 & 0 & 0 \\
0 & 0 & 1 & \cdots & 0 & 0 \\
0 & 1 & 0 & \cdots & 0 & 0 \\
1 & 0 & 0 & \cdots & 0 & 0  
\end{pmatrix}. 
\ee
The $Sp(2n; \mR)$ group condition  (\ref{sp2ndef}) 
can be expressed as 
\be
g^t \Omega^{\dagger}~ K~ \Omega g =\Omega^{\dagger}~ K~ \Omega. \label{matrixcondk}
\ee
Since $g$ is a real matrix, $g^t=g^{\dagger}$,  (\ref{matrixcondk}) is  rewritten as 
\be
(\Omega~ g ~\Omega^{\dagger})^{\dagger} ~K ~(\Omega ~g ~\Omega^{\dagger})=K. 
\ee
Therefore, 
\be
g'\equiv \Omega~ g ~\Omega^{\dagger} \label{defgdash}
\ee
realizes another representation of the $Sp(2n, \mathbb{R})$ group element that satisfies  
\be
{g'}^{\dagger} ~K ~g'=K. \label{condsp2nkco}
\ee
Notice that $g'$  no longer denotes a real matrix unlike $g$.  Since $g$ is a  $2n\times 2n$ real matrix,  $g'$ (\ref{defgdash}) with $\Omega$ (\ref{expomega}) is  parameterized as  
\be
g' =\begin{pmatrix}
U & V^* \\
V & U^{*}
\end{pmatrix},  \label{expuniformgda}
\ee
where each of $U$ and $V$ is a $n\times n$ complex matrix. 
For $g'$ subject to (\ref{condsp2nkco}), the blocks $U$ and $V$ must satisfy 
\be
U^{\dagger}U-V^{\dagger}V=1_n, ~~~~~~U^t V -V^t U=0_n.  
\label{sp2ngroupdconds}
\ee
The (real) number of constraints of (\ref{sp2ngroupdconds}) is $(n^2)+(n^2-n)=n(2n-1)$, and then the  real degrees of freedom $g'$ is obtained as  
\be
(2n)^2 - n(2n-1) =n(2n+1), 
\ee
which is indeed the dimension of the $sp(2n; \mathbb{R})$ algebra (\ref{dimsp2nrreal}). 
We can readily identify the form of the associated $sp(2n; \mR)$ generator $X'$   
\be
 \begin{pmatrix}
U & V^* \\
V & U^*
\end{pmatrix} =e^{iX'}  
\ee
as\footnote{$X'$ is related to  the original $sp(2n; \mathbb{R})$ (\ref{sp2ngeger}) as 
\be
\begin{pmatrix}
M & S_1 \\
S_2 & -M^t
\end{pmatrix} = -i\Omega^{\dagger}  \begin{pmatrix}
H & S^* \\
-S & -H^*
\end{pmatrix}\Omega
\ee
or 
\be
\begin{pmatrix}
H & S^* \\
-S & -H^*
\end{pmatrix}=i\Omega^{\dagger}  \begin{pmatrix}
M & S_1 \\
S_2 & -M^t
\end{pmatrix}\Omega. 
\ee
Then we can identify the block components of $X'$ as 
\bse
\begin{align}
&H=-\frac{1}{2}R(S_1-S_2-iM+iM^t)R,\\
&S=\frac{1}{2}R(S_1+S_2-iM-iM^t)R. 
\end{align}
\ese
}   
\be
X'=\begin{pmatrix}
H & S^* \\
-S & -H^*
\end{pmatrix},  \label{matrgenesp2nr}
\ee
where $H$ is arbitrary $n\times n$ Hermitian matrix and $S$ is arbitrary $n\times n$   symmetric complex matrix.\footnote{The real degrees of freedom of $X'$ is then counted as  $n^2+n(n+1)=n(2n+1)$.}   
Obviously, the maximal Cartan sub-algebra is given by 
\be
\begin{pmatrix}
H & 0 \\
0 & -H^*
\end{pmatrix}, 
\ee
which is the generator of $U(n)$. ($-H^*$ denotes the complex representation of $H$ and satisfies the same algebra of $H$.) $U(n)$ is the maximal Cartan group of the Cartan-Iwasawa decomposition of $Sp(2n; \mR)$ (see (\ref{polardecompsymp})).  
(\ref{condsp2nkco}) imposes the following condition on $X'$:  
\be
X'^{\dagger} K   - K X'=0 
\ee
or 
\be
X'^{\dagger} = K
X' K, 
\ee
and so the block matrices of $X$ must satisfy    
\be
H^{\dagger}=H, ~~~S^t=S. 
\ee
$H$ is  a $n\times n$ Hermitian matrix and $S$  a complex symmetric matrix. 
Notice that  
\be
K X'=\begin{pmatrix}
H & S^* \\
S & H^*
\end{pmatrix} 
\ee
denotes a Hermitian matrix. By sandwitching $KX$ by a Dirac spinor operator 
\be
\hat{\psi} =
\begin{pmatrix}
a_1 & 
\cdots & 
a_n & 
b_1^{\dagger} & 
\cdots &
b_N^{\dagger}
\end{pmatrix}^t \label{comppsiop}
\ee
and its conjugate, 
we can construct  Hermitian operators that satisfy $sp(2n; \mR)$ algebra. From the Hermitian operators,   independent operators are extracted as  
\bse
\begin{align}
&X_i^j =\frac{1}{2}\{a_i, a_j^{\dagger}\} + \frac{1}{2}\{b_i, b_j^{\dagger}\}={X_j^i}^{\dagger} ~~:~~n^2, \\
&X^{ij} =a_i^{\dagger}b_j^{\dagger}+a_j^{\dagger}b_i^{\dagger} =X^{ji}~~~~~~~~~~~~~~~:~~\frac{1}{2}n(n+1), \\
&X_{ij} =a_ib_j+a_jb_i=X_{ji}~~~~~~~~~~~~~~~~:~~\frac{1}{2}n(n+1). 
\end{align}\label{diracopsp2nr}
\ese
They indeed constitute  non-Hermitian operators for the $sp(2n; \mathbb{R})$ algebra (\ref{sp2nalgebraex}). In particular 
\be
X_{i}^j =a_j^{\dagger}a_i +b_j^{\dagger}b_i +\delta_i^j
\ee
satisfy the maximal Cartan $u(n)$  sub-algebra. 
The $Sp(2n; \mathbb{R})$ Casimir is derived as 
\be
C=X^{ij}X_{ij} +X_{ij}X^{ij} -2X^i_j X^j_i = -2(a_i^{\dagger}a_i -b^{\dagger}b_i+n)(a_i^{\dagger}a_i -b_i^{\dagger}b_i-n).
\ee

\subsection{Metaplectic algebra}\label{app:metgr}

The metaplectic group $Mp(2n; \mathbb{R})$ is the double cover of the symplectic group $Sp(2n; \mathbb{R})$: 
\be
Mp(2n; \mathbb{R})/\mathbb{Z}_2 \simeq Sp(2n; \mathbb{R}). 
\ee
Instead of the ``complex'' operator $\hat{\psi}$ (\ref{comppsiop}), we here introduce a ``real'' operator  
\be
\hat{\phi}=
\begin{pmatrix}
a_1 & 
\cdots & 
a_n   & 
a_{1}^{\dagger} & 
\cdots & 
a_{n}^{\dagger}
\end{pmatrix}^t, \label{realopphi}
\ee
that satisfies the real condition 
\be
\hat{\phi}^{*} =C\hat{\phi} ~~~~~(C\equiv \begin{pmatrix}
0 & 1_n \\
1_n & 0 
\end{pmatrix})
\ee
and 
\be
[\hat{\phi}_{\alpha}, \hat{\phi}_{\beta}]=J_{\alpha\beta}
\ee
with $J$ (\ref{sp2nrinvmat}). 
From (\ref{realopphi}), we can construct the following Hermitian operator 
\be
O_M = \frac{1}{2}~\hat{\phi}^t ~CK  X ~\hat{\phi}=  -\frac{1}{2}~\hat{\phi}^t ~J X ~\hat{\phi}. 
\ee
Here $X$ is given by (\ref{matrgenesp2nr}) and $J X$ is 
\be
J~X =-\begin{pmatrix}
S & H^* \\
H & S^* 
\end{pmatrix}=(JX)^t. 
\ee

From the original  $sp(2n; \mathbb{R})$ matrix $X$ (\ref{sp2ngeger}), we can also construct 
a symmetric matrix 
\be
J X =\begin{pmatrix}
S_2 & -M^t \\
-M & -S_1
\end{pmatrix}=(JX)^t \label{jxorigmat}
\ee
and associated non-Hermitian operator
\be
\hat{X}=-\frac{1}{2}~\hat{\phi}^t ~J X ~\hat{\phi}. 
\label{hatxijop}
\ee
The basis matrices of (\ref{jxorigmat}) are given by 
\begin{align}
&(JX^i_j)_{a,b} =(JX^i_j)_{b,a} =-\delta_{a,n+i}\delta_{b,j}-\delta_{a,j}\delta_{b,n+i}, \nn\\
&(JX^{ij})_{a,b} =(JX^{ji})_{a,b} =-\delta_{a,n+i}\delta_{b,n+j}-\delta_{a,n+j}\delta_{b,n+i}, \nn\\
&(JX_{ij})_{a,b} =(JX_{ji})_{a,b} =\delta_{a,i}\delta_{b,j}+\delta_{a,j}\delta_{b,i}, ~~~~~~~~~~(a,b=1,2,\cdots, 2n)  \label{hatxijopmat}
\end{align}
where $X$s are (\ref{matrixreasp2nr}). It is not difficult to verify  
that the corresponding operators satisfy the $sp(2n; \mathbb{R})$ algebra (\ref{sp2nalgebraex}) using the relations, $J^2=-1_{2n}$ and $(JX)^t=JX$.  
The basis operators corresponding to (\ref{hatxijopmat}) are obtained as 
\begin{align}
&\hat{X}_{j}^i=\frac{1}{2}\{ a^{\dagger}_i, a_j\}=a^{\dagger}_i a_j +\frac{1}{2}\delta_{ij}=(\hat{X}_i^j)^{\dagger}, ~~~~:~n^2 , \nn\\
&\hat{X}^{ij}=a_i^{\dagger}a_j^{\dagger} =\frac{1}{2}\{a_i^{\dagger}, a_j^{\dagger}\}=\hat{X}^{ji},~~~~~~~~~~~~~~~:~\frac{1}{2}n(n+1) \nn\\
&\hat{X}_{ij}=a_i a_j =\frac{1}{2}\{a_i, a_j\}=\hat{X}_{ji}, ~~~~~~~~~~~~~~~~:~\frac{1}{2}n(n+1)
\end{align}
which satisfy (\ref{sp2nalgebraex}). 
These operators for metaplectic representation can also be obtained from (\ref{diracopsp2nr}) with replacement of the operator $b$ by $a$ and by changing the overall scale factor by $1/2$.  

 The $Sp(2n; \mathbb{R})$ Casimir operator for the metaplectic representation  becomes a constant: 
\be
C=\hat{X}^{ij}\hat{X}_{ij} + \hat{X}_{ij}\hat{X}^{ij} -2 \hat{X}^i_j\hat{X}^j_i =n(n+\frac{1}{2}). 
\label{casimirsp2n} 
\ee

\subsection{Bogoliubov canonical transformations}\label{subsec:bogtrans}

We consider the canonical commutation relations for $n$ bosons: 
\be
[a_i, a_j^{\dagger}]=\delta_{ij}, ~~~[a_i, a_j]=0. ~~~~~(i,j=1,2,\cdots, n)
\ee
The transformation preserving the canonical commutation relations is called the Bogoliubov canonical transformation. 
Let us consider a general linear transformation of the bosonic operators:  
\be
a'_i=u_{ji}a_j +v_{ji}a^{\dagger}_j. 
\ee
For the canonical transformation, $u_{ij}$ and $v_{ij}$ should satisfy 
\be
[a'_i, {a'}_j^{\dagger}]=\delta_{ij}, ~~[a'_i, a'_j]=0~~~\rightarrow~~~u_{ki}u_{jk}^* -v_{ki}v_{kj}^*=\delta_{ij},~~u_{ki}v_{kj}-v_{ki}u_{kj}=0,  \label{const2}
\ee 
which, in matrix notation,  becomes  
\be
U^t U^*-V^t V^*=1_n,~~~~U^t V -V^t U=0_n. \label{constrauandvs2}
\ee 
(\ref{constrauandvs2}) is nothing but  the condition used in the definition of  the $Sp(2n;\mathbb{R})$ group (\ref{sp2ngroupdconds}). We thus find that the canonical transformation for bosons is described by the symplectic group \cite{Itzykson-1967,Perelomov-book}.

In a similar manner, we can readily identify the canonical transformation group for fermions.   
The commutation relations for $n$ fermions are given by  
\be
\{f_i, f_j^{\dagger}\}=\delta_{ij}, ~~~\{f_i, f_j\}=0.  ~~~~~(i,j=1,2,\cdots, n)
\ee
Among general linear transformations 
\be
f'_i=u_{ji}f_j +v_{ji}f^{\dagger}_j,  
\ee
the canonical transformation is realized as  
\be
\{f_i, {f'}_j^{\dagger}\}=\delta_{ij},~~\{f'_i, f'_j\}=0~~\rightarrow~~~u_{ki}u_{jk}^* +v_{ki}v_{kj}^*=\delta_{ij},~~u_{ki}v_{kj}+v_{ki}u_{kj}=0,  \label{fconst2}
\ee
which, in matrix notation, is given by
\be
U^t U^*+V^t V^*=1_n,~~U^t V +V^t U=0_n. \label{fconstrauandvs2}
\ee
Recall that the arbitrary $2n\times 2n$ real matrix $g$ is unitarily equivalent to  $g'=\Omega~ g~ \Omega^{\dagger} =\begin{pmatrix}
U & V^* \\
V & U^*
\end{pmatrix}$ (\ref{expuniformgda}). Therefore, the condition for the $SO(2n)$ group element $g$,  
\be
g^t g=1_{2n}, 
\ee
can be restated for $g'$ as 
\be
g'^{\dagger}g'=1_{2n}. \label{gdashcond}
\ee
(\ref{gdashcond}) imposes the following conditions on  the block components of $g'$ as 
\be
U^{\dagger} U+V^{\dagger} V=1_n,~~~U^t V +V^t U=0_n,  \label{fconstrauandvs22}
\ee
which is exactly the condition (\ref{fconstrauandvs2}), so we find that the canonical transformation for $n$ fermions is  the $SO(2n)$ transformation \cite{Berezin-1978,Perelomov-book}. 
For comparison to the expression  of the  $sp(2n; \mathbb{R})$ generator (\ref{matrgenesp2nr}),  we derive the matrix form of the $so(2n)$ generator $X'$ defined by   $g'=e^{iX'}$. (\ref{gdashcond}) implies 
\be
X'^{\dagger} =X' , 
\ee
or
\be
X'=\begin{pmatrix}
H & A^* \\
-A & -H^*
\end{pmatrix}. \label{geneoenu}
\ee
Here, $H$ and $A$ respectively denote arbitrary $n\times n$ Hermitian matrix and complex anti-symmetric matrix with $n^2$ and $n(n-1)$ real degrees of freedom.  
In total, $X'$ carries $n(2n-1)$ real degrees of freedom, $i.e.$, the dimension of the $so(2n)$ algebra.

\section{Useful formulas for the split-quaternions and $so(2, 3)$ algebra}\label{sec:split-so5}

\subsection{Algebra of split-quaternions}\label{appen:splitq}

The algebra of the split-quaternions (\ref{splitquaternonsbasalg}) is concisely expressed as 
\be
\{q^i,  q^j\}=-2g^{ij},~~~ 
[q^i, q^j]=-2\epsilon^{ijk}q_k, 
\ee
where 
\be
g_{ij}=g^{ij}=\text{diag}(-1, -1, +1), ~~~
\epsilon^{123}=1. 
\ee
The split quaternions  satisfy    
\begin{subequations}
\begin{align}
&q^{m}\bar{q}^{n}+q^{n}\bar{q}^{m}=\bar{q}^{m} q^{n}+\bar{q}^{n} q^{m}=2 g^{mn}, \\
&q^{m}\bar{q}^{n}-q^{n}\bar{q}^{m}=2{\eta}^{mn i} q_i, ~~~~~\bar{q}^{m} q^{n}-\bar{q}^{n} q^{m} =2\bar{\eta}^{mn i} q_i
\end{align}
\end{subequations}
and 
\begin{subequations}
\begin{align}
&q^i q^{m}=-{\eta}^{mn i} q_{n}, ~~~~q^{m} q^i =\bar{\eta}^{mn i} q_{n}, \\
&q^i\bar{q}^{m}=-\bar{\eta}^{mn i}\bar{q}_{n}, ~~~~\bar{q}^{m} q^i =\eta^{mn i}\bar{q}_{n}. 
\end{align}
\end{subequations}
Here, $g_{mn}$ is 
\be
g_{mn}=\text{diag}(-1, -1, +1, +1), 
\ee
and  $\eta^{mn i}$ and $\bar{\eta}^{mn i}$ are the 't Hooft symbols: 
\be
\eta^{mn i}=\epsilon^{m n i 4} +g^{m i}g^{n 4} -g^{n i}g^{m 4}, 
~~~
\bar{\eta}^{mn i}=\epsilon^{m n i 4} -g^{m i}g^{n 4} +g^{n i}g^{m 4}. 
\ee
They satisfy 
\begin{subequations}
\begin{align}
&\frac{1}{2}\epsilon_{mnpq}\eta^{pq i} =\eta_{mn}^{~~~ i},~~~~ \frac{1}{2}\epsilon_{mnpq}\bar{\eta}^{pq i} =-\bar{\eta}_{mn}^{~~~ i}, \\
&\eta^{m n i}\eta_{mn j} =4\delta^{i}_{~j},~~~~~~~\bar{\eta}^{m n i}\bar{\eta}_{mn j} =4\delta^{i}_{~j} ~~~~~~~~\eta^{mn i}\bar{\eta}_{mn j}=0.
\end{align}
\end{subequations}

\subsection{$so(2,3)$ generators and other variants}\label{append:so23impmatr}

Using the split-quaternions, we can realize the $SO(2,3)$ gamma matrices as 
\be
\gamma^{m}=\begin{pmatrix}
0 & \bar{q}^{m} \\
q^{m} & 0 
\end{pmatrix}, ~~~\gamma^5=\begin{pmatrix}
1 & 0 \\
0 & -1 
\end{pmatrix},  
\ee
which satisfy 
\be
\{\gamma^a , \gamma^b\} =2 g^{ab} 
\ee
with 
\be
g_{ab}=g^{ab}=\text{diag}(-1,-1,+1, +1, +1). 
\ee
The $so(2, 3)$ generators, $\sigma^{ab}=-i\frac{1}{4}[\gamma^a, \gamma^b]$, are obtained as 
\be
\sigma^{mn}=-i\frac{1}{4}[\gamma^{m}, \gamma^{n}]=-i\frac{1}{2}\begin{pmatrix}
\bar{\eta}^{mn i} q_i & 0 \\
0 & {\eta}^{mn i} q_i
\end{pmatrix}, ~~~~\sigma^{m 5} =-\sigma^{5 m} =-i\frac{1}{4}[\gamma^{m}, \gamma^5]=i\frac{1}{2}
\begin{pmatrix}
0 & \bar{q}^{m} \\
-q^{m} & 0 
\end{pmatrix}, 
\ee 
and they satisfy 
\be
[\sigma_{ab}, \sigma_{cd}] =i(g_{ac}\sigma_{bd} -g_{ad}\sigma_{cd} + g_{bd}\sigma_{ac} -g_{cd}\sigma_{ad}).
\ee
As maximal sub-algebra, the $so(2,3)$ algebra accommodates  $so(2,2)\simeq su(1,1)\oplus su(1,1)$  whose matrix representation is given by 
\be
{S}^i \equiv -\frac{1}{4}{\eta}_{mn}^{~~i}\sigma^{mn} =i\frac{1}{2}\begin{pmatrix}
0 & 0 \\
0 & q^i 
\end{pmatrix}, ~~~
\bar{S}^i \equiv -\frac{1}{4}\bar{\eta}_{mn}^{~~i}\sigma^{mn} =i\frac{1}{2}\begin{pmatrix}
q^i & 0 \\
0 & 0 
\end{pmatrix}.  
\ee
They satisfy 
\be
[S^i , S^j]=-i\epsilon^{ijk}S_k, ~~~
[\bar{S}^i , \bar{S}^j]=-i\epsilon^{ijk}\bar{S}_k, ~~~
[S^i , \bar{S}^j]=0.
\ee
The remaining algebraic relations of  $so(2,3)$  are  
\be
[S^i, \sigma^{m 5}] =-i\frac{1}{2}\eta^{mn i}\sigma_{n 5}, ~~~[\bar{S}^i, \sigma^{m 5}] =-i\frac{1}{2}\bar{\eta}^{mn i}\sigma_{n 5}. 
\ee
As in the main text,  we  express the split-quaternions as  (\ref{quaternionicq})
\be
\{q^1, q^2, q^3, q^4\} =\{\sigma_x, \sigma_y, -i\sigma_z, 1_2\}, 
\ee
and the $SO(2,3)$ gamma matrices become 
\be
\gamma^{1}=\begin{pmatrix}
0 & -\sigma_x \\
\sigma_x & 0 
\end{pmatrix},~~\gamma^{2}=\begin{pmatrix}
0 & -\sigma_y \\
\sigma_y & 0 
\end{pmatrix},~~\gamma^{3}=\begin{pmatrix}
0 & i\sigma_z \\
-i\sigma_z & 0 
\end{pmatrix},~~\gamma^4=\begin{pmatrix}
0 & 1_2 \\
1_2 & 0 
\end{pmatrix},~~\gamma^5=\begin{pmatrix}
1_2 & 0 \\
0 & -1_2 
\end{pmatrix}. \label{so23n1exmagammat}
\ee
The $so(2,3)$ matrices are  
\begin{align}
&\!\!\!\!\!\!\!\!\!\!\sigma^{12} =  -\frac{1}{2}\begin{pmatrix}
\sigma_z & 0 \\
0 & \sigma_z
\end{pmatrix}  ,~~~\sigma^{13}= -i\frac{1}{2}\begin{pmatrix}
\sigma_y & 0 \\
0 & \sigma_y
\end{pmatrix}    ,  ~~~\sigma^{14} = i\frac{1}{2}\begin{pmatrix}
\sigma_x & 0 \\
0 & -\sigma_x
\end{pmatrix}   ,~~~\sigma^{15}= -i\frac{1}{2}\begin{pmatrix}
0 & \sigma_x \\
\sigma_x & 0
\end{pmatrix},~~~
\sigma^{23}= i\frac{1}{2}\begin{pmatrix}
\sigma_x & 0 \\
0 & \sigma_x
\end{pmatrix} ,\nn\\
&\!\!\!\!\!\!\!\!\!\!\sigma^{24} =i\frac{1}{2}\begin{pmatrix}
\sigma_y & 0 \\
0 & -\sigma_y
\end{pmatrix}   ,~~~\sigma^{25}= -i\frac{1}{2}\begin{pmatrix}
0 & \sigma_y \\
\sigma_y & 0
\end{pmatrix}  ,~~~
\sigma^{34} =  \frac{1}{2}\begin{pmatrix}
\sigma_z & 0 \\
0 & -\sigma_z
\end{pmatrix}  ,~~~\sigma^{35}=-\frac{1}{2}\begin{pmatrix}
0 & \sigma_z  \\
\sigma_z & 0 
\end{pmatrix} ,~~~ \sigma^{45}=i\frac{1}{2}\begin{pmatrix}
0 & 1 \\
-1 & 0
\end{pmatrix} .   \label{so23eachgenemat}
\end{align}
They satisfy 
\be
(\gamma^a)^{\dagger}=\gamma_a,~~~~~(\sigma^{ab})^{\dagger}=\sigma_{ab}.  
\ee
The associated Hermitian matrices $k^a$ and $k^{ab}$ (\ref{kskabs}) are represented as 
\be
k^i =\begin{pmatrix}
0 & i\kappa^i \\
-i\kappa^i & 0 
\end{pmatrix}, ~~k^4=\begin{pmatrix}
0 & \sigma_z \\
\sigma_z & 0 
\end{pmatrix}, ~~k^5=\begin{pmatrix}
\sigma_z & 0 \\
0 & -\sigma_z
\end{pmatrix}, 
\ee
and 
\be
k^{mn} =-\frac{1}{2}\begin{pmatrix}
\bar{\eta}^{mn i}\kappa_i & 0 \\
0 & {\eta}^{mn i} \kappa_i 
\end{pmatrix},~~k^{i 5} =-\frac{1}{2} \begin{pmatrix}
0 & \kappa^i  \\
\kappa^i  & 0 
\end{pmatrix},~~
k^{45} =i\frac{1}{2}
\begin{pmatrix}
0 & \sigma_z \\
-\sigma_z & 0 
\end{pmatrix}. 
\ee
With the Pauli matrices, they are  given by 
\be
k^x =\begin{pmatrix}
0 & -i\sigma_y \\
i\sigma_y & 0 
\end{pmatrix}, ~~k^y =\begin{pmatrix}
0 & i\sigma_x \\
-i\sigma_x & 0 
\end{pmatrix}, ~~k^z =\begin{pmatrix}
0 & i 1_2 \\
-i1_2 & 0 
\end{pmatrix}, ~~k^4=\begin{pmatrix}
0 & \sigma_z \\
\sigma_z & 0 
\end{pmatrix}, ~~k^5=\begin{pmatrix}
\sigma_z & 0 \\
0 & -\sigma_z
\end{pmatrix}  \label{ksfiveexpli}
\ee
and 
\begin{align}
&k^{12} =  -\frac{1}{2}\begin{pmatrix}
1_2 & 0 \\
0 & 1_2
\end{pmatrix}  ,~~k^{13}= -\frac{1}{2}\begin{pmatrix}
\sigma_x & 0 \\
0 & \sigma_x
\end{pmatrix}   ,~~k^{14} = \frac{1}{2}\begin{pmatrix}
-\sigma_y & 0 \\
0 & \sigma_y
\end{pmatrix}   ,~~k^{15}= \frac{1}{2}\begin{pmatrix}
0 & \sigma_y \\
\sigma_y & 0
\end{pmatrix}, ~~k^{23}= -\frac{1}{2}\begin{pmatrix}
\sigma_y & 0 \\
0 & \sigma_y
\end{pmatrix},   \nn\\
&k^{24} =\frac{1}{2}\begin{pmatrix}
\sigma_x & 0 \\
0 & -\sigma_x
\end{pmatrix}   ,~~k^{25}= -\frac{1}{2}\begin{pmatrix}
0 & \sigma_x \\
\sigma_x & 0
\end{pmatrix}   ,~~k^{34} =  \frac{1}{2}\begin{pmatrix}
1_2 & 0 \\
0 & -1_2
\end{pmatrix}  ,~~k^{35}=-\frac{1}{2}\begin{pmatrix}
0 & 1_2  \\
1_2 & 0 
\end{pmatrix} ,~~ k^{45}=\frac{1}{2}\begin{pmatrix}
0 & i\sigma_z \\
-i\sigma_z & 0
\end{pmatrix} .  
\end{align}
$k^{ab}$ satisfy the relation 
\be
k^{ab} k k^{cd}- k^{cd} k k^{ab} = ig^{ac}k^{bd}-ig^{ad}k^{bc}+ig^{bd}k^{ac}-ig^{bc}k^{ad}.  
\ee
The antisymmetric matrices $m^a$ (\ref{antisymmamat}) and symmetric matrices $m^{ab}$ 
(\ref{defemab}) are given by 
\be
m^a
=\biggl\{ \begin{pmatrix}
0 & -\sigma_z \\
\sigma_z & 0 
\end{pmatrix},~   \begin{pmatrix}
0 & -i1_2 \\
i1_2 & 0 
\end{pmatrix},~\begin{pmatrix}
0 & -i\sigma_x \\
i\sigma_x & 0 
\end{pmatrix},~\begin{pmatrix}
0 & i\sigma_y \\
i\sigma_y & 0 
\end{pmatrix},~\begin{pmatrix}
i\sigma_y &  0\\
0 & -i\sigma_y
\end{pmatrix}  \biggr\} ,
\ee
and 
\be
m^{mn}=\frac{1}{2}
\begin{pmatrix}
\bar{\eta}^{mn i} m_i & 0 \\ 
0 & \eta^{mn i} m_i
\end{pmatrix}, ~~~~m^{i5} =\frac{1}{2}
\begin{pmatrix}
0 &  m^i \\
 m^i &0 
\end{pmatrix}, ~~
m^{45} = \frac{1}{2} 
\begin{pmatrix}
0 & \sigma_y \\
-\sigma_y & 0 
\end{pmatrix} .  
\ee
With the Pauli matrices, $m^{ab}$ are represented as 
\begin{align}
&m^{12}=-\frac{1}{2}\begin{pmatrix}
\sigma_x & 0 \\
0 & \sigma_x
\end{pmatrix}, ~~m^{13}=-\frac{1}{2}\begin{pmatrix}
1_2 & 0 \\
0 & 1_2
\end{pmatrix}, ~~m^{14}=i\frac{1}{2}\begin{pmatrix}
-\sigma_z & 0 \\
0 & \sigma_z
\end{pmatrix}, ~~m^{15}=i\frac{1}{2}\begin{pmatrix}
0 & \sigma_z\\
\sigma_z & 0 
\end{pmatrix},~~m^{23}=-i\frac{1}{2}\begin{pmatrix}
\sigma_z & 0 \\
0 & \sigma_z
\end{pmatrix},\nn\\
&m^{24}=\frac{1}{2}\begin{pmatrix}
1_2 & 0 \\
0 & -1_2
\end{pmatrix}, ~~m^{25}=-\frac{1}{2}\begin{pmatrix}
0 & 1_2 \\
1_2 & 0
\end{pmatrix}, ~~m^{34}=\frac{1}{2}\begin{pmatrix}
\sigma_x & 0 \\
0 & -\sigma_x
\end{pmatrix}, ~~m^{35}=-\frac{1}{2}\begin{pmatrix}
0 & \sigma_x \\
\sigma_x & 0 
\end{pmatrix},~~m^{45}=\frac{1}{2}\begin{pmatrix}
0 & \sigma_y \\
-\sigma_y & 0 
\end{pmatrix}. \label{sexpso23mab}
\end{align}
$m^{ab}$ satisfy the relation 
\be
m^{ab} \mathcal{E} m^{cd}- m^{cd} \mathcal{E} m^{ab} = ig^{ac}m^{bd}-ig^{ad}m^{bc}+ig^{bd}m^{ac}-ig^{bc}m^{ad} . 
\ee

\subsection{Transformation to the $sp(4; \mathbb{R})$ matrices}\label{sec:transfsp4r}

The $SO(2,3)$ invariant matrix $k=\begin{pmatrix}
\sigma_z & 0 \\
0 & \sigma_z
\end{pmatrix}$ can be transformed to the form (\ref{kmatsympano}):  
\be
K_{4} = W k W =\begin{pmatrix}
1_2 & 0 \\
0 & -1_2
\end{pmatrix}
\ee
with  
\be
W=\begin{pmatrix}
1 & 0 & 0 & 0 \\
0 & 0 & 1 & 0 \\
0 & 1 & 0 & 0 \\
0 & 0 & 0 & 1
\end{pmatrix}. 
\ee
The unitary matrix 
\be
W \Omega =  \begin{pmatrix}
1 & 0 & 0 & 0 \\
0 & 0 & 1 & 0 \\
0 & 1 & 0 & 0 \\
0 & 0 & 0 & 1
\end{pmatrix}\begin{pmatrix}
0 & 1 & 0 & -i \\
1 & 0 & -i & 0 \\
0 & 1 & 0 & i \\
1 & 0 & i & 0
\end{pmatrix}=\begin{pmatrix}
0 & 1 & 0 & -i \\
0 & 1 & 0 & i \\
1 & 0 & -i & 0 \\
1 & 0 & i & 0
\end{pmatrix}, 
\ee
transforms the $so(2,3)$ matrices $\sigma^{ab}$ (\ref{so23eachgenemat}) to $sp(4; \mR)$ matrices as 
\be
t^{ab} = (W\Omega)^{\dagger} \sigma^{ab} (W\Omega) 
\label{transtabsab}
\ee
where 
\begin{align}
&t^{12} = i\frac{1}{2}\begin{pmatrix}
0 & 1_2 \\
-1_2 & 0 
\end{pmatrix}, ~t^{13} = i\frac{1}{2}\begin{pmatrix}
0 & -1_2 \\
-1_2 & 0 
\end{pmatrix}, ~t^{14} = -i\frac{1}{2}\begin{pmatrix}
\sigma_z & 0 \\
0 & -\sigma_z 
\end{pmatrix}, ~t^{15} = -i\frac{1}{2}\begin{pmatrix}
\sigma_x & 0 \\
0 & -\sigma_x 
\end{pmatrix}, \nn\\
&t^{23} = i\frac{1}{2}\begin{pmatrix}
1_2 & 0  \\
0 & -1_2  
\end{pmatrix}, ~t^{24} = -i\frac{1}{2}\begin{pmatrix}
0 & \sigma_z \\
\sigma_z & 0 
\end{pmatrix}, ~t^{25} = -i\frac{1}{2}\begin{pmatrix}
0 & \sigma_x \\
\sigma_x & 0 
\end{pmatrix}, \nn\\
&t^{34} = -\frac{1}{2}\begin{pmatrix}
0 & \sigma_z \\
-\sigma_z & 0 
\end{pmatrix}, ~t^{35} = \frac{1}{2}\begin{pmatrix}
0 & i\sigma_x \\
-i\sigma_x & 0 
\end{pmatrix}, ~t^{45} = \frac{1}{2}\begin{pmatrix}
\sigma_y & 0 \\
0 & \sigma_y
\end{pmatrix}.  
\end{align}
They are pure imaginary matrices satisfying the condition of the $sp(4; \mathbb{R})$ algebra (\ref{sprelatgene}).  (\ref{transtabsab}) implies the isomorphism 
\be
sp(4; \mathbb{R})~\simeq ~so(2,3). 
\ee

\section{$Sp(4; \mR)$ squeeze matrix from the non-compact Hopf spinor}\label{app:relhopfmat}
 
 The non-compact 2nd Hopf  spinor $\psi_{(L)}\equiv \psi$ (\ref{2ndhopfexaplit}) satisfies 
\be
x^a \gamma_a \psi_{(L)} =+\psi_{(L)}
\ee
and 
\be
{\psi_{(L)}}^{\dagger}k \psi_{(L)}=1. \label{condpsill}
\ee
(\ref{condpsill}) implies that the geometry of $\psi_{(L)}$ is  
$H^{4,3}$. 
We introduce its charge conjugation spinor as 
\be
\psi_{(L)}^{C} \equiv C {\psi_{(L)}}^*, 
\ee
whose ``chirality'' is also +: 
\be
x^a \gamma_a \psi_{(L)}^C=+\psi^C_{(L)}. \label{chargeconpsi}
\ee
In (\ref{chargeconpsi}), we utilized the property of the charge conjugation matrix $C=\begin{pmatrix}
\sigma_x & 0 \\
0 & \sigma_x
\end{pmatrix}$ : 
\be
C\gamma^a C={\gamma^a}^*.
\ee
The original spinor $\psi_{(L)}$ and its charge conjugation is orthogonal: 
\be
{\psi_{(L)}}^{\dagger}k \psi_{(L)}^C=0, 
\ee
with $k=\begin{pmatrix}
\sigma_z & 0 \\
0 & \sigma_z
\end{pmatrix}$. 
The ``normalization'' is given by 
\be
{\psi^C_{(L)}}^{\dagger} k \psi^C_{(L)}=-1. 
\ee
The corresponding negative chirality spinor can also be constructed as 
\be
\psi_{(R)}  =\gamma^4 \psi'=\begin{pmatrix}
\psi'_3\\
\psi'_4 \\
\psi'_1 \\
\psi'_2
\end{pmatrix} , 
\ee
where 
\be
\psi'(\theta, \rho,  \chi, \phi) \equiv \psi(\theta,  \rho, \chi, -\phi) . 
\ee
Using the explicit form of $\psi$ (\ref{2ndhopfexaplit}), we can readily show  
\be
x^a \gamma_a \psi_{(R)} =-\psi_{(R)}. 
\ee
Similarly, $\psi_{(R)}^C \equiv C {\psi_{(R)}}^*$  satisfies 
\be
x^a \gamma_a \psi_{(R)}^C =-\psi_{(R)}^C. 
\ee
``Normalizations'' of the negative chirality spnors are given by 
\be
{\psi^C_{(R)}}^{\dagger} k \psi_{(R)}=1,~~{\psi^C_{(R)}}^{\dagger} k \psi^C_{(R)}=-1. \label{northogophopf}
\ee
Arbitrary  pairs of four spinors, $\psi_{(L)}$, $\psi^C_{(L)}$, $\psi_{(R)}$, $\psi^C_{(R)}$, are orthogonal in the sense 
\be
  {\psi_{(L)}}^{\dagger}k {\psi_{(R)}} =  {\psi_{(L)}}^{\dagger}k \psi_{(R)}^C={\psi_{(L)}^C}^{\dagger}k\psi_{(R)} = \cdots =0. \label{orthogophopf}
\ee
With the four $SO(2,3)$ spinors, the following $4\times 4$ matrix is constructed: 
\be
\mathcal{M}(\xi, \chi, \rho, \phi) = \begin{pmatrix} 
\psi_{(L)} & \psi^C_{(L)} & \psi_{(R)} & \psi_{(R)}^C
\end{pmatrix}=\begin{pmatrix}
\psi_1  & \psi_2^* & \psi'_3 & {\psi_4'}^* \\
\psi_2  & \psi_1^* & \psi'_4 & {\psi_3'}^* \\
\psi_3  & \psi_4^* & \psi'_1 & {\psi_2'}^* \\
\psi_4  & \psi_3^* & \psi'_2 & {\psi_1'}^* \\ 
\end{pmatrix} ,   \label{44matspinmat}
\ee
which satisfies 
\be
\det (\mathcal{M})=1. \label{detm1}
\ee
Though  $\mathcal{M}$ may be a $4\times 4$ matrix, its four columns are eventually obtained by $\psi_{(L)}$, and so $\mathcal{M}$ carries the same degrees of freedom of $\psi_{(L)}$, which is $H^{4,3}$.    
The coordinates on $H^{2,2}$ are extracted by the map    
\be
x^a =\frac{1}{4}\tr(k^5 \mathcal{M}^{\dagger}k_a \mathcal{M})  =\frac{1}{4}\tr(\mathcal{M}^{-1}\gamma^a\mathcal{M}). \label{xfromxmathm}
\ee
Obviously $x^a$ (\ref{xfromxmathm}) are invariant under the $SU(1,1)$ transformation 
\be
\mathcal{M}~~\rightarrow~~\mathcal{M} \begin{pmatrix}
g(\sigma, \omega, \psi) & 0 \\
0 & g(\sigma,\omega,-\psi)
\end{pmatrix}. ~~~~~~(g \in SU(1,1))
\ee
This implies that the map (\ref{xfromxmathm}) realizes the 2nd non-compact Hopf map:   
\be
H^{2,2} ~\simeq~H^{4,3}/SU(1,1) ~\simeq ~H^{4,3}/H^{2,1}. 
\ee

Another way to find the degrees of freedom  of  $\mathcal{M}$  is as follows.   
With respect to $4\times 4$ matrix ${M}$, the orthonormal properties, (\ref{northogophopf}) and (\ref{orthogophopf}),  are expressed as 
\be
{M}^{\dagger}k {M}=k, \label{prommnor}
\ee
and the 2nd non-compact Hopf map (\ref{2ndnoncomhopfmap}) is given by 
\be
x^a \gamma_a M =M\gamma^5. \label{propM}
\ee
From (\ref{prommnor}) and (\ref{propM}), we have 
$M^{-1} =k M^{\dagger}k$   
 and 
\be
x^a M^{\dagger} k^a M=k^5.  
\label{xmkmkg5}
\ee
Here $k^a$ are given by (\ref{ksfiveexpli}). 
Both (\ref{prommnor}) and (\ref{xmkmkg5}) are  invariant under the right action  $H$ to $M$ :  
\be
M ~~\rightarrow ~~M\cdot H, 
\ee
where $H$ is a 4$\times$4 matrix subject to    
\be
H^{\dagger} k H=k,~~~~H^{\dagger} k^5 H=k^5, \label{hbacondhk}
\ee
which is identified with the $SU(1,1)_L\otimes SU(1,1)_R$  
group element 
\be
H=\begin{pmatrix}
H_L &  0 \\
0 & H_R
\end{pmatrix}.~~~~~~~({H_L}^{\dagger}\sigma_z H_L=\sigma_z, ~~~{H_R}^{\dagger}\sigma_z H_R=\sigma_z)
\ee
Meanwhile, $SO(2,3)$ transformations act to $M$ from the left. This situation occurs in  non-linear sigma model with  $SO(2,3)$  global symmetry and  
 $SU(1,1)\otimes SU(1,1) \simeq SO(2,2)$  gauge symmetry. In the language of field theory,  the coset manifold for the non-linear field $M$ is accounted for by   
\be
SO(2,3)/SO(2,2) \simeq H^{2,2}. 
\ee
Again, we thus find that $M$ represents the $H^{2,2}$ geometry. 
 
The $SO(2,3)$ group element that signifies  $H^{4,3}$  is given by\footnote{The explicit form of (\ref{hinvu1h}) is 
\begin{align}
&g_{H^{4,3}}(\rho,\chi,\phi,\theta, \sigma,\omega,\psi) =\nn\\
&\begin{pmatrix}
e^{-i\frac{1}{2}(\phi+\chi-\psi-\omega)}~\text{c}\frac{\theta}{2}~\text{ch}\frac{\rho-\sigma}{2} & ie^{-i\frac{1}{2}(\phi+\chi+\psi+\omega)}~\text{c}\frac{\theta}{2}~\text{sh}\frac{\rho-\sigma}{2} & -ie^{-i\frac{1}{2}(\phi+\chi+\psi-\omega)}~\text{s}\frac{\theta}{2}~\text{ch}\frac{\rho+\sigma}{2} & -e^{-i\frac{1}{2}(\phi+\chi-\psi-\omega)}~\text{s}\frac{\theta}{2}~\text{sh}\frac{\rho+\sigma}{2} \\
-ie^{i\frac{1}{2}(\phi+\chi+\psi+\omega)}~\text{c}\frac{\theta}{2}~\text{sh}\frac{\rho-\sigma}{2} & e^{i\frac{1}{2}(\phi+\chi-\psi-\omega)}~\text{c}\frac{\theta}{2}~\text{ch}\frac{\rho-\sigma}{2} & -e^{i\frac{1}{2}(\phi+\chi-\psi+\omega)}~\text{s}\frac{\theta}{2}~\text{sh}\frac{\rho+\sigma}{2} & ie^{i\frac{1}{2}(\phi+\chi+\psi-\omega)}~\text{s}\frac{\theta}{2}~\text{ch}\frac{\rho+\sigma}{2} \\
-ie^{i\frac{1}{2}(\phi-\chi+\psi+\omega)}~\text{s}\frac{\theta}{2}~\text{ch}\frac{\rho+\sigma}{2} & -e^{i\frac{1}{2}(\phi-\chi-\psi-\omega)}~\text{s}\frac{\theta}{2}~\text{sh}\frac{\rho+\sigma}{2} & e^{i\frac{1}{2}(\phi-\chi-\psi+\omega)}~\text{c}\frac{\theta}{2}~\text{ch}\frac{\rho-\sigma}{2} & ie^{i\frac{1}{2}(\phi-\chi+\psi-\omega)}~\text{c}\frac{\theta}{2}~\text{sh}\frac{\rho-\sigma}{2} \\
-e^{-i\frac{1}{2}(\phi-\chi-\psi-\omega)}~\text{s}\frac{\theta}{2}~\text{sh}\frac{\rho+\sigma}{2} & ie^{-i\frac{1}{2}(\phi-\chi+\psi+\omega)}~\text{s}\frac{\theta}{2}~\text{ch}\frac{\rho+\sigma}{2} & -ie^{-i\frac{1}{2}(\phi-\chi+\psi-\omega)}~\text{c}\frac{\theta}{2}~\text{sh}\frac{\rho-\sigma}{2} & e^{-i\frac{1}{2}(\phi-\chi-\psi+\omega)}~\text{c}\frac{\theta}{2}~\text{ch}\frac{\rho-\sigma}{2} 
\end{pmatrix}, 
\end{align}
where $\text{c}\frac{\theta}{2}\equiv \cos\frac{\theta}{2}$,  $\text{s}\frac{\theta}{2}\equiv \sin\frac{\theta}{2}$,  $\text{ch}\frac{\rho}{2}\equiv \cosh\frac{\rho}{2}$ and  $\text{sh}\frac{\rho}{2}\equiv \sinh\frac{\rho}{2}$. }  
\be
g_{H^{4,3}}(\rho,\chi,\phi,\theta, \sigma,\omega,\psi) =H(\rho,\chi,\phi)^{-1}~ U_1(\theta)~ H(\sigma,\omega,\psi) \label{hinvu1h}
\ee
where 
\be
U_1(\theta) =e^{i\theta\Sigma^{34}}, ~~~~H(\rho,\chi,\phi) =e^{i\rho\Sigma^{13}}e^{-i\chi\Sigma^{12}} e^{i\phi \Sigma^{34}}, 
\ee
and the Dirac-type (\ref{coorh22xsspolar}) and Schwinger-type (\ref{polarmathcalmmat}) squeeze matrices are respectively realized as  
\begin{align}
&M=g_{H^{4,3}}( \rho, \chi, \phi, \theta, \sigma, \omega, \psi )|_{(\sigma, \omega, \psi)=-(\rho, \chi, \phi)},  \nn\\
&\mathcal{M}=g_{H^{4,3}}( \rho, \chi, \phi, \theta, \sigma, \omega, \psi )|_{\sigma= \omega= \psi=0}. 
\end{align}


\end{document}